\documentclass[sigconf]{acmart}
\AtBeginDocument{%
  }
\setcopyright{acmlicensed}
\copyrightyear{2026}
\acmYear{2026}
\setcopyright{cc}
\setcctype{by}
\acmConference[KDD 2026] {Proceedings of the 32nd ACM SIGKDD Conference on Knowledge Discovery and Data Mining V.2}{August 9--13, 2026}{Jeju Island, Republic of Korea.}
\acmBooktitle{Proceedings of the 32nd ACM SIGKDD Conference on Knowledge Discovery and Data Mining V.2 (KDD 2026), August 9--13, 2026, Jeju Island, Republic of Korea}
\acmISBN{979-8-4007-2259-2/2026/08}
\acmDOI{10.1145/3770855.3817497}
\usepackage{float}
\usepackage{longtable}
\usepackage[utf8]{inputenc} % allow utf-8 input
\usepackage[T1]{fontenc}    % use 8-bit T1 fonts
\usepackage{booktabs}       % professional-quality tables
\usepackage{amsfonts}       % blackboard math symbols
\usepackage{nicefrac}       % compact symbols for 1/2, etc.
\usepackage{microtype}      % microtypography
\usepackage{enumitem}
\usepackage{subfigure}
\usepackage{multirow}
\usepackage{float}
\usepackage{graphicx}
\usepackage{rotating,bm}
\usepackage{algorithm}
\usepackage{algorithmic}
\usepackage{physics}
\usepackage{pifont}
\usepackage{amsmath}
\usepackage{bm}
\usepackage{graphicx}
\usepackage{subcaption}
\usepackage{wrapfig,lipsum,booktabs}
\usepackage[table]{xcolor} % 提供颜色功能并支持表格颜色
\usepackage{colortbl}      % 提供表格颜色功能
\usepackage{wrapfig,lipsum}
\usepackage{tabularx}

\definecolor{LightSteelBlue4}{RGB}{96,123,139}
\definecolor{DodgerBlue4}{RGB}{16,78,139}
\definecolor{Turquoise4}{RGB}{0,134,139}
\definecolor{Green4}{RGB}{0,139,0}
\definecolor{Brown3}{RGB}{205,85,85}
\definecolor{Azure3}{RGB}{193,205,205}
\usepackage{subcaption}
\usepackage{tikz}
\usepackage{caption}
\usepackage{subcaption}
\usetikzlibrary{calc}
\usepackage{hyperref}
\hypersetup{
    colorlinks=true,
    linkcolor={red},
    citecolor={blue},
    urlcolor={red}
}

\usepackage[labelfont={color=DodgerBlue4}]{caption}

\tikzstyle{prompt} = [rectangle,
text centered, 
minimum width = 2cm,
minimum height = 0.3cm,
font=\fontfamily{CascadiaCode-TLF}\selectfont, %
fill=LightSteelBlue4,
text=white
]

\tikzstyle{llm} = [rectangle, rounded corners,
text centered, 
draw=DodgerBlue4,
text=DodgerBlue4,
minimum width = 2cm,
minimum height = 0.6cm,
font=\small\sffamily, %
line width=1.0pt, 
fill={rgb,255:red,223;green,236;blue,248} 
]

\tikzstyle{resp} = [rectangle, %
text centered, 
draw=none,
font=\fontfamily{CascadiaCode-TLF}\selectfont,
fill=Turquoise4,
text=white
]

\tikzstyle{correct} = [rectangle, inner sep=4pt, 
text centered, 
draw=Green4,
text=Green4,
line width=1.0pt,
font=\large,
minimum width = 1.65cm,
minimum height = 0.3cm,
fill={rgb,255:red,240;green,255;blue,240}
]

\tikzstyle{wrong} = [rectangle, inner sep=4pt,
text centered, 
draw=Brown3,
text=Brown3,
line width=1.0pt,
font=\large,
minimum width = 1.65cm,
minimum height = 0.3cm,
fill={rgb,255:red,255;green,240;blue,240}
]

\tikzstyle{arrow} = [->,>=stealth,
line width=1.0pt,
draw=Azure3,
fill=Azure3
]

\tikzstyle{textlabel} = [font=\footnotesize\itshape]

\tikzstyle{sresp}=[resp,rotate=90,font=\small\fontfamily{CascadiaCode-TLF}\selectfont,inner sep=2pt]

\newcommand{\cmark}{\textcolor{green!70!black}{\checkmark}}
\newcommand{\xmark}{\textcolor{red}{\ding{55}}}
\definecolor{textpink}{HTML}{F18078}
\newcommand{\sol}{\mathcal{S}}
\def\method{FD-Bench}
\newcounter{baselines}
\renewcommand{\arraystretch}{0.85}

\begin{document}

\title{\method: A Modular and Fair Benchmark for Data-driven \\ Fluid Simulation}


\author{Haixin Wang}
\authornote{These authors contributed equally to this research.}
\email{whx@ucla.edu}
\affiliation{%
  \institution{University of California, Los Angeles}
  \country{Los Angeles, CA, USA}
}

\author{Ruoyan Li}
\authornotemark[1]
\email{liruoyan2002@cs.ucla.edu}
\affiliation{%
  \institution{University of California, Los Angeles}
  \country{Los Angeles, CA, USA}}

\author{Fred Xu}
\authornotemark[1]
\email{xuzeyuanfred@gmail.com}
\affiliation{%
  \institution{University of California, Los Angeles}
  \country{Los Angeles, CA, USA}}

\author{Fang Sun}
\email{fts@cs.ucla.edu}
\affiliation{%
  \institution{University of California, Los Angeles}
  \country{Los Angeles, CA, USA}}

\author{Kaiqiao Han}
\email{kqhan@cs.ucla.edu}
\affiliation{%
  \institution{University of California, Los Angeles}
  \country{Los Angeles, CA, USA}
  }

\author{Zijie Huang}
\email{zijiehjj@gmail.com}
\affiliation{%
  \institution{Meta}
  \country{Menlo Park, CA, USA}}

\author{Chi Chang}
\email{blacksnail789521@gmail.com}
\affiliation{%
  \institution{University of California, Los Angeles}
  \country{Los Angeles, CA, USA}}

\author{Xiao Luo}
\email{xiao.luo@wisc.edu}
\affiliation{%
  \institution{University of Wisconsin–Madison}
  \country{Madison, WI, USA}}

\author{Wei Wang}
\email{weiwang@cs.ucla.edu}
\affiliation{%
  \institution{University of California, Los Angeles}
  \country{Los Angeles, CA, USA}}

\author{Yizhou Sun}
\email{yzsun@cs.ucla.edu}
\affiliation{%
  \institution{University of California, Los Angeles}
  \country{Los Angeles, CA, USA}}

\renewcommand{\shortauthors}{Haixin Wang \textit{et al.}}

\newcommand{\nop}[1]{}

\begin{abstract}

Data-driven modeling of fluid dynamics has advanced rapidly with neural PDE solvers, yet a fair and strong benchmark remains fragmented due to the absence of unified PDE datasets and standardized evaluation protocols. Although architectural innovations are abundant, fair assessment is further impeded by the lack of a clear disentanglement between spatial, temporal and loss modules.
In this paper,
we introduce \textbf{\method{}}, a large-scale, modular benchmarking infrastructure that unifies datasets, evaluation protocols, and reproducible pipelines for data-driven fluid simulation.
\method{} systematically reviews and decomposes mainstream baseline models reported across recent publications, extracting and standardizing their key architectural and training components for fair, unified comparison across 10 representative flow scenarios.
It provides four key contributions: (1) a modular design that enables fair comparisons across spatial, temporal, and loss function modules; (2) the first systematic framework for direct comparison with traditional numerical solvers; (3) fine-grained generalization analysis across resolutions, initial conditions, and prediction time window; 
and (4) a user-friendly, extensible codebase to support future research. Through rigorous empirical studies, \method{} establishes the most comprehensive leaderboard to date, resolving long-standing issues in reproducibility and comparability, and laying a foundation for robust evaluation of future data-driven fluid models. The code is open-sourced at \underline{\url{https://github.com/WillDreamer/FD-Bench}}.
\end{abstract}




\maketitle

\section{Introduction}

The accurate simulation of complex fluid dynamical systems governed by partial differential equations (PDEs) is a cornerstone of scientific and engineering applications, ranging from aerodynamics~\citep{o2022neural,deng2023prediction,wu2023spatio,mufti2024shock,wu2025advanced}, chemical engineering~\citep{cao2018liquid,cao2019liquid,zhang2022multiscale,sun2024graph,hu2025omni}, biology~\citep{yin2022simulating,voorter2023improving,shen2023multiple}, and environmental science~\citep{wen2022u,pathak2022fourcastnet,bi2023accurate,rajagopal2023evaluation,zhang2024survey}. Traditional numerical solvers have achieved remarkable success in offering high-fidelity solutions grounded in well-established mathematical formulations and rigorous convergence guarantees.
Recent advancements in machine learning have introduced numerous neural solvers for data-driven fluid simulation~\citep{wang2024recent}, which have enabled efficient prediction of complex dynamics with reduced computational cost. These methods have shown strong potential in capturing nonlinear behaviors beyond the reach of traditional solvers. 

Despite rapid progress, the field faces significant challenges in establishing consistent evaluation standards.
Researchers in both the field of AI and traditional Computational Fluid Dynamics (CFD) often face three pressing questions. 
\textbf{First}, how effective are current neural solvers, comparing against traditional numerical solvers in terms of accuracy? \textbf{Second}, how efficient are current neural solvers, and to what extent can they accelerate simulations compared to traditional numerical methods?
\textbf{Third}, and perhaps most critically, how well do these neural solvers generalize when applied to more practical and realistic scenarios? Therefore, as recently emphasized by \citep{brandstetter2025envisioning}, recent analyses reveal persistent challenges in making fair comparisons to prior methods. Advancing the field will require stronger and more consistent benchmark problems.

\begin{figure}[!t]
    \centering
\includegraphics[width=0.48\textwidth]{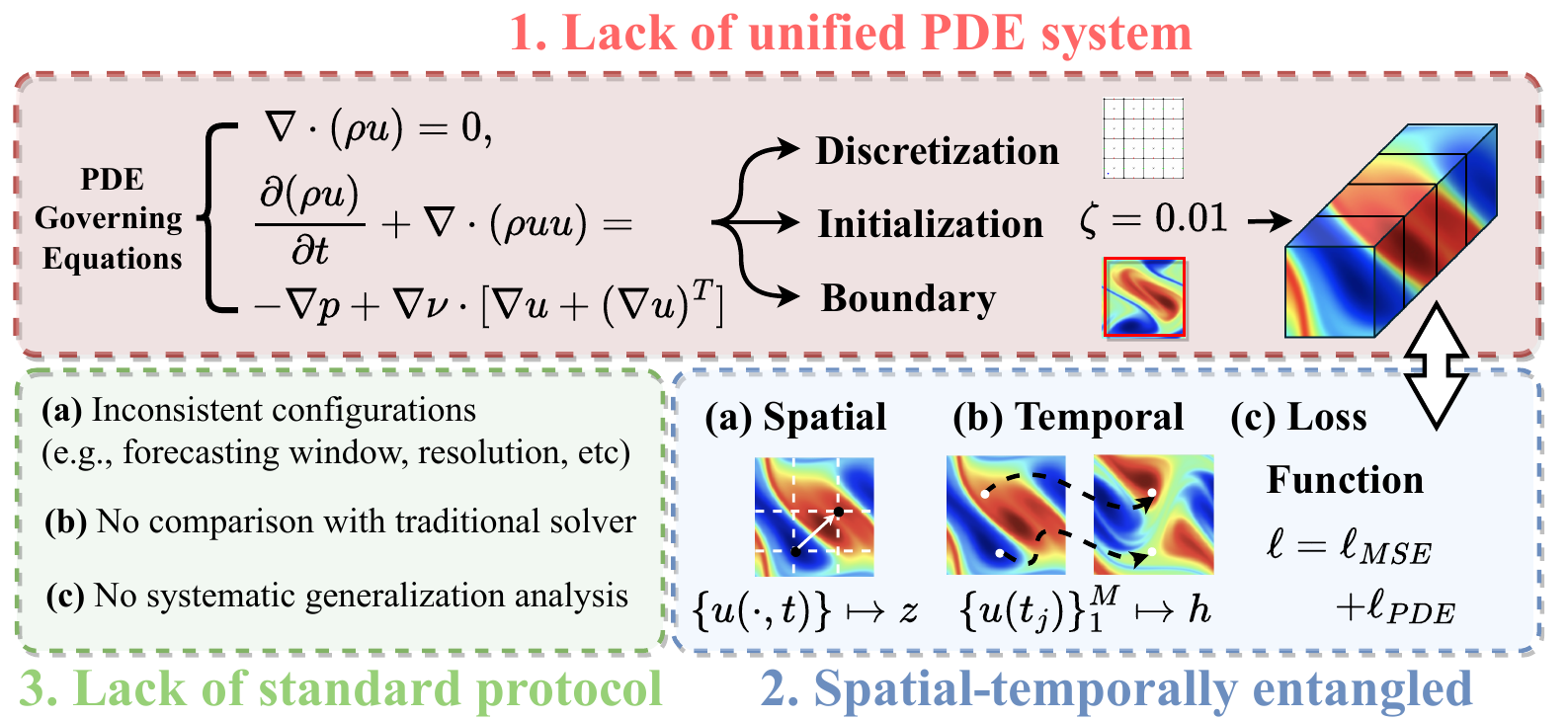}
    \caption{Limitations identified in three key areas.} 
    \label{fig:limitation}
\end{figure}

To date, the three aforementioned questions remain unanswered, largely due to the inherent infinite-dimensional continuity and variability of fluid dynamics, which makes definitive analysis challenging.
Significant limitations remain in three key areas, as illustrated in Figure~\ref{fig:limitation}, and are detailed below.

\begin{table*}[t]
\centering
\small
\setlength{\tabcolsep}{0.5mm}{
\rowcolors{2}{gray!5}{white}
\begin{tabular}{lcccccccc}
\toprule
\textbf{Benchmark} 
& \begin{tabular}[c]{@{}c@{}}\textbf{1.} Different \\ grid discretization\end{tabular} 
& \begin{tabular}[c]{@{}c@{}}\textbf{2.} Long-term \\ rollout \end{tabular} 
& \begin{tabular}[c]{@{}c@{}}\textbf{3.} Numerical \\ method comp.\end{tabular} 
& \begin{tabular}[c]{@{}c@{}}\textbf{4.} Temporal \\ modular comp. \end{tabular} 
& \begin{tabular}[c]{@{}c@{}}\textbf{5.} Spacial \\ modular comp. \end{tabular} 
& \begin{tabular}[c]{@{}c@{}}\textbf{6.} Loss \\ modular comp. \end{tabular}
& \begin{tabular}[c]{@{}c@{}}\textbf{7.} Generalization \\ study on conditions\end{tabular}
& \# Baselines \\
\midrule
PDEBench~\citep{takamoto2022pdebench}     & \xmark & \cmark & \cmark & \xmark & \xmark & \xmark & \xmark & 3 \\
CFDBench~\citep{luo2023cfdbench}          & \xmark & \cmark & \xmark & \xmark & \xmark & \xmark & \cmark & 9 \\
DiffBench~\citep{kohl2023benchmarking}    & \xmark & \cmark & \xmark & \xmark & \xmark & \xmark & \xmark & 6 \\
PDENNEval~\citep{wei2024pdenneval}        & \xmark & \cmark & \xmark & \xmark & \xmark & \cmark & \cmark & 12 \\
PINNacle~\citep{hao2023pinnacle}          & \xmark & \cmark & \xmark & \xmark & \xmark & \xmark & \cmark & 12 \\
Well~\citep{ohana2024well}                & \xmark & \cmark & \xmark & \xmark & \xmark & \xmark & \xmark & 4 \\
Posiden~\citep{herde2024poseidon}         & \xmark & \cmark & \xmark & \xmark & \cmark & \xmark & \cmark & 7 \\
ML-PDE~\citep{mcgreivy2024weak}           & \xmark & \xmark & \cmark & \xmark & \xmark & \xmark & \xmark & 10 \\
InverseBench~\citep{zheng2025inversebench}& \xmark & \cmark & \xmark & \xmark & \xmark & \xmark & \cmark & 14 \\
\textbf{FD-Bench}                         & \cmark & \cmark & \cmark & \cmark & \cmark & \cmark & \cmark & \textbf{\arabic{baselines}} \\
\bottomrule
\end{tabular}}

\caption{ A comprehensive comparison between our modular benchmarking protocol and existing benchmarks. Each column represents a key evaluation dimension, where ``comp." stands for ``comparison". \# Baselines indicates the number of existing methods reproduced and evaluated within each benchmark.
}
\label{tab:taxnomy}
\vspace{-0.5cm}
\end{table*}

\noindent\textcolor[HTML]{A63C3C}{\textbf{(1) Lack of unified PDE systems.}} From a data perspective, the diversity of benchmark settings, ranging in governing equations, domain geometries, spatial resolutions, temporal discretizations, and flow conditions, hinders fair comparison across methods. This variability often results in fragmented and non-reproducible findings. 
As shown in Table \ref{tab:taxnomy}, the benchmarks rely on largely different datasets, which makes it difficult to draw reliable and directly comparable conclusions.

\noindent\textcolor[HTML]{6C8EBF}{\textbf{(2) Entangled spatial-temporal modeling and loss design.}}
Capturing fluid dynamics requires architectures that jointly model spatial structures and temporal evolution while maintaining physical consistency. However, many neural solvers entangle components from diverse domains without clear attribution, and lack systematic ablations, making it difficult to isolate the source of performance gains. For example, FNO~\citep{li2020fourier} learns spatiotemporal features by stacking the time dimension in the frequency domain, while MP-PDE~\citep{brandstetter2022message} introduces temporal bundling to enhance temporal evolution. These differing design choices make it nearly impossible to directly compare their individual contributions.

\noindent\textcolor[HTML]{82B366}{\textbf{(3) Lack of standardized evaluation protocol.}} From an evaluation standpoint, the dynamical nature of fluid systems, characterized by sensitivity to initial conditions and parameter settings. This reveals three primary limitations:
\textcolor[HTML]{82B366}{\textbf{(3.1)}} Inconsistent experimental configurations. It is common for different studies to adopt distinct experimental setups, including variations in discretization, flow field conditions, and forecasting windows. Moreover, the evaluation metrics are inconsistent across studies, which further limits the comparability of different methods.
\textcolor[HTML]{82B366}{\textbf{(3.2)}} Limited benchmarking against traditional solvers. Many neural solvers lack rigorous comparisons with classical numerical solvers, particularly under settings where accuracy, stability, and computational efficiency must be jointly evaluated.
\textcolor[HTML]{82B366}{\textbf{(3.3)}} Absence of systematic generalization analysis. Few studies examine how models generalize across different resolutions, conditions, or parameter regimes.

To overcome these limitations, we introduce \textbf{\method{}}, a rigorously designed and fully reproducible benchmark. 
Rather than merely re-implementing prior work, \method{} systematically reviews and decomposes existing baseline models reported across recent publications (see Appendix \ref{sec:decomposition}), extracting and standardizing their key architectural and training components for fair, unified comparison across 10 representative flow scenarios.

\begin{figure*}[!h]
 \begin{minipage}{0.4\textwidth} 
 \centering
 \fontsize{8.2pt}{\baselineskip}\selectfont %
 \renewcommand\tabcolsep{1.0pt} %
 \renewcommand\arraystretch{0.8} %
 \resizebox{\textwidth}{!}{\begin{tabular}{l | c | c }

\toprule 

\bf Name & \bf Shape & \bf Size \\

\midrule

Incompressible N-S & $\{$1200, 1000, 256, 256$\}$ & 2.3T \\
Compressible N-S & $\{$42k, 4, 4, 128, 128 $\}$ & 222G  \\

Stochastic N-S & $\{$100, 1k, 2, 128,128$\}$ & 12G \\
Kolmogorov Flow & $\{$20k, 21, 5, 128, 128$\}$ &17G  \\
Diffusion-Reaction &$\{$1k, 100, 2, 128,128$\}$ & 13G \\
Taylor-Green Vortex & $\{$204, 126, 2, 10000$\}$ & 1.8G \\
Reverse Poiseuille Flow & $\{$1, 30000, 2, 12800$\}$ & 2.9G \\
Advection & $\{$1200, 1000, 256, 256$\}$ & 2.3T \\
Lid-driven Cavity Flow & $\{$1, 20000, 2, 11236$\}$ & 3.81T \\
Burgers & $\{$1200, 1000, 2, 256, 256$\}$ & 3.7T \\
\midrule
Total & - & \textbf{8.6T} \\
\bottomrule
\end{tabular}}
\captionof{table}{\textbf{Statistics} of flows. The ``shape” column represents, in order: the trajectory sample number, time steps, feature channels, and the flow field resolution. }
\label{tab:main_statistics}
\end{minipage} 
\vspace{-0.3cm}
\hfill
\begin{minipage}{0.55\textwidth}
\centering
\vspace{-4mm}
\hspace{-5mm}
\includegraphics[width=\textwidth]{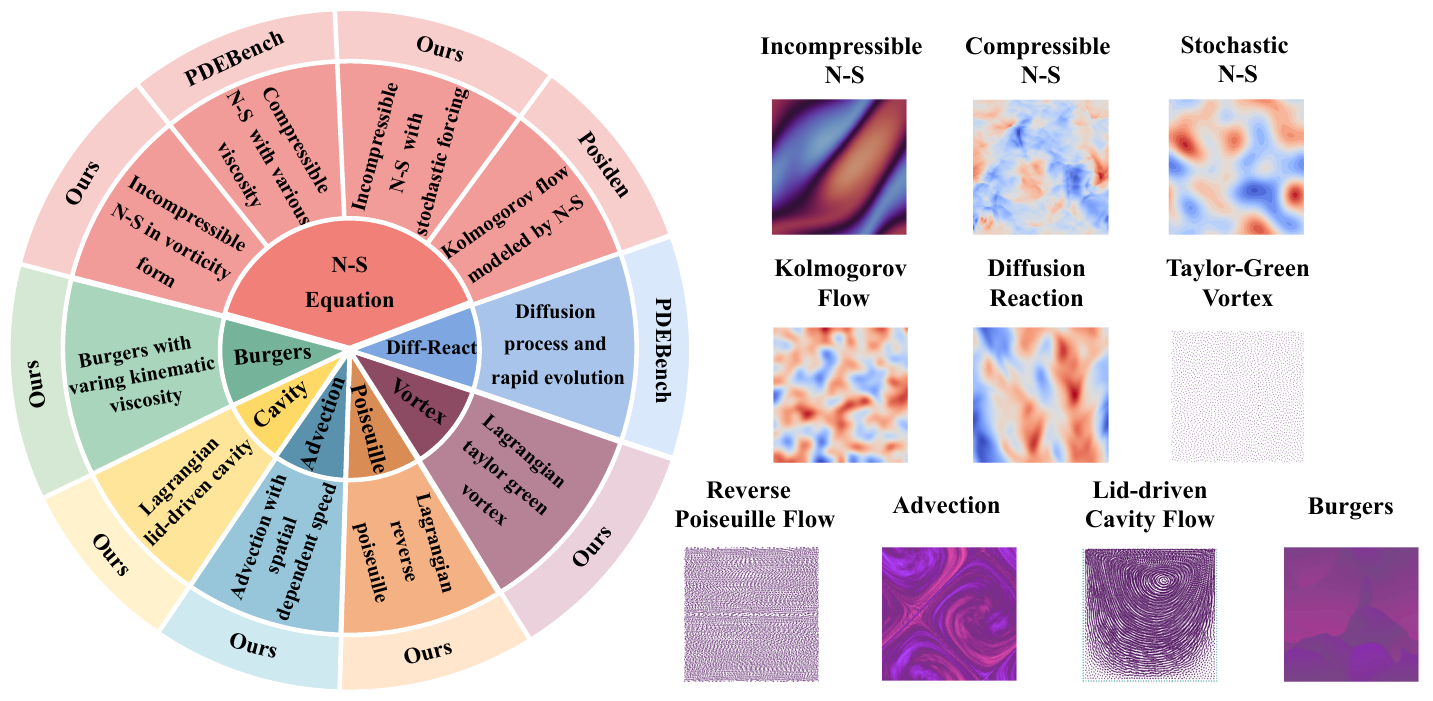}
\vspace{-2.75mm}
\caption{We collect and generate 10 representative fluid flow scenarios that span a diverse range of physical conditions. We also present the corresponding visualizations.}
\label{fig:source_dataset}
\end{minipage}
\vspace{-0.3cm}
\end{figure*}

\method{} brings three key innovations to the field, each directly addressing the limitations above. 
\textbf{\underline{First}}, we collect and generate 10 representative fluid flow scenarios that span a diverse range of physical conditions shown in Figure~\ref{fig:source_dataset}. Building on these datasets, we enable fair and direct comparisons among diverse neural solvers as well as against traditional numerical solvers for the first time. Besides, this unified system makes it possible to conduct fine-grained generation analysis, revealing how variations in initial conditions, spatial resolution, prediction time window, and model capacity affect performance.
\textbf{\underline{Second}}, \method{} enables modular and fair comparisons by isolating spatial, temporal, and loss function modules, thereby eliminating confounding factors introduced by implementation-specific details. We construct the most rigorous and comprehensive leaderboard to date, providing a strong foundation for fair and reproducible evaluation in the field.
\textbf{\underline{Third}}, \method{} addresses the long-standing lack of standardized evaluation protocols by providing a fully modular, easy-to-use, and reproducible codebase. Researchers can simply specify the dataset and select the desired combination to automatically run experiments under a unified and controlled setup. This design not only ensures that all baseline methods can be faithfully reproduced with minimal configuration but also enables seamless extension to novel architectures or loss functions.

\section{Related Work and Position of \method{}}

PDEBench~\citep{takamoto2022pdebench} first significantly advanced the field by introducing standardized datasets. CFDBench~\citep{luo2023cfdbench} extended this idea to a wider set of CFD problems and emphasized generalization across flow conditions. More specialized efforts such as PINNacle~\citep{hao2023pinnacle} and PDENNEval~\citep{wei2024pdenneval} target physics-informed neural networks. More recent efforts such as Well~\citep{ohana2024well} and Posiden~\citep{herde2024poseidon} further enriched the landscape by introducing their own curated datasets and conducting detailed experimental analyses, offering deeper insights into model behavior.
However, most of them remain limited in scope: they typically focus on a narrow class of PDE systems or solver families. They just treat each baseline method as a monolithic whole, without probing the sources of its contributions, and consequently lack a fully modular decomposition of spatial, temporal, and loss components. Moreover, they seldom provide rigorous head-to-head comparisons with classical numerical solvers~\citep{mcgreivy2024weak}.

Table~\ref{tab:taxnomy} provides an overview of existing benchmarks. As shown in Table~\ref{tab:taxnomy}, most prior benchmarks focus on a limited subset of evaluation axes, typically emphasizing long-term rollout performance while overlooking other critical factors such as grid discretization variability, modular comparisons across temporal and spatial components, and systematic loss function analysis. Moreover, comparisons with traditional numerical methods and structured generalization studies are often conducted in isolation, if at all, and vary significantly across benchmarks. In contrast, \method{} is designed as a fully modular and extensible benchmarking protocol that simultaneously covers all listed dimensions. It enables consistent evaluations across different dimensions while also supporting comprehensive generalization studies under diverse conditions. As a result, \method{} not only includes the largest set of reproduced baselines among existing benchmarks but also provides a unified framework for conducting controlled, fine-grained comparisons that are difficult to achieve with prior benchmark designs.

\nop{
\noindent\textbf{Q1. Does an optimal neural architecture exist under equal computational budgets?} 
Modeling complex fluid dynamical systems often suffers from the ``coupling curse”~\citep{wu2024spatio} where spatio-temporal flow conditions are inextricably intertwined, making it difficult to pinpoint the true source of performance gains. 
\method{} proposes to systematically decouple fluid dynamical system modeling into three orthogonal dimensions to break the coupling curse, enabling direct attribution of model behavior to specific design choices. It also yields taxonomy-driven insights by aligning baselines within a unified schema. 

\noindent\textbf{Q2. Can neural solvers truly replace traditional numerical solvers?}
To obtain a fair comparison of both computational efficiency and predictive accuracy, we propose benchmarking neural solvers against traditional numerical solvers operating on coarser discretizations or using lower-order time-integration schemes (\textit{e.g.}, Euler’s method instead of fourth-order Runge–Kutta) that produce equivalent error. Thus, we extend the idea in \citet{mcgreivy2024weak} to compare the rollout error of neural solvers to that of these coarse‐grid solvers. We perform experiments on three subsets with varying parameters, evaluate their long‐horizon rollout performance, and ensure a fair comparison under identical system environments.}

\nop{
\noindent\textbf{Q3. Which spatial discretization scheme better supports their representative neural solvers?}
Neural solvers employ one of two discretization schemes: Eulerian or Lagrangian. In Eulerian approaches (grid- or mesh-based)~\citep{subramaniam2013lagrangian,10.1145/311535.311548}, the spatial domain is discretized by a fixed arrangement of nodes. In Lagrangian (particle-based) methods~\citep{muller2003particle}, discretization is realized through material points that move with the local deformation of the continuum. Existing works only compare models with the same discretization schemes.
To this end, we generate three subsets to train their representative neural solvers and compare their rollout performance.

\noindent\textbf{Q4. Whether knowledge learned from different systems boost generalization?} 
In practical applications, evaluating a neural solver’s performance in out-of-distribution scenarios is of greater significance. We provide a detailed investigation into how different modular designs perform under varying initial conditions, resolutions, model parameters, and longer rollout horizons. Specifically, we examine the zero-shot generalization ability of the models to unseen initial conditions, their convergence behavior when trained at different resolutions, their scalability with increased parameter size, and their ability to extrapolate in zero-shot rollout.
}

\vspace{-0.3cm}
\section{PDE Datasets Curation}
\label{sec:Datasets_Curation}

\subsection{Data Selection Criteria}
To ensure a comprehensive evaluation, our benchmark incorporates datasets that rigorously satisfy the following requirements. (1) \textbf{Diverse PDE types}: Inclusion of PDE families with real-world physical significance, such as the N-S equations for fluid dynamics and Burgers’ equation for shock waves. This ensures the universality of tested methods across mathematical formulations.
(2) \textbf{Heterogeneous data distributions}: Coverage of varying initial conditions (\textit{e.g.,} smooth, discontinuous, stochastic) and boundary conditions to rigorously evaluate generalization capabilities under distribution shifts.  
(3) \textbf{Multi-scale field settings}: Datasets spanning coarse-to-fine spatial resolutions and temporal discretizations, coupled with diverse numerical discretization schemes to assess resolution scalability and discretization invariance.  
(4) \textbf{Complex temporal dynamics}: Inclusion of both short-term transient behaviors (\textit{e.g.,} rapid instabilities) and long-term equilibrium states, as well as systems with stiff dynamics or temporal discontinuities, to test robustness in temporal representation.  
(5) \textbf{Irregular geometries}: Domains with non-Cartesian boundaries (\textit{e.g.,} curved and fractured) and mixed boundary types.

\subsection{Data Collection \& Generation} 

\noindent\textbf{For modular comparison}, we collect three representative flow problems from existing benchmarks, namely \underline{\texttt{Compressible N-S}} ~\citep{takamoto2022pdebench}, \underline{\texttt{Diffusion-Reaction}}~\citep{takamoto2022pdebench}, and \underline{\texttt{Kolmogorov Flow}}~\citep{takamoto2022pdebench} for fair comparison. In addition, we generate a \underline{\texttt{Stochastic N-S}} subset, driven by incompressible flow with stochastic forcing and initialized with 10 distinct initial conditions, to further enhance the diversity of our experiments. For more details, please refer to Appendix \ref{sec:modular_data}.

\noindent\textbf{For comparison with traditional numerical solvers}, we generate three subsets with varying parameters using a pseudo-spectral solver \citep{canuto1988spectral}. To simulate \underline{\texttt{Incompressible N-S}} in vorticity form, we sample five Reynolds numbers: 50, 200, 500, 1k, and 2k. To simulate \underline{\texttt{Burgers}}, we use five kinematic viscosities: $5 \times 10^{-4}$, $1 \times 10^{-3}$, $5 \times 10^{-3}$, $1 \times 10^{-3}$, and $5 \times 10^{-2}$. Finally, to simulate \underline{\texttt{Advection}}, we consider five spatially varying advection speeds. Details on the solver implementation and the PDE formulations are provided in Appendix~\ref{appendix:datasets_solver}.

\noindent\textbf{For comparison on different discretizations}, we use Smoothed Particle Hydrodynamics (SPH) \citep{gingold1977smoothed} to generate three Lagrangian particle subsets, \underline{\texttt{Taylor–Green vortex}} (Re = 100), \underline{\texttt{Lid-driven}} \underline{\texttt{cavity flow}} (Re = 100), and \underline{\texttt{Reverse Poiseuille flow}} (Re = 10), each governed by the compressible N-S equations. Particle positions are recorded every 100 time steps, and neural models are trained to predict the flow evolution over these intervals. We convert Lagrangian particle datasets into Eulerian form. For grid representations, we generate uniformly spaced nodes and aggregate particle velocity information onto these nodes. For mesh representations, we randomly sample Eulerian nodes and perform the same velocity‐aggregation procedure. Aggregation is performed with a quintic smoothing kernel. See Appendix~\ref{appendix:datasets_discretization} for additional details.

\section{FDBench System: Decoupling in Modular Design}

\label{sec:modular}
To enable a unified and fair comparison across baselines, each originally proposed under different datasets and modeling paradigms, we creatively decompose each method into four key design dimensions inspired by a Taylor expansion: Design $=$ ``\textbf{Spatial representation}'' $+$ ``\textbf{Temporal representation}'' $+$ ``\textbf{Loss function}'' $+$ Additional Technique, where the “Additional Technique” captures the distinctive architectural innovation introduced by each work.

\subsection{Spatial Representation}
\label{sec:spatial}

A key challenge is how to encode the continuous spatial field
\(\displaystyle u \colon \Omega\times[0,T]\to\mathbb{R}^{d}\).
Let
$\Omega\subset\mathbb{R}^d$ denote the physical domain,
and let $u(\bm{x},t)\in\mathbb{R}^c$ be the physical variable (\textit{e.g.,} velocity, pressure), so that $u(\bm{x},t)\in\mathbb{R}^c$ represents the value of the field at spatial location $\bm{x}\in\Omega$ and time \(t\). $c\in\mathbb{N}$ is the number of channels (for example, three velocity components and pressure). % \YS{explain c}
We collect \(N\) spatial samples
\(\{\bm{x}_i\}_{i=1}^N\subset\Omega\)
(\textit{e.g.,} grid points, mesh nodes, or particles) and denote $u(\cdot,t)= \bigl[u(\bm{x}_1,t),\,u(\bm{x}_2,t),\,\dots,\,u(\bm{x}_N,t)\bigr]^\top
  \in\mathbb{R}^{N\times c}$.
Spatial representation is any map that encodes the continuous field \(u(\bm{x},t)\) at time \(t\)
into a finite-dimensional feature \(z\) as ${\mathcal S}_{\theta}: 
  \bigl\{u(\cdot,t)\bigr\}
  \;\mapsto\;
  z
  \;\in\;
  \mathbb{R}^{D}$.
Here, we consider the commonly used spatial representation paradigms.

\noindent\textbf{Fourier.}
It transforms the spatial field to the spectral domain using frequency transform~\citep{lifourier,li2023fourier,zhang2025beyond} (\textit{e.g.,} Discrete Fourier Transform), yielding spectral coefficients $\hat u(\mathbf{k},t)=\mathcal{F}(u)$, where $\mathbf{k} = (k_1,\dots,k_d)$ denotes the discrete frequency. A learnable spectral filter $\phi_\theta(\mathbf{k})$ is applied by element-wise multiplication to the Fourier coefficients $\hat u(\mathbf{k},t)=\mathcal{F}(u(\cdot,t))$, thereby efficiently capturing long-range correlations via global mixing in physical space while decoupling spectral modes to learn coherent structures. $\mathcal{S}_{\mathrm{Fourier}}(u(\cdot,t)) = \mathcal{F}^{-1}(\phi_\theta \odot \mathcal{F}(u(\cdot,t))) \in \mathbb{R}^{N\times c}.$

\noindent\textbf{Self-attention.} It allows each spatial location to attend to all others by computing weighted combinations of features based on learned similarity scores~\citep{wu2024transolver,geneva2022transformers}, thereby capturing global interactions adaptively.
Let $Q=u(\cdot,t)W_Q$, $K=u(\cdot,t)W_K$, $V=u(\cdot,t)W_V$.
Then
\begin{equation}
\mathcal{S}_{\mathrm{SA}}(u(\cdot,t))
=\mathrm{softmax}\!\left(\frac{QK^\top}{\sqrt{d_k}}\right)V .
\end{equation}

\noindent\textbf{Convolution.}
Convolution~\citep{raonic2023convolutional} aggregates information from a local neighborhood using a shared kernel, leveraging spatial locality and translation equivariance. Let $\mathcal{H}\subset\mathbb{Z}^d$ be the set of stencil offsets. Let $c_{\mathrm{in}}=c$ and $c_{\mathrm{out}}$ be the number of input and output channels. For each offset $h\in\mathcal{H}$, the kernel weight is $\mathbf{W}_h\in\mathbb{R}^{c_{\mathrm{out}}\times c_{\mathrm{in}}}$. Then, for each spatial index $i$:
\begin{equation}
    \mathcal{S}_{\mathrm{Conv}}(u(\cdot,t))_i = \sum_{h\in\mathcal{H}} \mathbf{W}_h\;u\bigl(\bm{x}_{i+h},t\bigr) \;\in\;\mathbb{R}^{c_{\mathrm{out}}}
\end{equation}

\noindent\textbf{Graph.}
Graph convolutions~\citep{brandstetter2022message,lino2022multi,wang2024beno} generalize standard convolutional filtering to irregular domains by aggregating neighbor features weighted by graph connectivity, enabling modeling on meshes or point clouds. A typical form is:
$\mathcal{S}_{\mathrm{Graph}}(u(\cdot,t)) = \mathrm{softmax}(\tilde D^{-\tfrac12} \tilde A \tilde D^{-\tfrac12} \, u(\cdot,t) \, W) \in \mathbb{R}^{N\times D},$
where $A$ is the binary adjacency, $\tilde D^{-1/2}\tilde A\tilde D^{-1/2}$ symmetrically normalizes the adjacency to control spectral properties, and $\sigma(\cdot)$ is a pointwise nonlinearity.

\noindent\textbf{Reduced-Order Modeling (ROM).}
Typical work like Proper Orthogonal Decomposition~\citep{rojas2021reduced,wentland2023scalable} approximates $u(\cdot,t)$ by projecting onto the $K$ leading modes $\{\phi_k\}$ found via the covariance operator
\begin{equation}
    C(\bm{x},\bm{x}') = \frac{1}{M}\sum_{i=1}^M u(\bm{x},t_i)\,u(\bm{x}',t_i)
\end{equation}
and solving $\int_\Omega C(\bm{x},\bm{x}')\phi_k(\bm{x}')\,\mathrm{d}\bm{x}'$
% \begin{equation}
%      = \lambda_k\phi_k(\bm{x})
% \end{equation}
under the inner product $\langle f,g\rangle=\int_\Omega f(\bm{x})g(\bm{x})\,\mathrm{d}\bm{x}$. The projection coefficients
$\alpha_k(t) = \langle u(\cdot,t),\phi_k\rangle$
give $\mathcal{S}_{\mathrm{ROM}}(u(\cdot,t)) = [\alpha_1(t),\dots,\alpha_K(t)]^\top\in\mathbb{R}^K,$
where $M$ is the number of snapshots, $N$ the spatial samples, and $K\ll N$ the retained modes.

\noindent\textbf{Implicit Neural Representation.}
It yields a continuous, differentiable approximation~\citep{chen2023implicit} of the field with sub-grid resolution and compact storage. The continuous field \(u(\cdot,t)\) is modeled by implementing a coordinate-to-value mapping with sampled 2D coordinates \(\{({x}_i,{y}_i)\}_{i=1}^N\) as:
\begin{equation}
    \mathcal{S}_{\mathrm{INR}}\bigl(u(\cdot,t)\bigr)
  = \arg\min_{\theta\in\mathbb{R}^D}
\frac{1}{N}\sum_{i=1}^N \bigl\|f_\theta(({x}_i,{y}_i)) - u(({x}_i,{y}_i)),t)\bigr\|^2
\end{equation}

\begin{figure*}[!t]
    \centering
    \includegraphics[width=\linewidth]{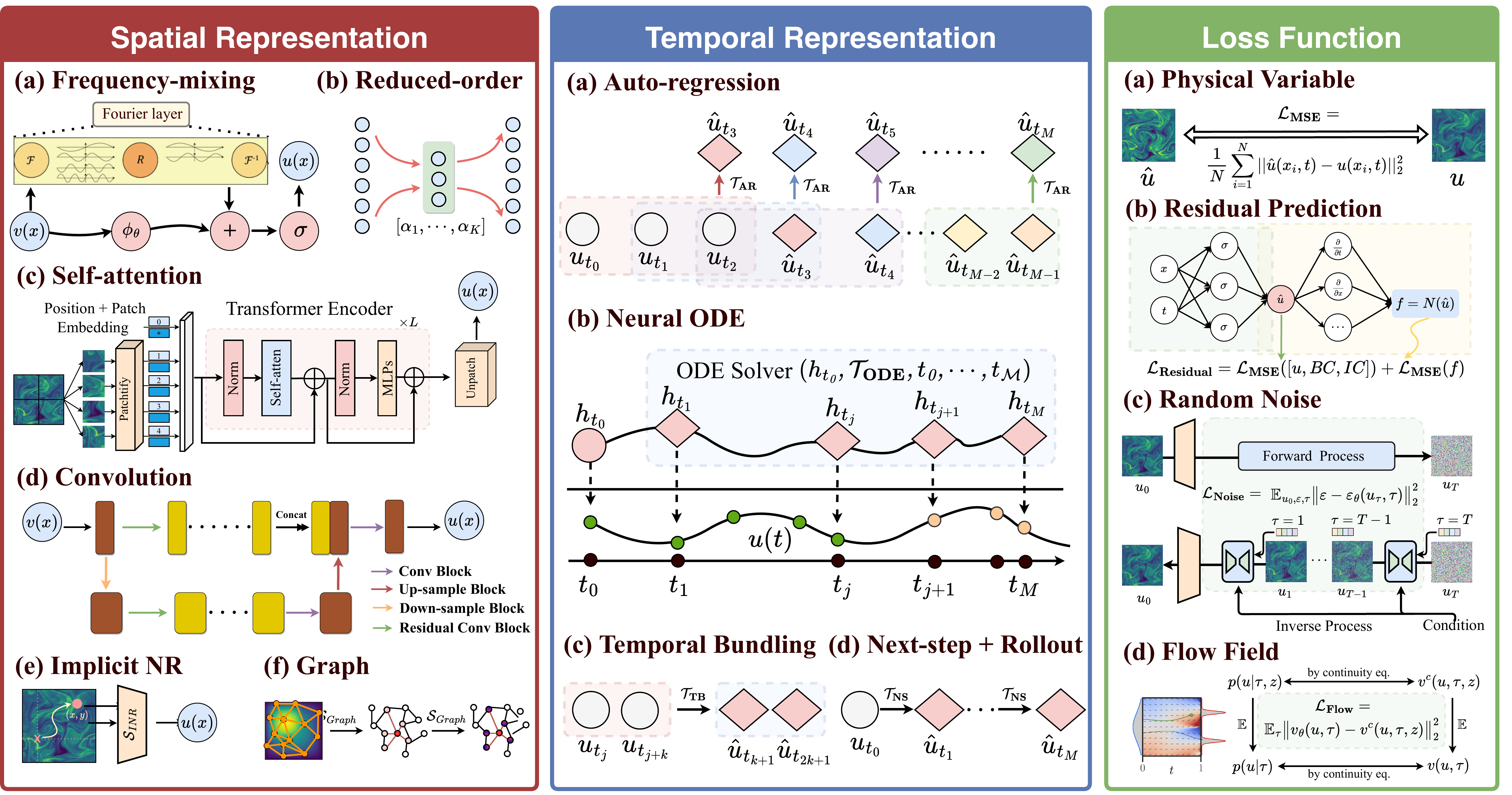}
    \caption{A schematic illustration of common approaches for each key module in data-driven neural PDE solvers. Note that self-attention is also applicable to temporal representation.}
    \label{fig:main}
\end{figure*}

\subsection{Temporal Representation}

Temporal representation encodes the evolution of the fluid field over time, providing essential features for predicting the future state of the flow field.  Following the definition in Section \ref{sec:spatial}, a discrete sequence of \(M\) successive snapshots $\{u(t_j)\}_{j=1}^M
\;\subset\;
\mathbb{R}^{N\times c}$ are learned with a mapping $\mathcal{T}_{\theta} \bigl(\{u(t_j)\}_{j=1}^M\bigr)
  = h
  \;\in\;
  \mathbb{R}^{D}$,
where \(\theta\) parameterizes the encoder and \(h\) is a \(D\)-dimensional feature.

\noindent\textbf{Autoregression.}  
Autoregressive approaches~\citep{li2020fourier,li2020multipole} decompose the sequence model into successive predictions using a sliding history window of length \(k\). If the input window at step \(j\) is as $\bigl[u(t_{j-k}),\,u(t_{j-k+1}),\,\dots,\,u(t_{j-1})\bigr]\;\in\;\mathbb{R}^{k\times N\times c}$.
Then for $j$ from $k+1$ to $M$, we have
$\hat{u}(t_{j+1}) = \mathcal{T}_{\mathrm{AR}}\bigl( \bigl[u(t_{j-k+1}),\,\dots,\,u(t_{j-1}),\,\hat{u}(t_j)\bigr]\bigr), $
so that each new prediction is appended to the window while the oldest state is discarded.

\noindent\textbf{Next‐Step.}
Unlike the continuous prediction used in autoregression, next-step prediction~\citep{cao2023efficient,fortunato2022multiscale} predicts only a single step ahead based on the most recent state and is not affected by teacher forcing. It aims to generate long-horizon forecasts as $\hat{u}(t_{j})
  = \mathcal{T}_{\mathrm{Next}}\bigl(u(t_{j-1})\bigr),$ $j=\{1,\dots,M\}.$

\noindent\textbf{Temporal Bundling.}
It is inspired by MP-PDE~\citep{brandstetter2022message}, which stacks a fixed window of \(k\) frames and applies a feed‐forward or convolutional encoder to predict \(k\) future steps. Thus a new vector $\mathbf{d} = \mathcal{T}_{\mathrm{TB}}(\mathbf{d}^0) = (\mathbf{d}^1,\cdots,\mathbf{d}^k)$ is used to update the solution $u\bigl(t_{k+j}\bigr)
= u\bigl(t_{k}\bigr)
+ \bigl(t_{k+j} - t_{k}\bigr)\,\mathbf{d}^j, j=\{1,\dots,k\}$.

\noindent\textbf{Self-attention.} It parallels the spatial mechanism; however, whereas attention scores are originally computed over spatial patches, they are here computed along the temporal dimension.

\noindent\textbf{Neural ODE.}  
Neural ordinary differential equations~\citep{chen2018neural,sun2024graph} provide a continuous‐time framework for modeling temporal evolution in a latent space~\citep{han2024brainode,wang2025conditional}. We define a hidden state \(h(t)\in\mathbb{R}^D\) that evolves according to the parameterized dynamics:
$
\frac{\mathrm{d}h}{\mathrm{d}t}
  = \mathcal{T}_{\mathrm{ODE}}\bigl(h(t), u(t_{0:t-1}),  t\bigr),
  h(t_0) = \zeta(u({t_0})), u(t) = \xi (h(t))
$,
where \(\zeta(\cdot)\) and \(\xi(\cdot)\) are parameterized to encode the dynamics of the hidden state. The sequence of latent states \(\{h(t_j)\}_{j=1}^M\) is obtained by solving ODE with a numeric integrator, and the final state is
\begin{equation}
    h(t_1),\cdots, h(t_j) = \mathrm{ODESolver}(\mathcal{T}_{\mathrm{ODE}}, h(t_0),(t_0,\cdots,t_j))
\end{equation}
This continuous formulation naturally captures irregular time sampling and allows for arbitrary‐horizon forecasting. 

\begin{table*}[t]
\centering
\scriptsize
\setlength{\tabcolsep}{3mm}{
\begin{tabular}{lc|ccc|ccc|ccc|cc}
\toprule[1.1pt]
 \multicolumn{1}{l}{}& \multicolumn{1}{l}{}& \multicolumn{3}{c}{\textbf{Compressible N-S}} & \multicolumn{3}{c}{\textbf{Diffusion-Reaction}} & \multicolumn{3}{c}{\textbf{Kolmogorov Flow}} & \multicolumn{2}{c}{Efficiency} \\ \cmidrule(lr){3-5} \cmidrule(lr){6-8} \cmidrule(lr){9-11} \cmidrule(lr){12-13}
\textsc{Model}
& \textsc{Param}
& \textsc{RMSE}$\downarrow$  & \textsc{nRMSE}$\downarrow$  &
\textsc{fRMSE}$\downarrow$  &
\textsc{RMSE}$\downarrow$  & \textsc{nRMSE}$\downarrow$  &
\textsc{fRMSE}$\downarrow$  &
\textsc{RMSE}$\downarrow$ & \textsc{nRMSE}$\downarrow$ &
\textsc{fRMSE}$\downarrow$  &
\textsc{Mem } & \textsc{GFLOPs} \\
\midrule
\rowcolor[RGB]{250, 220, 220} \multicolumn{13}{l}{\textbf{$\mathcal{X}$} + Next-step + Physical variable \textit{(\textbf{$\mathcal{X}$} can be graph, ROM, Fourier, Self-attention, Convolution)}} \\
Graph & 18.6M & 0.2648 & 0.3013 & 0.0097 & 0.0128 & 0.0251 & 0.0011 & 0.0709 & 0.0240 & 0.0044 & 1.17G & 79.38 \\
% Graph & 18.6
% & 0.266\textsubscript{\tiny $\pm$0.02}
% & 0.298\textsubscript{\tiny $\pm$0.03} 
% & 0.009\textsubscript{\tiny $\pm$0.01} 
% & 0.013\textsubscript{\tiny $\pm$0.01} 
% & 0.025\textsubscript{\tiny $\pm$0.01} 
% & 0.001\textsubscript{\tiny $\pm$0.00} 
% & 0.071\textsubscript{\tiny $\pm$0.01} 
% & 0.023\textsubscript{\tiny $\pm$0.01} 
% & 0.004\textsubscript{\tiny $\pm$0.00} 
% & 1.17 & 79.38 \\
ROM & 89.0M & 0.1043 & 0.1542 & 0.0088 & 0.0523 & 0.0208 & 0.0019 & 0.0151 & 0.0151 & 0.0012 & 1.52G & 81.96 \\
Fourier & 62.0M & 0.0779 & 0.1019 & 0.0085 & 0.0077 & 0.0150 & 0.0008 & 0.0049 & 0.0049 & 0.0005 & 1.67G & 79.46 \\
Self-attention & 76.2M & \textcolor{textpink}{0.0589} & \textcolor{textpink}{0.0714} & \textcolor{textpink}{0.0042} & \textcolor{textpink}{0.0065} & \textcolor{textpink}{0.0150} & \textcolor{textpink}{0.0008} & \textcolor{textpink}{0.0032} & \textcolor{textpink}{0.0031} & \textcolor{textpink}{0.0002} & 1.41G & 82.62 \\
Convolution & 64.8M & 0.1325 & 0.1528 & 0.0074 & 0.0123 & 0.0274 & 0.0017 & 0.0675 & 0.0227 & 0.0041 & 1.39G & 80.60 \\
\midrule
\rowcolor[RGB]{250, 220, 220} \multicolumn{13}{l}{\textbf{$\mathcal{X}$} + Next-step + Noise \textit{(\textbf{$\mathcal{X}$} can be Fourier, Self-attention, Convolution, ROM)}} \\
  Fourier & 62.0M & 0.3387 & 0.3610 & 0.0962 & 0.0822 & 0.0948 & 0.0249 &0.1161 & 0.1108 & 0.0424 & 2.10G & 85.44 \\
  Self-attention & 79.8M & 0.3002 & 0.3354 & 0.0598 & \textcolor{textpink}{0.0302} & \textcolor{textpink}{0.0594} & \textcolor{textpink}{0.0051} & \textcolor{textpink}{0.0841} & \textcolor{textpink}{0.0816} & \textcolor{textpink}{0.0176} & 1.82G & 82.59 \\
  Convolution & 48.9M & \textcolor{textpink}{0.1981} & \textcolor{textpink}{0.3348} & \textcolor{textpink}{0.0242} & 0.0437 & 0.0814 & 0.0089 & 0.1279 & 0.1155 & 0.0496 &0.99G & 74.92\\
  ROM & 75.9M& 0.3518 & 0.3762 & 0.1006 & 0.0759 & 0.0962 &0.0238 & 0.1345 & 0.1283 & 0.0559 & 1.07G & 80.72 \\
\midrule
\rowcolor[RGB]{220, 230, 245} \multicolumn{13}{l}{Self-attention representation + $\mathcal{Y}$ + Physical variable \textit{(\textbf{$\mathcal{Y}$} can be Self-attention, Neural ODE, Autoregression, Next + Rollout, Temporal Bund)}} \\
Self-attention & 30.9M & 0.2319 & 0.3094  & 0.0280 & 0.0443 &0.1083 &0.0067 & 0.0137 & 0.0145 & 0.0021 &1.07G &91.37 \\
Neural ODE & 32.9M & 0.3362 & 0.4137  &0.0399 &0.0135 &0.0311 &0.0014 & 0.0297 & 0.0307 & 0.0038 &1.09G &87.52 \\
Autoregression  & 24.9M & 0.3943 & 0.4576 &  0.0478 & 0.0785 & 0.1398 & 0.0087 &0.0182 &0.0198 & 0.0026 &1.24G &87.58 \\
   Next + Rollout & 36.6M & 0.2563 & 0.3177 & 0.0309 & 0.0454 & 0.1382 & 0.0041 &0.0085 &0.0086 &0.0008 &0.82G &79.15 \\
   Temporal Bund & 36.3M & \textcolor{textpink}{0.1357} &\textcolor{textpink}{0.1729} &\textcolor{textpink}{0.0122} &\textcolor{textpink}{0.0081} &\textcolor{textpink}{0.0205} & \textcolor{textpink}{0.0006} & \textcolor{textpink}{0.0049} & \textcolor{textpink}{0.0051} & \textcolor{textpink}{0.0005} &1.71G &97.33 \\
\midrule
\rowcolor[RGB]{220, 230, 245} \multicolumn{13}{l}{ Fourier representation + $\mathcal{Y}$ + Physical variable \textit{ (\textbf{$\mathcal{Y}$} can be Neural ODE, Autoregression, Next + Rollout, Temporal Bund)}} \\
   Neural ODE & 8.42M & 0.2172 & 0.2884 & 0.0293 &0.0124 & 0.0381& 0.0010 & 0.0545 & 0.0589 & 0.0022 & 0.38G &29.16 \\
   Autoregression  & 18.0M & 0.2032 & 0.2676 & 0.0291 & 0.0734 & 0.2089 &0.0113 & 0.0849 & 0.0871 & 0.0095 & 1.69G & 16.61 \\
   Next + Rollout & 22.1M & 0.1942 & 0.2581 & 0.0277 &0.0165 &0.0531 &0.0014 &0.0073 &0.0081 &0.0008 &1.21G & 20.45 \\
   Temporal Bund & 17.4M & \textcolor{textpink}{0.1842} & \textcolor{textpink}{0.2468} & \textcolor{textpink}{0.0187} &\textcolor{textpink}{0.0091} &\textcolor{textpink}{0.0193} &\textcolor{textpink}{0.0008}  & \textcolor{textpink}{0.0072} &\textcolor{textpink}{0.0074} & \textcolor{textpink}{0.0007} & 1.14G & 20.85 \\
\midrule
\rowcolor[RGB]{234, 238, 234}\multicolumn{13}{l}{ Self-attention representation + Next-step + \textbf{$\mathcal{Z}$} \textit{(\textbf{$\mathcal{Z}$} can Physical variable, Random Noise, Flow velocity, PDE residual)}} \\    Physical var  & 24.7M & \textcolor{textpink}{0.0997} & \textcolor{textpink}{0.0891} & \textcolor{textpink}{0.0094} & \textcolor{textpink}{0.0145} & \textcolor{textpink}{0.0277} & \textcolor{textpink}{0.0019} & \textcolor{textpink}{0.0082} & \textcolor{textpink}{0.0075} & \textcolor{textpink}{0.0009} &0.72G&33.74 \\
Noise  & 46.4M & 0.3211 & 0.3367 & 0.0769 &0.0310 &0.0602 &0.0048 & 0.0916 & 0.0859 & 0.0200 & 0.97G & 35.46\\
Flow  & 42.6M & 0.1232 & 0.1494 & 0.0118 & 0.0629 & 0.1039 & 0.0098 & 0.0609 & 0.0613 & 0.0050 & 1.15G & 20.19\\
PDE Residual  & 29.6M & 0.1370 & 0.1752 & 0.0118 & 0.1364 & 0.1644 & 0.0102 & 0.0302 & 0.0306 & 0.0021 & 0.67G & 31.53 \\
\bottomrule[1.1pt]
\end{tabular}
}
\caption{
The performance arises from three design factors: \textcolor[rgb]{0.99,0.82,0.82}{\textbf{spatial representation ($\mathcal{X}$)}}, \textcolor[rgb]{0.83,0.87,0.96}{\textbf{temporal representation ($\mathcal{Y}$)}}, and the \textcolor[rgb]{0.82,0.93,0.82}{\textbf{loss function ($\mathcal{Z}$)}}. Thus, we decouple them into modular designs and conduct comparisons under controlled variables on three datasets. MEM is the maximum memory.
}
\label{tab:main}
\vspace{-0.7cm}
\end{table*}

\subsection{Loss Function}

\noindent\textbf{Physical Variable Prediction Loss.}
The mean squared error (MSE)~\citep{cao2023lno,hao2024dpot,choubineh2023fourier,fortunato2022multiscale} between predicted and ground truth fields is commonly used in neural solvers given by $\mathcal{L}_{\mathrm{MSE}}
= \frac{1}{N}\sum_{i=1}^N \bigl\|\hat u(\bm{x}_i,t) - u(\bm{x}_i,t)\bigr\|_2^2.$

\noindent\textbf{Diffusion Denoising Loss.}
Under the diffusion framework~\citep{ho2020denoising,song2020denoising}, one perturbs the true field \(u_0\) by a noise schedule
$u_{\tau} = \sqrt{\bar\alpha_{\tau}}\,u_0 + \sqrt{1-\bar\alpha_{\tau}}\,\varepsilon,\varepsilon\sim\mathcal{N}(0,I)$,
and trains a neural network \(\varepsilon_\theta\) to estimate the added noise. The corresponding loss is $\mathcal{L}_{\mathrm{Noise}}
= \mathbb{E}_{u_0,\varepsilon,{\tau}}\bigl\|\varepsilon - \varepsilon_\theta(u_{\tau},{\tau})\bigr\|_2^2$.
This approach captures the full distribution of flow fields, enabling stochastic sampling of complex outcomes. 

\noindent\textbf{Flow Matching Loss.}
Flow matching~\citep{lipman2022flow} directly learns the instantaneous vector field by minimizing the discrepancy between a predicted velocity and a numerical approximation of the true flow with a simple (invertible) affine map $\psi$ as:
$\mathcal{L}_{\mathrm{Flow}}
= \mathbb{E}_{u,\tau}\Bigl\|v_\theta\bigl(\psi_{\tau}(u_0),\tau\bigr) - \frac{d}{d\tau}\psi_{\tau}(u_0)
\Bigr\|_2^2$.
$\psi_{\tau}(u)$ can be $\mu_{\tau}(u_1) + \sigma_{\tau}(u_1)u, \tau \in [0,1]$. By training \(v_\theta\) to match true dynamics, this loss facilitates stable multi‐step and continuous forecasting, often producing sharper trajectories.

\noindent\textbf{Physics‐Informed Residual Loss.}
Physics‐informed neural networks~\citep{raissi2019physics,wang2024pinn} enforce the governing PDE by penalizing the residual of its differential operator. Denoting the learned field \(u_\theta\), the residual is $\mathcal{R}[u_\theta](x,t)
= \partial_t u_\theta + \mathcal{N}(u_\theta) - g(x,t)$,
and the loss reads $\mathcal{L}_{\mathrm{Residual}}
= \mathbb{E}_{x,t}\bigl\|\mathcal{R}[u_\theta](x,t)\bigr\|_2^2.$
Here \(\mathcal{N}\) represents spatial operators (\textit{e.g.,}\ advection) and \(g\) is any source term. It enforces exact satisfaction of physical laws in the continuous limit and often generalizes well beyond training data.

\section{Experiments}
\label{sec:exp}

\subsection{Experimental Setup}

\noindent\textbf{Evaluation Metrics.}
To ensure a comprehensive and objective evaluation, we employ a diverse suite of metrics. First, we assess the global performance using the root-mean-squared-error (RMSE) and its normalized version (nRMSE). However, these measures do not capture local performance nuances. We further incorporate additional metrics that focus on specific failure modes, the RMSE in Fourier space (fRMSE) evaluated separately in low, middle, and high-frequency regions. Moreover, we extend our assessment to include efficiency evaluations, such as computational time cost and memory consumption. More details can be seen in Appendix \ref{sec:imple_details}.

\noindent\textbf{Experimental Setting.}
We create a brand-new codebase covering all of the decompositions. We use PyTorch to implement experiments on 8 $\times$ NVIDIA A6000 GPUs. Each dataset is by default split into 80\% for training, 10\% for testing, and 10\% for validation, with a default spatial resolution of 128$\times$128. Extensive experiments with varying conditions will be introduced correspondingly.

To establish a fair and rigorous evaluation protocol, we address the concerns outlined above from three complementary aspects.
(1) We adopt consistent experimental configurations across all model families. In particular, we standardize the hyperparameter tuning process by fixing the optimizer type, search ranges for learning rate and weight decay, and scheduler choices. All methods are tuned under the same grid-search budget, and we release the complete set of tuned configuration files in the codebase, including random seeds, to ensure full reproducibility.
(2) We conduct fair comparisons with traditional numerical solvers by using high-resolution simulations as ground truth and evaluating all methods at matched physical simulation times. Classical solvers are implemented with standard numerical schemes and adaptive time stepping governed by the Courant–Friedrichs–Lewy (CFL) condition, reflecting common practice in computational physics.
(3) To systematically assess generalization, we perform a series of experiments covering zero-shot initial conditions, varying spatial resolutions, and long-term temporal rollouts. In the appendix, we further examine generalization under additional settings, including different model sizes, mixed boundary conditions, non-Cartesian mesh discretizations, 3D PDEs, and multi-physics PDE systems.

\begin{figure*}[ht]
    \centering
    \scalebox{1.0}{
        \begin{minipage}{\textwidth}
            \centering
            \begin{subfigure}
                \centering
                \includegraphics[width=0.19\linewidth]{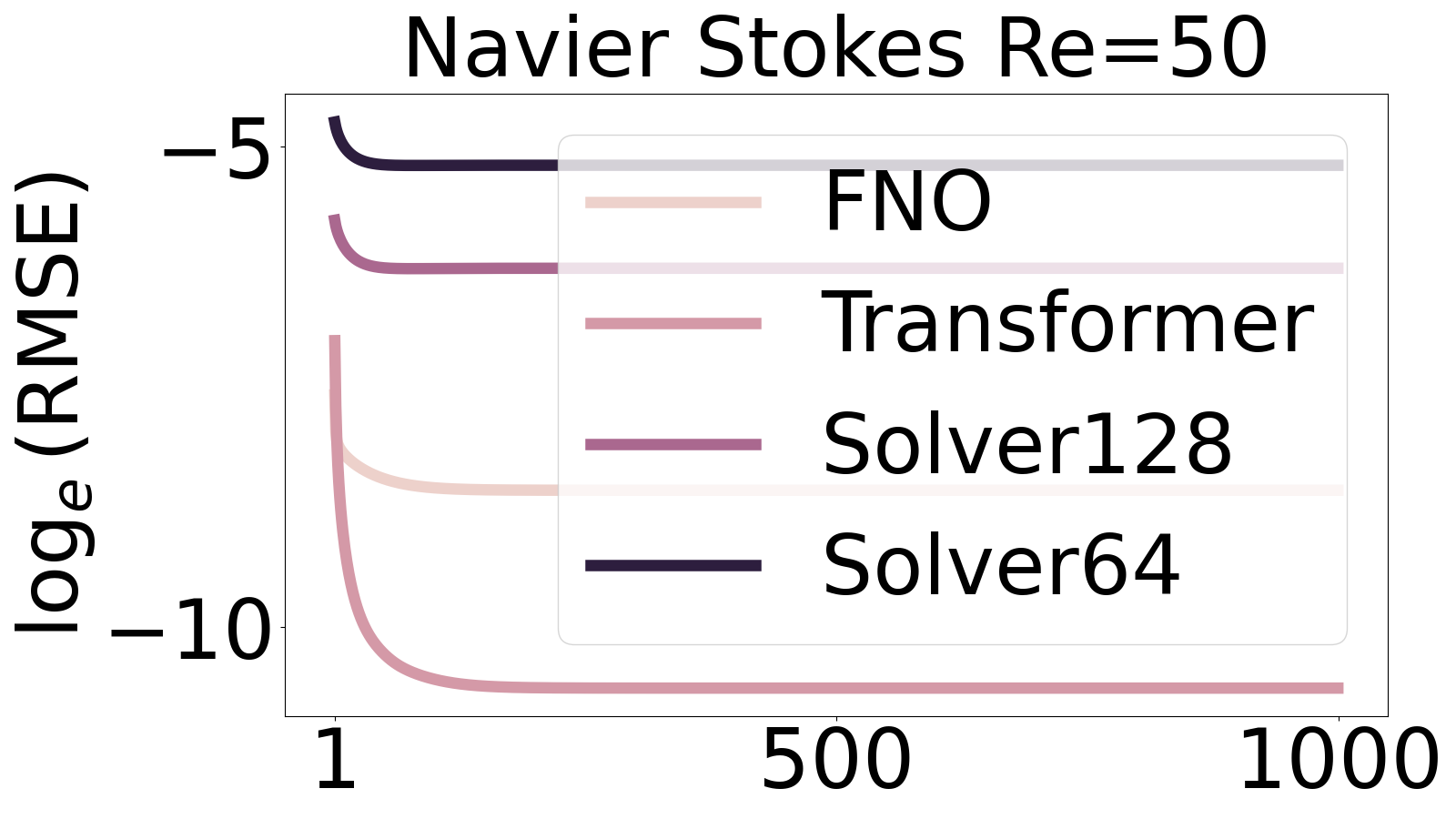}
            \end{subfigure}
            \hfill
            \begin{subfigure}
                \centering
                \includegraphics[width=0.19\linewidth]{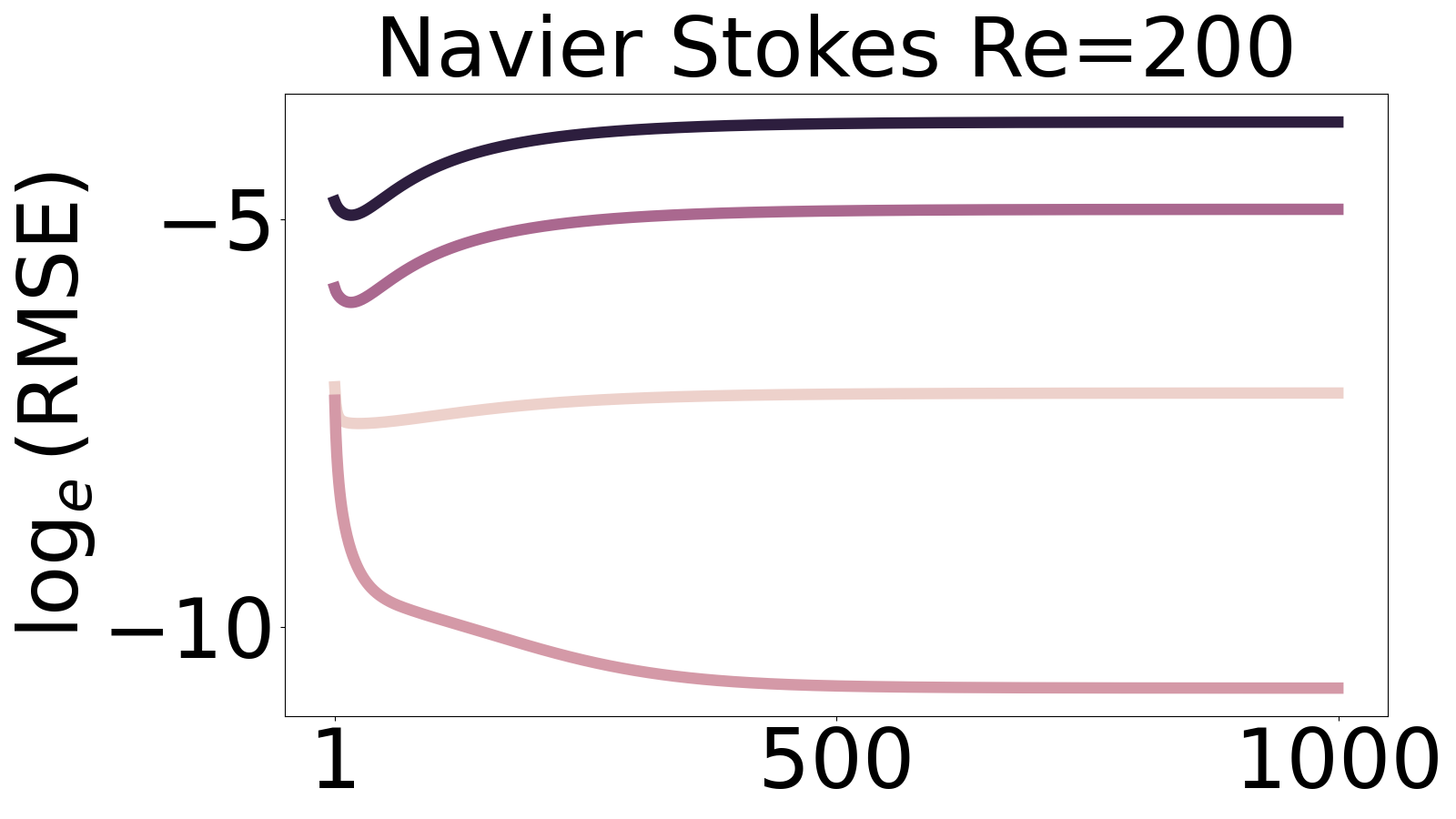}
            \end{subfigure}
            \hfill
            \begin{subfigure}
                \centering
                \includegraphics[width=0.19\linewidth]{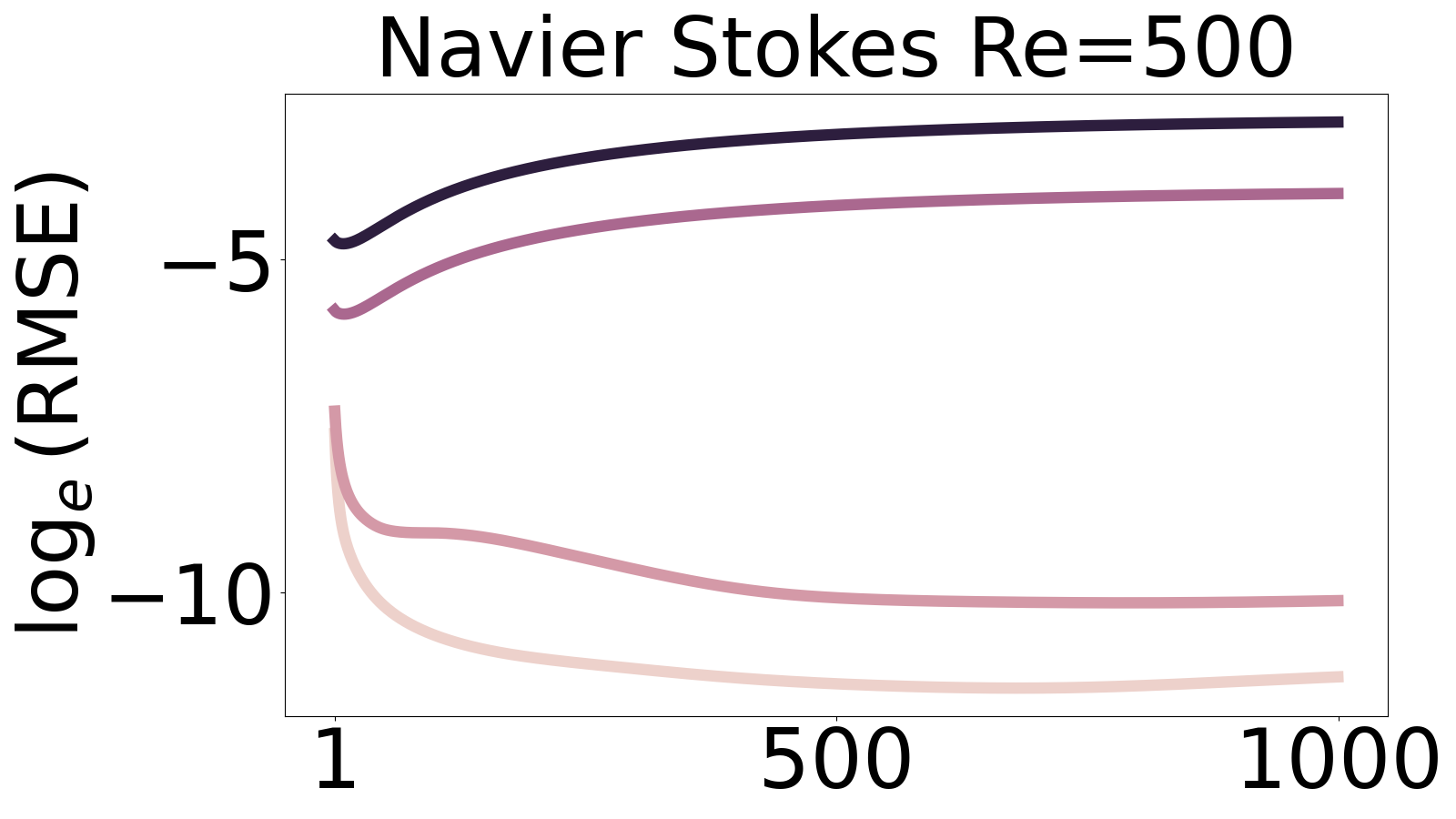}
            \end{subfigure}
            \hfill
            \begin{subfigure}
                \centering
                \includegraphics[width=0.19\linewidth]{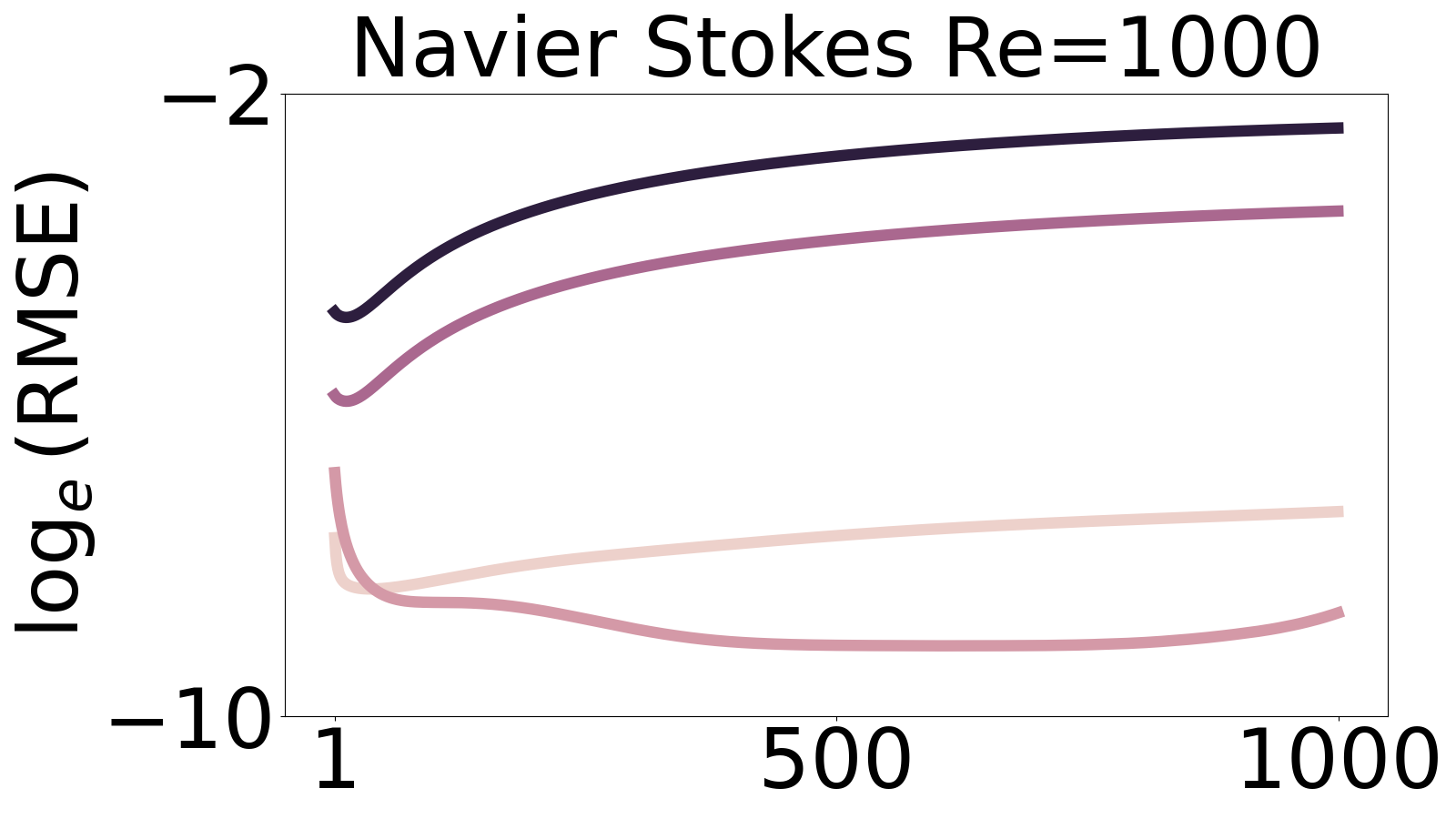}
            \end{subfigure}
            \hfill
            \begin{subfigure}
                \centering
                \includegraphics[width=0.19\linewidth]{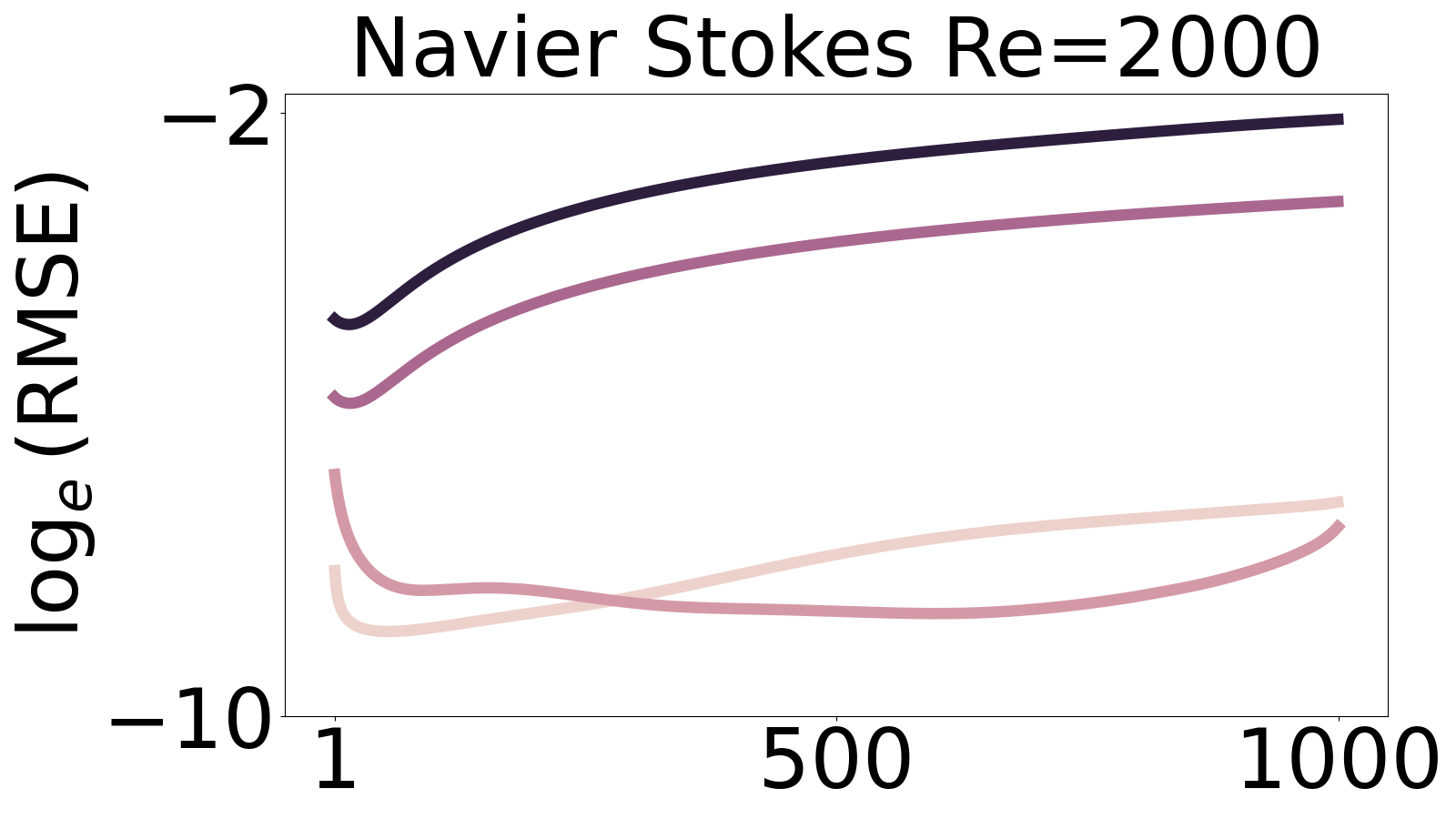}
            \end{subfigure}
        \end{minipage}
    }
    \vspace{-0.5cm}
    \caption{We compare neural solvers against traditional solver (\textit{i.e.,} pseudo-spectral solver) operating at lower resolutions on N-S equation across five Reynolds numbers, spanning laminar to turbulent regimes. Solver $x$ implies a pseudo-spectral solver operating at $x \times x$ resolution.}
    \label{fig:solver_ns}
    \vspace{-0.1cm}
\end{figure*}

\subsection{How Should We Define and Search for Optimal Neural Architectures?}

To ensure fairness, all results are reported under approximately matched computational costs (\textit{e.g.,} GFLOPs). In the main text, we evaluate three representative $128 \times 128$ two-dimensional flow scenarios. Beyond these, we also consider more complex settings, including 3D flows, multi-physics problems, mixed boundary conditions, and complex geometries.

To investigate the effect of spatial representations, we fix the prediction modes to \texttt{$\mathcal{Y}$=next-step}, and compare the performance of different methods. As shown in Table~\ref{tab:main}, the self-attention-based representation consistently achieves the best results across multiple tasks, followed by the Fourier-based representation. It is worth noting that, due to the high computational cost of training convolutional models, we were unable to evaluate their performance under larger parameter budgets. Additionally, for the graph-based model, we apply sampling strategies to accommodate the specific data topology before training. More analysis details (\textit{e.g.,} results on Stochastic N-S) are in Appendix \ref{sec:module_more}.

To investigate temporal evolution, we fix the spatial encoder to either \texttt{$\mathcal{X}$=Self-attention} or \texttt{Fourier}. As shown in Table~\ref{tab:main}, temporal bundling consistently achieves the best performance across tasks while maintaining relatively low computational overhead. This highlights the importance of developing modules that explicitly capture the temporal dynamics of fluid evolution. Notably, neural ODEs approach the performance of temporal bundling on several tasks, suggesting a promising direction for future research. 

Lastly, with respect to the loss, physical variable prediction remains the mainstream approach. However, generative models such as diffusion-based and flow-based methods also demonstrate competitive performance, indicating promising directions for the future.

In conclusion, our results show that \textbf{(1)} self-attention is particularly effective for spatial encoding, and its performance can likely be further improved via techniques such as sparse attention, receptive field modulation, or transformer regularization. Meanwhile, frequency-based methods offer strong performance at minimal cost, highlighting the promise of embedding mathematical priors into neural solvers. \textbf{(2)} Simple yet effective temporal bundling strategies achieve strong performance under constrained compute budgets. This suggests a valuable research direction: accelerating temporal evolution modules like neural ODE through lightweight designs, potentially yielding a more favorable trade-off between predictive fidelity and efficiency in real-world applications.

\subsection{Can neural solvers truly replace traditional numerical solvers?}

\begin{figure}[ht]
    \centering
    \includegraphics[width=0.4\textwidth]{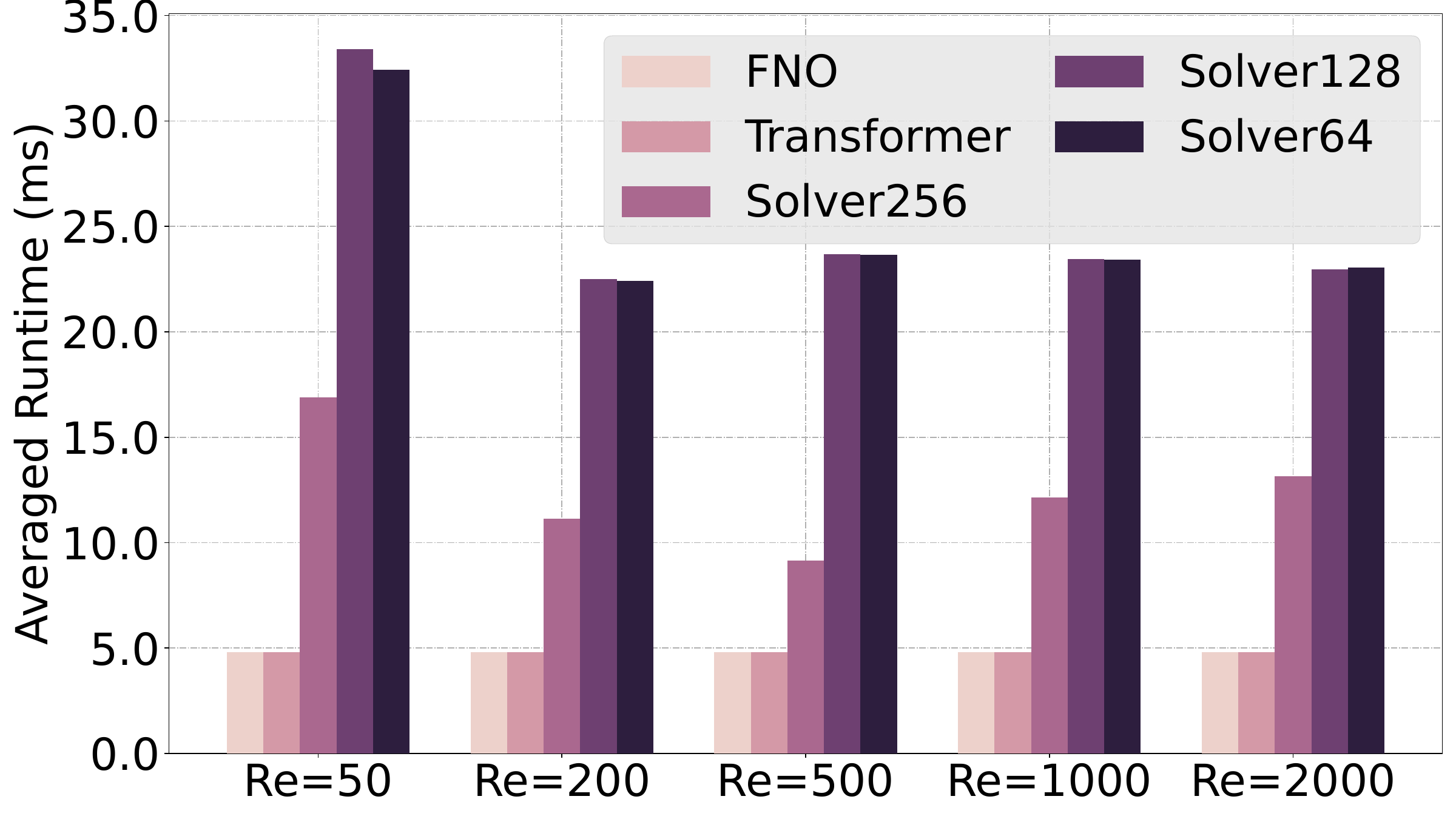}
  \caption{Runtime of neural solvers against numerical solver operating at lower resolutions on the incompressible N-S for predicting up to $T=\frac{1}{32}$. Solver $x$ implies $x \times x$ resolution of pseudo-spectral solver. }
  \label{fig:ns_runtime}
  \vspace{-0.3cm}
\end{figure}

In terms of both performance and efficiency, the answer may be yes.
Our experiments are conducted under carefully controlled and reproducible settings on incompressible N-S with results shown in Figure~\ref{fig:solver_ns}. 
Additional results are in Appendix~\ref{appendix:comp_solvers}. Across most benchmarks, Transformer achieves substantially lower prediction error, especially at longer rollout steps. In terms of runtime, which is shown in Figure \ref{fig:ns_runtime}, Transformer outperforms both the high-fidelity simulator and coarse-grid solvers, delivering $10 \times$ to $48 \times $ speedups in the Burgers’ equation experiments. Data generated at $256 \times 256$ is considered ground truth. All traditional methods, including those on coarser grids, employ adaptive time stepping governed by the Courant–Friedrichs–Lewy (CFL) condition to maintain numerical stability. On a coarser grid like $64 \times 64$, the solution can become less smooth and more prone to numerical instability. This can lead to smaller allowed $\Delta t$ to maintain stability. As a result, even though each prediction with step size $\Delta t$ is cheaper, the total number of steps required to reach the final simulation time can increase. We attribute the speedup of neural solvers to their ability to predict solutions with larger step size, allowing them to take fewer prediction steps than conventional numerical solvers. To examine this property in detail, we include extensive experiments in the Appendix~\ref{appendix:comp_solvers_timesteps}, analyzing the neural solver’s performance across varying step sizes. The results indicate that, unlike numerical solvers where increasing $\Delta t$ often causes severe numerical instabilities, neural solvers experience only minor performance degradation when trained to predict with a larger step size. They can still surpass numerical solvers operating at coarser resolution, enabling speedups of several hundred times.
% Notably, we notice that neural operators can be further accelerated. 
We refer the readers to Appendix~\ref{appendix:datasets_solver} for details of solvers and Appendix~\ref{sec:imple_details} for details of neural solvers.

\begin{figure*}[h]
    \centering
    \scalebox{1.0}{
        \begin{minipage}{\textwidth}
            \centering
            \begin{subfigure}
                \centering
                \includegraphics[width=0.32\linewidth]{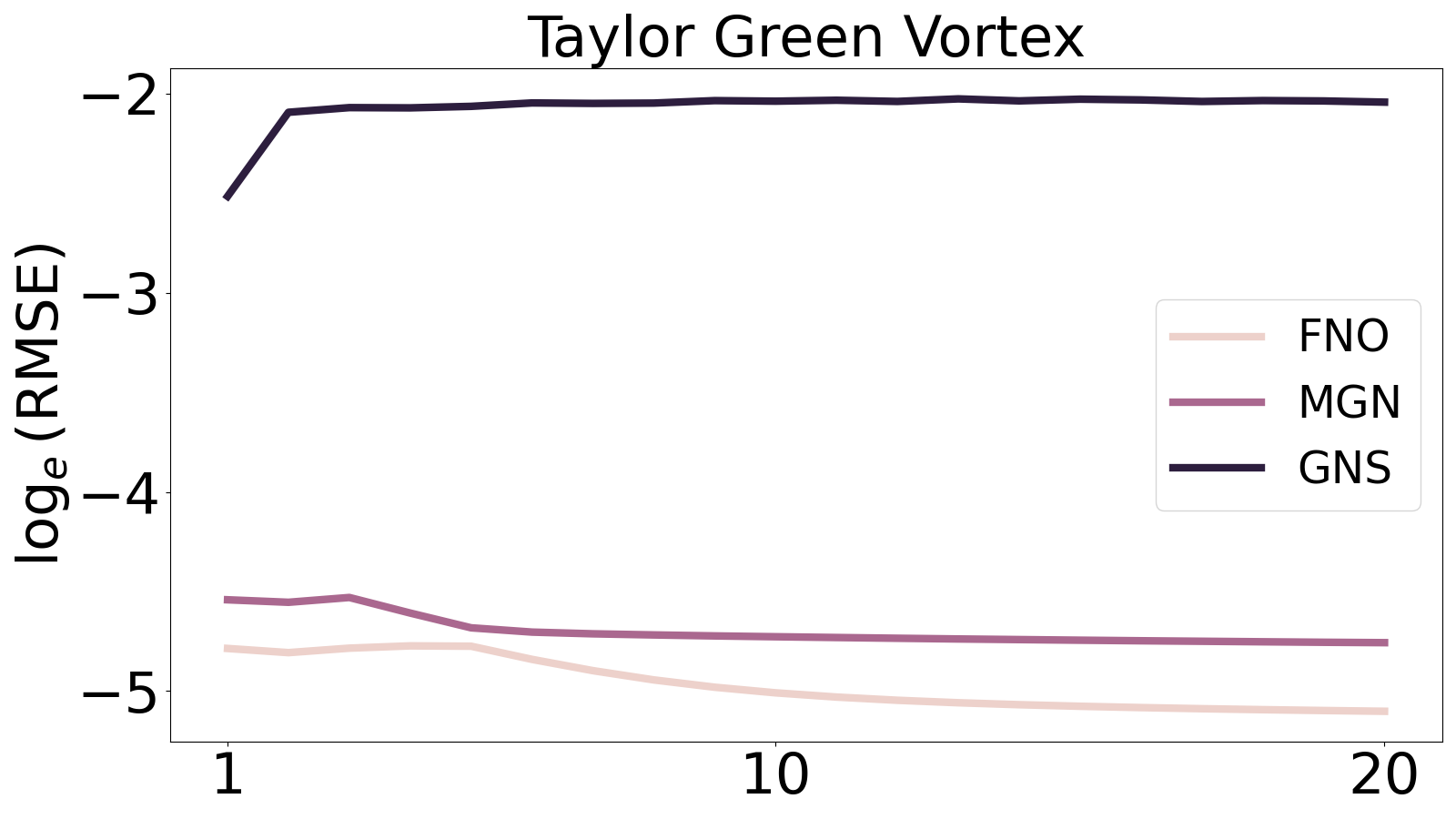}
            \end{subfigure}
            \hfill
            \begin{subfigure}
                \centering
                \includegraphics[width=0.32\linewidth]{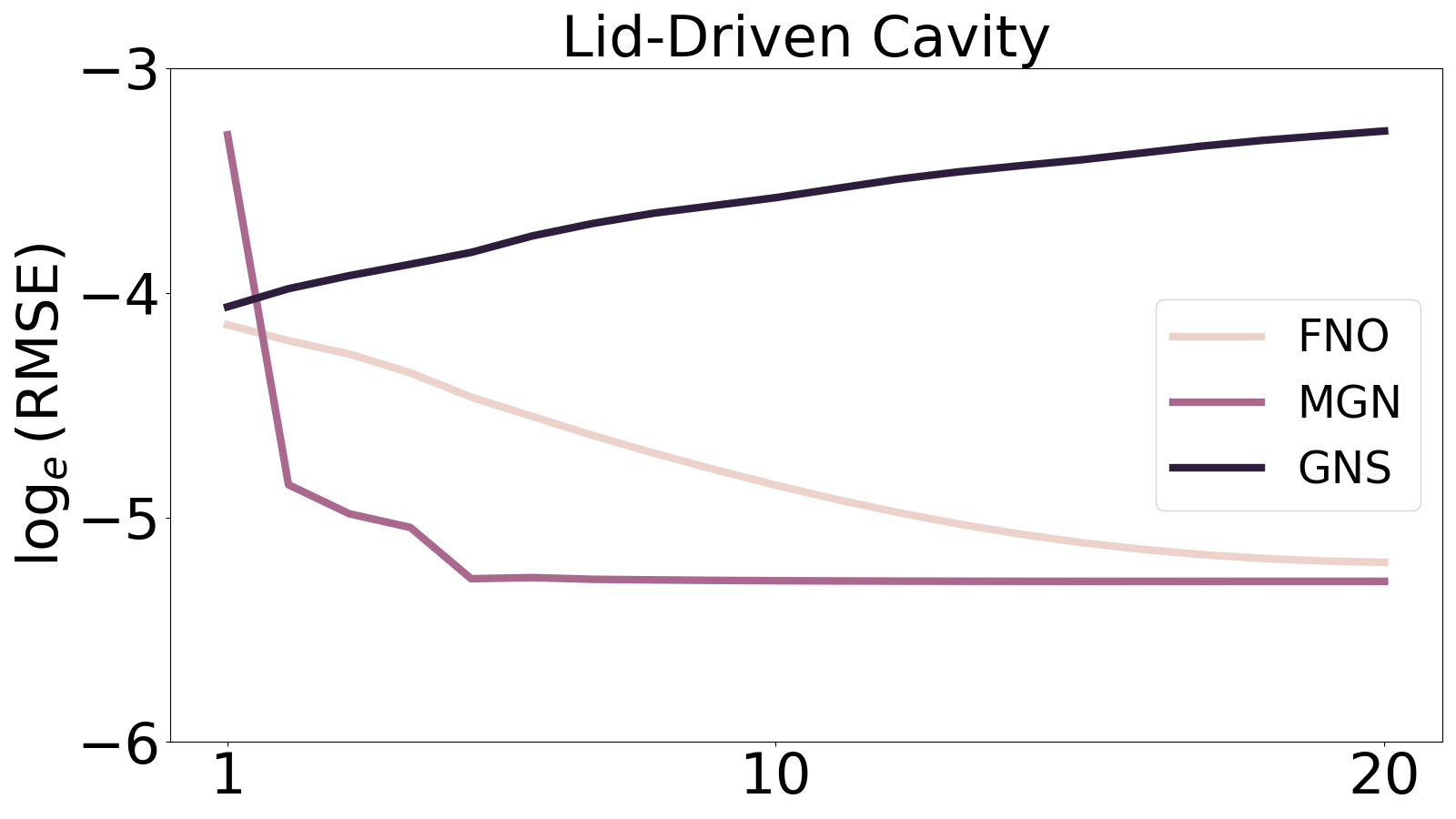}
            \end{subfigure}
            \hfill
            \begin{subfigure}
                \centering
                \includegraphics[width=0.32\linewidth]{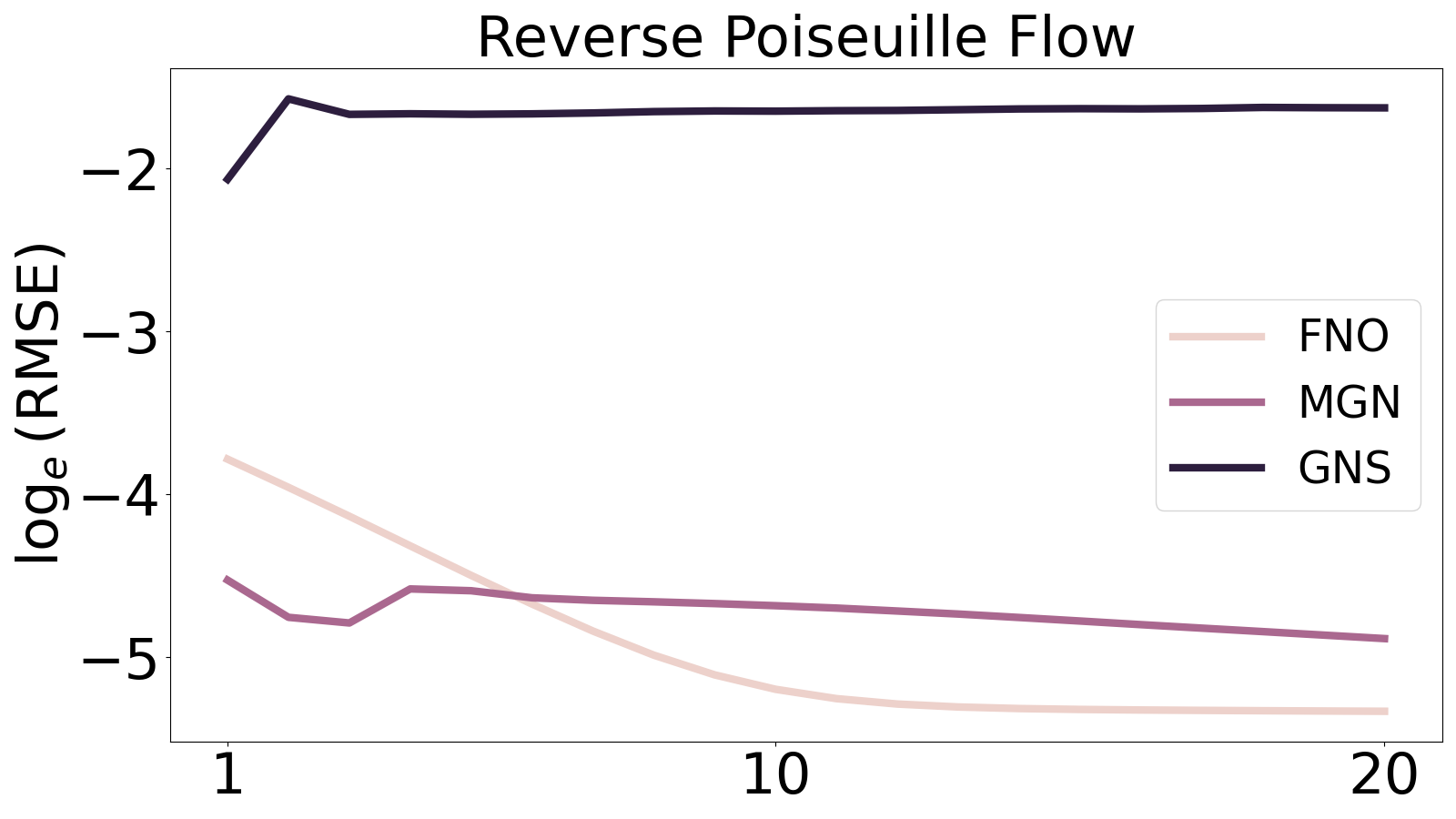}
            \end{subfigure}
        \end{minipage}
    }
    \vspace{-0.5cm}
    \caption{Comparison of rollout performance between FNO (grid data), MeshGraphNets (mesh data), and GNS (particle data). We transform particle and mesh predictions to grid data to evaluate.}
    \label{fig:eulerian_mse}
    \vspace{-0.3cm}
\end{figure*}

\subsection{Which discretization better supports their representative neural solvers?}

% model-based comparison
% data-level trasnfer + comparison
To ensure a rigorous and fair comparison, we evaluate each solver on the discretization scheme it is specifically designed for: 
FNO \citep{li2020fourier} is trained on regular Eulerian grids, MeshGraphNets (MGN) \citep{pfaff2020learning} on unstructured meshes, and GNS \citep{sanchez2020learning} on particle-based Lagrangian representations. 
All experiments are conducted under matched grid resolutions, comparable computational budgets, and carefully tuned hyperparameters (see Appendix~\ref{sec:imple_details}) to isolate the effect of discretization rather than model capacity or training regime.  
Figure~\ref{fig:eulerian_mse} reports the rollout MSE of the three solvers. 
Across all experiments, the Eulerian discretizations achieve significantly lower rollout MSE than the Lagrangian scheme.

In conclusion, we observe that models employing Eulerian schemes more accurately capture fundamental fluid dynamics. In contrast, Lagrangian representations must model fine‐grained details, which impedes their ability to learn robust representations and renders them susceptible to error accumulation over extended rollout time steps. Their emphasis on simulating high-resolution local interactions leads to severe error accumulation and propagation across particles over time. To mitigate this, future directions may involve hierarchical or multi-scale architectures, such as U-Net-based designs, that can dynamically balance fine-grained resolution with global structural coherence, thus improving both stability and generalization. We refer the readers to Appendix~\ref{sec:imple_details} for more details of model and training hyperparameters.

\begin{figure*}[!t]
    \centering
    \includegraphics[width=\linewidth]{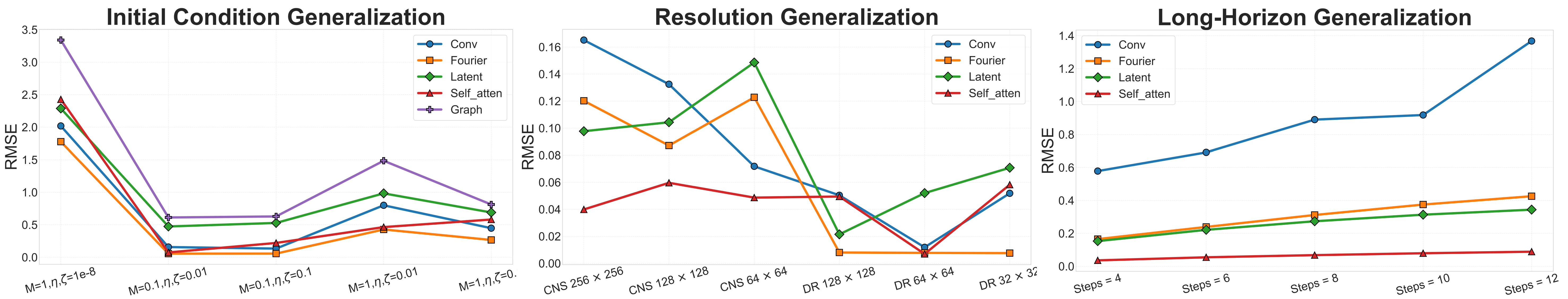}
    \vspace{-0.3cm}
    \caption{Extrapolation evaluation on zero-shot generalization across diverse initial conditions, resolution generalization across spatial scales, and long-horizon generalization. All evaluations are performed on the Compressible N-S system using different methods for next-step prediction. }
    \label{fig:genera}
    \vspace{-0.1cm}
\end{figure*}

\subsection{Whether knowledge from different systems can boost generalization?}
\label{sec:generalization}

\noindent\textbf{Generalization on zero-shot initial conditions.}
We first pre-train all candidate models on Compressible N–S with a moderate Mach number $M=0.1$, where both shear and bulk viscosities are fixed at $\eta=\zeta=10^{-8}$. Then we evaluate the trained models directly on test data with OOD initial conditions, which exhibit qualitatively different flow structures and turbulence patterns. This setting is designed to mimic real-world scenarios where the governing dynamics remain the same, but the flow configurations encountered may deviate significantly from those seen during training.

As shown in the first panel of Figure~\ref{fig:genera}, all models experience varying degrees of performance degradation under these OOD shifts, confirming the challenge of prediction across diverse initial conditions. Interestingly, models leveraging global spatial representations, such as Fourier-based operators and self-attention mechanisms, exhibit significantly better robustness. Their ability to capture long-range correlations and encode global flow structures appears to confer a clear advantage when extrapolating to turbulent regimes with unseen coherent structures.

\noindent\textbf{Generalization on different resolutions.}
We also train models using various methods on both the Compressible N-S and Diffusion-Reaction subsets across different spatial resolutions and evaluate them on their corresponding test resolutions. This controlled setup allows us to disentangle how each model family benefits from increasing grid fidelity and whether their inductive biases are well-aligned with the underlying physics at different scales. As shown in the middle panel of Figure~\ref{fig:genera}, self-attention and latent-based models benefit from higher-resolution training, leveraging richer spatial information to improve performance. In contrast, Fourier-based representations exhibit stronger performance at lower resolutions. 

\noindent\textbf{Generalization on long-horizon rollout.}
We also evaluate the long-horizon generalization ability of different methods under the setting \texttt{$\mathcal{Z}$ = variable}, by rolling out predictions over extended time steps. This experiment is designed to probe whether the learned representations can remain numerically stable and physically consistent when error accumulation becomes a dominant factor. As shown on the right of Figure~\ref{fig:genera}, all methods experience some degree of error accumulation as the rollout length increases. However, the spatial representation based on self-attention remains the most stable, exhibiting the lowest performance degradation. 
Additionally, models trained with a denoising loss and subsequently sampled exhibit substantially greater stability compared to deterministic models. Detailed comparison is in Appendix~\ref{sec:supp_generalization_diff}.

\noindent\textbf{Generalization on scaling ability.} Please refer to Appendix \ref{sec:supp_generalization_diff_scale} for comparisons of scaling abilities.

\vspace{-0.3cm}
\section{Extensive explorations and analysis}

We conduct additional evaluations spanning mixed boundary conditions, non-Cartesian meshes, three-dimensional PDEs, and genuinely coupled multiphysics systems (Appendix~\ref{sec:mixedbc}--\ref{sec:supp_mpde}), showing that the main architectural trends persist under substantially more realistic conditions.

\noindent\textbf{Mixed boundary conditions and wall physics.}
Appendix~\ref{sec:mixedbc} introduces a compressible Euler airfoil dataset with curved geometry and mixed boundary conditions, including a per-boundary-condition error analysis (Appendix~\ref{sec:bc-analysis}), where Fourier and self-attention methods consistently outperform convolution and graph across wall, far-field, and interior regions.

\noindent\textbf{Non-Cartesian mesh discretizations.}
Appendix~\ref{sec:supp_mesh} evaluates representative models on the Pipe~\cite{li2022fourier} with non-Cartesian meshes, demonstrating that self-attention achieves the best performance while Fourier methods remain competitive, even if they are not in the regular grid structure.

\noindent\textbf{3D PDEs.}
Appendix~\ref{sec:supp_3d} extends the benchmark to 3D compressible MHD turbulence~\cite{ohana2024well}, where self-attention continues to outperform convolutional and Fourier alternatives across all error metrics despite increased dimensionality.

\noindent\textbf{Genuinely coupled multiphysics PDEs.}
Appendix~\ref{sec:supp_mpde} evaluates all methods on a two-phase boiling dataset~\cite{hassan2023bubbleml}, showing that Fourier operators retain advantages for high-frequency interfacial dynamics while self-attention achieves the best overall accuracy in a tightly coupled multiphysics regime.

\noindent\textbf{Qualitative study.} Additional qualitative results and detailed visualization analyses are provided in Appendix~\ref{sec:vis} to further illustrate the model behavior and support our quantitative findings.

\section{Conclusion}

Taken together, findings of \method{} not only establish a rigorous and fair benchmark but also chart a forward-looking research agenda for the community. We envision that \method{} will serve as a catalyst for developing neural PDE solvers that are not only more accurate and efficient but also more robust and scalable, paving the way toward practical, real-time, and general-purpose simulators. Moreover, our codebase is easy to integrate with future research.

\section*{Acknowledgments}
This work was partially supported by National Science Foundation (2106859, 2200274, 2303037, 2312501, 2531008), National Institutes of Health (OT2OD038003, U54HG012517, U54DK097771, U24DK097771, U54OD036472), DARPA HR001126CE054, Amazon, NEC, Optum AI, SRC JUMP 2.0 Center, UCLA CDSC Center, Amazon Research Awards, Snapchat, and Google Gifts. 

\bibliography{reference}
\bibliographystyle{ACM-Reference-Format}

\appendix

\clearpage
\onecolumn
\section*{Appendix}

\begin{description}[leftmargin=3.5cm, style=nextline]
  \item[Section \ref{sec:notation}] Symbol Notation \dotfill \pageref{sec:notation}
  \item[Section \ref{sec:decomposition}] Decompositions Corresponding to Existing Works \dotfill \pageref{sec:decomposition}
  \item[Section \ref{appendix:add_tech}] Supplementary Results on Additional Techniques \dotfill \pageref{appendix:add_tech}
  \item[Section \ref{appendix:comp_solvers}] Supplementary Results on Comparison with Traditional Solvers \dotfill \pageref{appendix:comp_solvers}
  \item[Section \ref{sec:module_more}] Supplementary Results on Different Modules \dotfill \pageref{sec:module_more}
  \item[Section \ref{sec:supp_generalization}] Supplementary Results on Generalization Study \dotfill \pageref{sec:supp_generalization}
  \item[Section \ref{sec:mixedbc}] Supplementary Results on Mixed Boundary Conditions \dotfill \pageref{sec:mixedbc}
  \item[Section \ref{sec:supp_mesh}] Supplementary Results on Non-Cartesian mesh \dotfill \pageref{sec:supp_mesh}
  \item[Section \ref{sec:supp_3d}] Supplementary Results on 3D PDE \dotfill \pageref{sec:supp_3d}
\item[Section \ref{sec:supp_mpde}] Supplementary Results on Multi-physics PDE \dotfill \pageref{sec:supp_mpde}
  \item[Section \ref{sec:imple_details}] More Implementation Details \dotfill \pageref{sec:imple_details}
  \item[Section \ref{appendix:leader}] Easy Integration of Existing Leaderboard \dotfill \pageref{appendix:leader}
  \item[Section \ref{sec:supp_data}] More Dataset Details \dotfill \pageref{sec:supp_data}
  \item[Section \ref{sec:vis}] Visualization Study \dotfill \pageref{sec:vis}
  \item[Section \ref{appendix:limit}] Limitations and Broad Impact \dotfill \pageref{appendix:limit}
\end{description}

\section{Notation}
\label{sec:notation}
Table~\ref{tab:notation} summarizes the key mathematical symbols and their meanings used throughout our benchmark framework and avoids symbol ambiguity across different methods, covering core components such as spatial representation, temporal modeling, and loss functions.

\begin{table}[h]
  \centering
  \begin{tabular}{ll}
    \toprule
    Symbol & Meaning \\
    \hline
    \(\Omega\)                & Spatial domain with dimension \(d\), \(\Omega\subset\mathbb{R}^d\) \\
    \(T\)                     & Temporal horizon, \(t\in[0,T]\) \\
    \(u(\bm{x},t)\)           & Field variable at location \(\bm{x}\) and time \(t\) with \(c\) channels \\
    \(\{\bm{x}_i\}_{i=1}^N\)  & Spatial locations of \(N\) samples, \(\bm{x}_i\in\Omega\) \\
    \hline
    \(M\)                     & Number of temporal snapshots \\
    \(\mathcal{S}_\theta, \mathcal{T}_\theta\)    & Spatial and temporal encoder parameterized by \(\theta\) \\
    \(z,h\)            & Feature representation in spatial and temporal domain \\
    \hline
    \(\mathcal{F}, \mathcal{F}^{-1}\) & Fourier and inverse Fourier transforms \\
    \(\hat{u}(\mathbf{k},t)\) & Spectral coefficient at frequency \(\mathbf{k}\) \\
    \(\phi_\theta(\mathbf{k})\) & Learnable spectral filter of frequency index \(\mathbf{k}\)  \\
    \(\odot\)                 & Elementwise (Hadamard) product \\
    \hline
    % Self-attention
    \(\mathbf{W}_Q,\mathbf{W}_K,\mathbf{W}_V\) & Query, key and value projection matrices, \(\in\mathbb{R}^{c\times d_k}\) \\
    \(d_k,d_k,d_v\)                        & Query/key and Value dimension in self-attention \\
    \hline
    % Convolution
    \(\mathcal{H}\)            & Set of stencil offsets for convolution \\
    \(c_{\mathrm{in}},c_{\mathrm{out}}\)        & Number of input and output channels to convolution \\
    \(\mathbf{W}_h\)           & Convolution kernel at offset \(h\), \(\in\mathbb{R}^{c_{\mathrm{out}}\times c_{\mathrm{in}}}\) \\
    \hline
    % Graph
    \(\tilde{A},\tilde{D}\)              & Adjacency with self-loops, \(\tilde{A}=A+I\), and degree matrix of \(\tilde{A}\) \\
    \hline 
    % ROM
    \(K\)                     & Number of retained modes in ROM, \(K \ll N\) \\
    \(C(\bm{x},\bm{x}')\)      & Covariance kernel between spatial locations \\
    \(\phi_k(\bm{x})\)         & \(k\)-th POD basis function \\
    \(\lambda_k\)              & Eigenvalue associated with \(\phi_k\) \\
    \(\alpha_k(t)\)            & POD projection coefficient at time \(t\) \\
    \(\langle f,g\rangle\)     & \(L^2\) inner product over \(\Omega\) \\
    \hline
    % Temporal models
    \(\mathbf{d}^j\)           & Predicted increment for bundling step \(j\) \\
    \(h(t)\)                   & Latent state in Neural ODE \\
    \(\zeta(\cdot), \xi(\cdot)\)& Encoder/decoder for latent ODE state \\
    \hline
    % Losses
    \(\varepsilon_\theta\)              & Network estimating noise in diffusion models \\
    \(u_\tau, \bar\alpha_\tau\)                          & Noisy version of field and noise schedule coefficient in diffusion \\
    \(v_\theta, \psi_\tau(u)\)                        & Predicted velocity field and interpolated affine flow \\
    \(\mathcal{R}[u_\theta](x,t)\)      & Residual of the PDE operator \\
    \(g(x,t)\)                          & Source term in PDE \\
    \bottomrule
  \end{tabular}
  \caption{Unified notation across spatial, temporal, and loss components in our paper.}
  \label{tab:notation}
\end{table}

\section{Modular Decomposition}
\label{sec:decomposition}

We systematically review \arabic{baselines} representative data-driven fluid simulation methods reported across major venues (ICLR, NeurIPS, ICML, JMLR, Nat. Mach. Intell., etc.) and decompose each method into four key design dimensions under a Taylor-expansion-inspired formulation:
Design $=$ ``\textbf{Spatial representation}'' $+$ ``\textbf{Temporal representation}'' $+$ ``\textbf{Loss function}'' $+$ Additional Technique,
where the “Additional Technique” captures the distinctive architectural or training innovation introduced by each work.

This decomposition allows us to (i) standardize diverse neural solver designs into a comparable format, (ii) isolate the contribution of each design choice (\textit{e.g.}, spatial vs. temporal module), and (iii) facilitate cross-method analysis across 10 canonical flow scenarios under a unified lens. Instead of re-running all models, we extract the essential architectural and training components from each publication and categorize them in Table~\ref{tab:method_comparison}, thereby enabling a principled, reproducible, and cost-efficient comparison framework. Nowadays, more architectures are being applied to fluid simulation, such as LLMs~\cite{luo2025large}; however, due to page limitations, they are not discussed in this paper.

\begin{center}
\footnotesize
\setlength{\tabcolsep}{3mm}
\renewcommand{\arraystretch}{1.1}

\begin{longtable}{c|l|c|c|c|c|l}
\caption{We propose that the design of data-driven fluid simulation models can be analogized to a Taylor expansion. Design $=$ ``\textbf{Spatial representation}'' $+$ ``\textbf{Temporal representation}'' $+$ ``\textbf{Loss function}'' $+$ Additional Technique, where the additional item represents the key innovations introduced in each method.}
\label{tab:method_comparison}
\\
\toprule
\textbf{\#} & \textbf{Method} & \textbf{Publication} & \textbf{Spatial Repre.} & \textbf{Temporal Repre.} & \textbf{Loss} & \textbf{Additional Technique} \\
\midrule
\endfirsthead

\multicolumn{7}{c}{{\bfseries Table~\thetable{} -- continued from previous page}} \\
\toprule
\textbf{\#} & \textbf{Method} & \textbf{Publication} & \textbf{Spatial Repre.} & \textbf{Temporal Repre.} & \textbf{Loss} & \textbf{Additional Technique} \\
\midrule
\endhead

\midrule \multicolumn{7}{r}{{Continued on next page}} \\
\endfoot

\bottomrule
\endlastfoot

1 & \tiny{FNO~\citep{lifourier}} & ICLR 2021 & Fourier & Autoregression & Variable & Fourier integral operator \\
2 & \tiny{AFNO~\citep{guibas2021efficient}} & ICLR 2022 & Fourier & Autoregression & Variable & Mixing of tokens \\
3 & \tiny{Geo-FNO~\citep{li2023fourier}} & JMLR & Fourier & Autoregression & Variable & Data deformation \\
4 & \tiny{PINO~\citep{li2021physics}} & ACM Data Sci. & Fourier & Autoregression & PDE loss & Efficient derivative compute \\ 
5 & \tiny{U-NO~\citep{rahman2022u}} & TMLR & Fourier & Autoregression & Variable & U-net architecture \\ 
6 & \tiny{F-FNO~\citep{tran2022factorized}} & ICLR 2023 & Fourier & Autoregression & Variable & Factorize Fourier transform \\ 
7 & \tiny{CFNO~\citep{brandstetter2022clifford}} & ICLR 2023 & Fourier & Autoregression & Variable & Clifford algebras \\ 
8 & \tiny{CMWNO~\citep{xiao2022coupled}} & ICLR 2023 & Fourier & Autoregression & Variable & Multiwavelet decomposition \\ 
9 & \tiny{$G$-FNO~\citep{helwig2023group}} & ICML 2023 & Fourier & Autoregression & Variable & Group equivariant layers \\
10 & \tiny{GINO~\citep{li2023geometry}} & NeurIPS 2023 & Graph/Fourier & Autoregression & Variable & Discretization convergence \\ 
11 & \tiny{DAFNO~\citep{liu2023domain}} & NeurIPS 2023 & Fourier & Autoregression & Variable &  Smoothed characteristic func\\ 
12 & \tiny{PAC-FNO~\citep{jeon2024pac}} & ICLR 2024 & Fourier & Autoregression & Variable &  Parallel structure \\ 
13 & \tiny{AM-FNO~\citep{xiao2024amortized}} & NeurIPS 2024 & Fourier & Autoregression & Variable &  Amortized parameterization \\ 
14 & \tiny{DE-DON~\citep{qiu2024derivative}}& NeurIPS 2024 & MLP/Fourier & Autoregression &  Derivative  &   Dimension reduction \\ 
15 & \tiny{MCNP~\citep{zhang2023monte}} & TPAMI & Fourier & Autoregression & Neural MC & Mathematical expectation \\ 
16 & \tiny{ST-FNO~\citep{cao2025spectral}} & ICLR 2025 & Fourier & Time-depth Conv & Variable & Spectral fine-tuning \\ 
17 & \tiny{scOT~\citep{herde2024poseidon}}& NeurIPS 2024 & Self-atten & Temporal Bund & Variable & Large-scale pretraining \\ 
18 & \tiny{DiffPDE~\citep{huang2024diffusionpde}}& NeurIPS 2024 & Convolution & Autoregression & Noise & Guided Diffusion \\ 
19 & \tiny{\citet{shysheya2024on}} & NeurIPS 2024 & Convolution & Autoregression & Noise & Flexible pre- and post-training \\
20 & \tiny{\citet{valencia2025learning}} & ICLR 2025 & Graph & Autoregression & Noise & Latent multi-scale GNN diffusion  \\
21 & \tiny{\citet{bastek2025physicsinformed}} & ICLR 2025 & Convolution & Autoregression & Noise & Physics informed diffusion \\
22 & \tiny{TA-BN~\citep{zheng2024improving}} & NeurlPS 2024 &Convolution&Neural ODE&Variable&Temporal
adaptive BatchNorm\\
23 &\tiny{\citet{sholokhov2023physics}}&Sci. Rep.&MLP&Neural ODE&variable& Collocation Points\\
24 & \tiny{\citet{ghanem2025learning}}&AAAI 2025&MLP&Neural ODE&Variable&Physical-informed loss\\
25 &\tiny{PGODE~\citep{luo2024pgodehighqualitydynamicsmodeling}}&ICML2024&Self-atten&Neural  ODE&Variable&Environment Parameter\\
26 &\tiny{Treat~\citep{huang2024physicsinformedregularizationdomainagnosticdynamical}}&NeurlPS 2024&Self-atten&Neural ODE&Vairable&Physical-informed regularization\\
27 &\tiny{Pure\citep{NEURIPS2024_bd92deba}}&NeurlPS 2024&Self-atten&Neural ODE&Vairable& Multi-view context information\\
28&\tiny{EGODE\citep{yuan2024egode}}&NeurlPS 2024&Self-atten&Neural ODE&Vairable&Event-attended information\\
29&\tiny{Prometheus\citep{wu2024prometheus}}&ICML 2024&Self-atten&Neural ODE&Vairable&disentangled representations\\

30&\tiny{HOPE\citep{pmlr-v202-luo23f}}&ICML 2023&Self-atten&Neural ODE&Vairable&High order information\\

31&\tiny{LGODE\citep{LG-ODE}}&NeurlPS 2020&Self-atten&Neural ODE&Vairable&Temporal Self-attention\\

32 & \tiny{\citet{brandstetter2022message}} & ICLR 2022 & Graph & Temporal Bund & Variable & Model training stability \\

33 & \tiny{BENO\citep{wang2024beno}} & ICLR 2024 & Graph & Autoregression & Variable & Complex geometry embedding \\

34 & \tiny{UPT\citep{alkin2024universal}} & NeurIPS 2024 & Self-atten & Autoregression & Variable & Scalability across discretization \\

35 & \tiny{Transolver\citep{wu2024transolver}} & ICML 2024 & Self-atten & Autoregression & Variable & Physics-Attention \\

36 & \tiny{OMO\citep{pmlr-v235-xiao24c}} & ICML 2024 & Self-atten & Autoregression & Variable & Orthogonal Attention \\

37 & \tiny{LSM\citep{wu2023LSM}} & ICML 2023 & Self-atten & Autoregression & Variable & Latent Propagation \\

38 & \tiny{GNOT\citep{hao2023gnot}} & ICML 2023 & Self-atten & Autoregression & Variable & Linear Attention \\

39 & \tiny{Galerkin-Trans\citep{cao2021choose}} & NeurIPS 2021 & Self-atten & Autoregression & Variable & Petrov-Galerkin projection \\

40 & \tiny{MWT\citep{gupta2021multiwaveletbased}} & NeurIPS 2021 & Convolution & Autoregression & Variable & Multi-wavelet filters \\

41 & \tiny{CARE~\citep{luo2023care}} & NeurIPS 2023 & Graph & ODE & Variable & Context acquirement \\

42 & \tiny{\citet{bryutkin2024hamlet}} & ICML 2024 & Graph$+$SA & Next-step & Variable & Modular input encoders\\

43 & \tiny{LAMP~\citep{wu2023learning}} & ICLR 2023 & Graph & Next-step & Variable & GNN-based actor-critic for policy\\

44 & \tiny{DINo~\citep{yin2022continuous}} & ICLR 2023 & Implicit &  Neural ODE & Variable &  Extrapolates at arbitrary loc\\
45 & \tiny{\citet{zhao2023computationally} }& Adv Water Resour & Graph Conv & Next-step & Variable & GNN Grad-CAM\\
46 & \tiny{FCN~\citep{he2022flow}}  & PHYS FLUIDS. & Graph & Next-step & Variable & Vortex force contribution\\
47 & \tiny{GNODE~\citep{bishnoi2022enhancing}} & ICLR 2023 & Graph & Neural ODE & Variable & Encode the constraints explicitly \\
48 & \tiny{MAgNet~\citep{boussif2022magnet}} & NeurIPS 2022 & Graph & Next-step & Variable & Implicit neural representation\\
49 & \tiny{TIE~\citep{shao2022transformer}} & ECCV 2022 & Self-atten & Next-step & Variable & Implicit Edges\\
50 & \tiny{\citet{han2022predicting}} & ICLR 2022 & Self-atten & Self-atten & Variable & Encoder-decoder structure\\
51 & \tiny{MGN~\citep{pfaff2020learning}} & ICLR 2021 & Graph & Next-step & Variable & Mesh graph representation\\
52 & \tiny{GNS~\citep{sanchez2020learning}} & ICML 2022 & Graph & Nest-step & Variable & Message passing layers\\
53 & \tiny{RSteer~\citep{wang2022approximately}} & ICLR 2023 & Graph & Next-step & Variable & Approximately equivariant
networks\\
54 & \tiny{EquNet~\citep{wang2020incorporating}} & ICLR 2021 & Conv & Temporal Bund & Variable & Incorporate symmetry \\
55 & \tiny{TF-Net~\citep{wang2020towards}} & KDD 2020 & Conv & Next-step & Variable & Marry two simulation \\
56 & \tiny{DPUF~\citep{lee2019data}} & J. Fluid Mech. & Conv & Next-step & Physical & Conservation of mass, momentum\\
57 &  \tiny{\citet{wessels2020neural}} & CoM Appl M & MLP & Next-step & Variable & Updated Lagrangian\\
58 & \tiny{FGN~\citep{li2022graph}} & Comput Graph & Graph & Next-step & Variable & Node-focused and edge-focused\\
59 & \tiny{MCC~\citep{liu2023fast}} & AAAI 2023 & Conv & Next-step & Variable &  Dynamic multi-scale gridding \\
60 & \tiny{LFlows~\citep{torres2023lagrangian}} & ICLR 2024 & MLP & Temporal Bund & Flow & Satisfy the continuity equation\\
61 & \tiny{\citet{li2024synthetic}}& Nat. Mach. Intell. & Conv & Next-step & Noise & Two different classes of DM\\
62 & \tiny{DeepONet~\citep{lu2021learning}} & Nat. Mach. Intell. & MLP & Next-step & Variable & Two sub-nets\\
63 & \tiny{\citet{wang2021learning}} & Sci. Advection & MLP & Next-step & Physics &  Soft penalty constraints for law \\
64 & \tiny{MIONet~\citep{jin2022mionet}} & Siam J Sci Comput & MLP & Next-step & Physics & A low-rank approximation\\
65 & \tiny{\citet{geneva2022transformers}} & Neural Networks & Self-atten & Self-atten & Variable & Koopman dynamics\\
66 & \tiny{NOMAD~\citep{seidman2022nomad}} & NeurIPS 2022 & MLP & Next-step & Variable &  Nonlinear decoder map\\
67 & \tiny{HyperDON~\citep{lee2023hyperdeeponet}} & ICLR 2023 & MLP & Next-step & Variable & Use a hypernetwork\\
68 & \tiny{B-DON~\citep{lin2023b}} & J. Comput. Phy. & MLP & Next-step & Variable & Bayesian network \\
69 & \tiny{SVD-DON~\citep{venturi2023svd}} & CoM Appl M & MLP & Next-step & Variable & Orthogonal decomposition\\
70 & \tiny{L-DON~\citep{kontolati2023learning}} & Arxiv 2023 & MLP & Next-step & Variable &  Low-dimensional latent space\\
71 & \tiny{CNO~\citep{raonic2023convolutional}} & NeurIPS 2023 & Conv & Autoregression & Variable & Adaptations for convolution\\
72 & \tiny{FactFormer~\citep{li2024scalable}} & NeurIPS 2023 & Self-atten & Self-atten & Variable & Axial factorized kernel\\
73 & \tiny{LNO~\citep{cao2023lno}} & Nat. Mach. Intell. &  Laplace & Autoregression & Variable & Pole-residue relationship\\
74 & \tiny{KNO~\citep{xiong2023koopman}} & APL ML & Koopman & Autoregression & Variable & Approximate the Koopman operator\\
75 & \tiny{ICON~\citep{yang2023context}} & PNAS & Self-atten & Self-atten & variable & In-context learning\\
76 & \tiny{FCG-NO~\citep{rudikov2024neural}}  & ICML 2024 & Frequency & Autoregression & Variable & Flexible conjugate gradient\\
77 & \tiny{PINN~\citep{raissi2019physics}} & J. Comput. Phy. & MLP & Next-step & Residual &  - \\
78 &  \tiny{PINN-SR~\citep{chen2021physics}} & Nat. Commun. & MLP & Next-step & Residual & Regularization loss\\
79 & \tiny{NSFnet~\citep{jin_nsfnets_2021}} & J. Comput. Phys. & MLP & Next-step & Residual & For N-S Eq.\\
80 & \tiny{MAD~\citep{huang2022meta}} & NeurIPS 2022 & MLP & Next-step & Variable & Meta-learning\\
81 & \tiny{\citet{toloubidokhti2023dats}} & ICLR 2024 & MLP & Next-step & Residual & Difficulty-aware task sampler \\
82 & \tiny{\citet{zhao2023pinnsformer}} & ICLR 2024 & Self-atten & Next-step & Residual & Pseudo sequences\\
83 & \tiny{PeRCNN~\citep{rao2023encoding}} & Nat. Mach. Intell. & Conv & Next-step & Variable & Forcibly encodes physics structure\\
84 & \tiny{FFM~\citep{kerrigan2023functional}} & AISTATS 2024 & Conv & Next-step & Flow & Function space generation\\
85 & \tiny{ECI~\citep{cheng2024gradient}} & ICLR 2025 & Fourier & Next-step & Flow & Integration of constraint\\
86 & \tiny{FourierFlow~\citep{wang2025fourierflow}} & Arxiv 2025 & Fourier & Temporal Bund & Flow & For spectral bias \& common-mode noise.\\
87 & \tiny{FunDS~\citep{yao2025guided}} & Arxiv 2025 & Fourier & Next-step & Noise & Function-space posterior sampling\\
88 & \tiny{FunDIff~\citep{wang2025fundiff}} & Arxiv 2025 & Self-atten  & Next-step & Noise & Latent diffusion process\\
89 & \tiny{SG-Diff~\citep{li2026self}} & KDD 2026 & Self-atten  & Next-step & Noise & mportance Weight strategy\\
\end{longtable}
\end{center}

\section{Supplementary Results on Additional Techniques}
\label{appendix:add_tech}

In fact, when conducting experiments for each method, researchers often incorporate techniques from the field of deep learning to achieve better performance, smoother optimization, and more stable training. Considering that, in practice, different users may add different additional techniques, we provide examples in our code showing how to quickly add any desired technique.
When designing the code repository, we reserve kwargs for each model to support various extensions. Using `\texttt{$\mathcal{X}$ = self-attention}' as an example, we list common techniques, including \texttt{GlobalFilter}, \texttt{Long-short term attention}, and \texttt{Adaptive Fourier attention}. Specifically, we show the experimental results on \texttt{Compressible N-S} with `\texttt{$\mathcal{Y}$ = next-step}' and `\texttt{$\mathcal{Z}$ = variable}' are shown below.

\begin{table}[h]
\centering
\setlength{\tabcolsep}{3mm}{
\begin{tabular}{lccc}
\toprule[1.1pt]
\textbf{Additional Technique} & \textbf{RMSE} & \textbf{nRMSE} & \textbf{fRMSE} \\
\midrule
\rowcolor[RGB]{234, 238, 234} \multicolumn{4}{l}{Self-attention + Next-step + Physical variable + \textit{\textbf{Additional Techniques}}} \\
% \rowcolor[RGB]{234, 238, 234} \multicolumn{5}{l}{\textit{\textbf{$\mathcal{X}$} + Next-step + Physical variable}} \\
Vanilla Self-attention       & 0.0571 & 0.0695 & 0.0035 \\
GlobalFilter                 & 0.0642 & 0.0711 & 0.0052 \\
Long-short term attention    & 0.0559 & 0.0709 & 0.0043 \\
Adaptive Fourier attention   & 0.0597 & 0.0699 & 0.0047 \\
\bottomrule[1.1pt]
\end{tabular}}
\caption{Performance comparison of various additional techniques.}
\label{tab:additional-techniques}
\end{table}

\begin{figure*}[htbp]
    \centering
    \scalebox{1.0}{
        \begin{minipage}{\textwidth}
            \centering
            \begin{subfigure}
                \centering
                \includegraphics[width=0.19\linewidth]{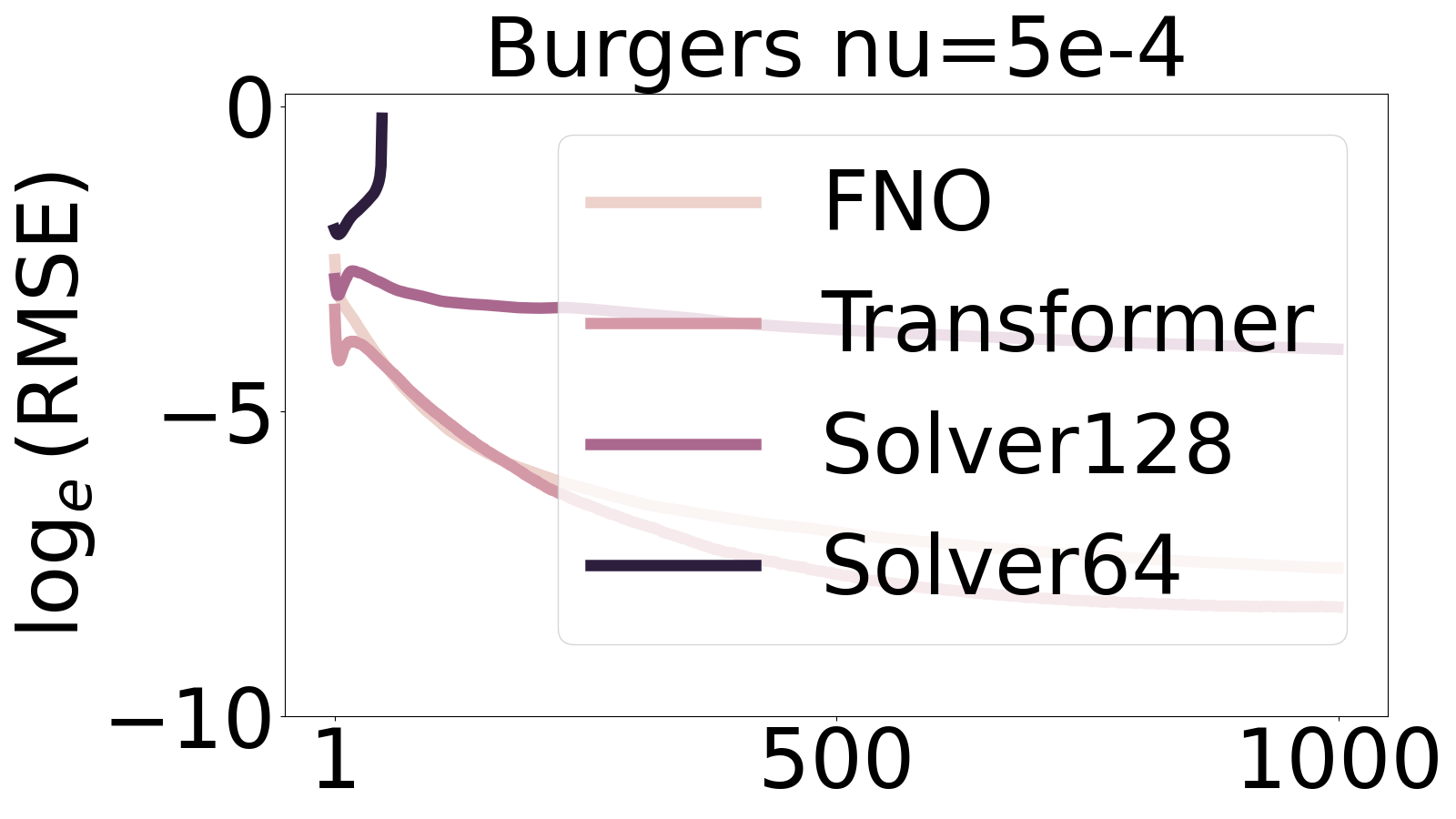}
            \end{subfigure}
            \hfill
            \begin{subfigure}
                \centering
                \includegraphics[width=0.19\linewidth]{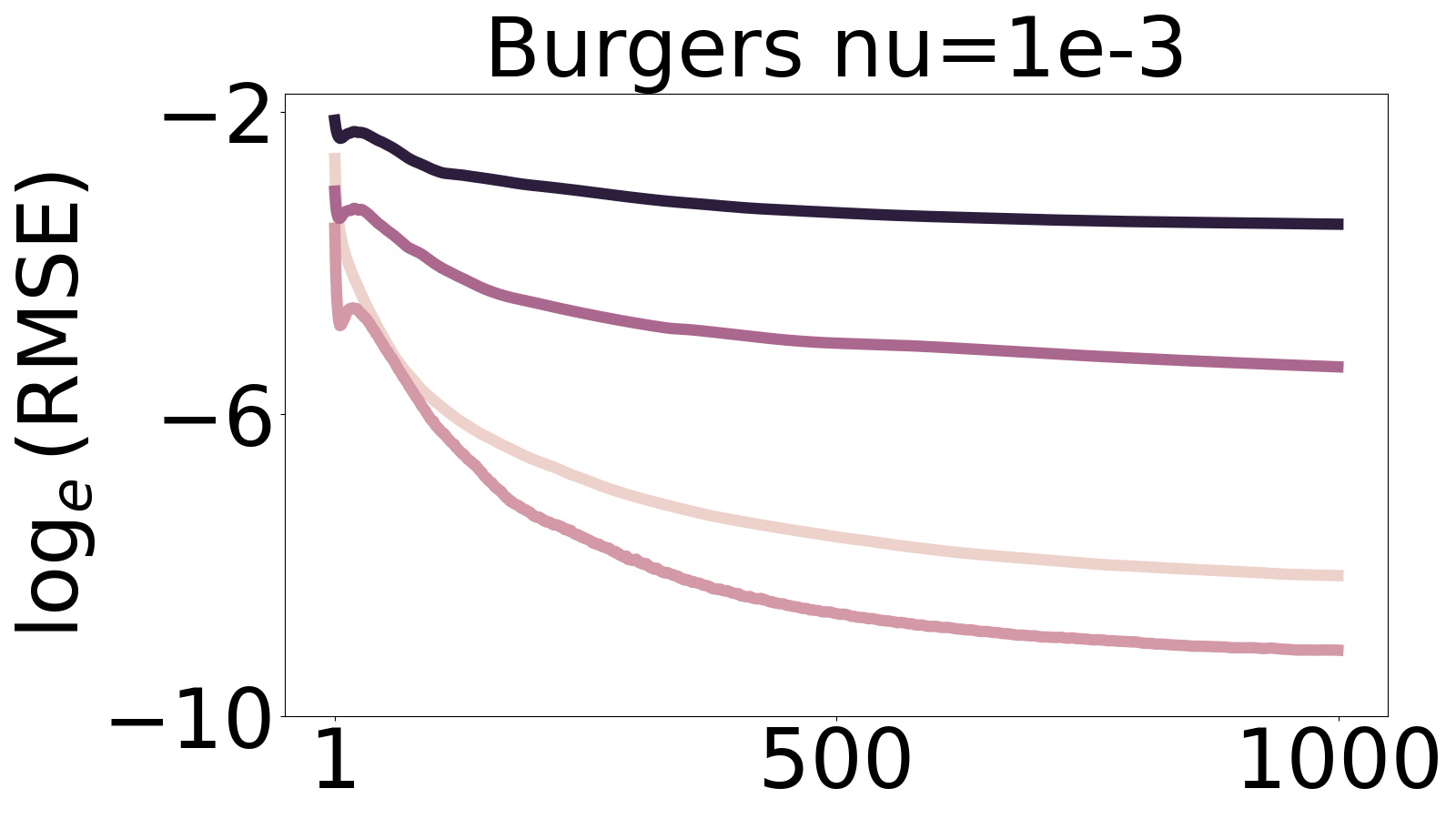}
            \end{subfigure}
            \hfill
            \begin{subfigure}
                \centering
                \includegraphics[width=0.19\linewidth]{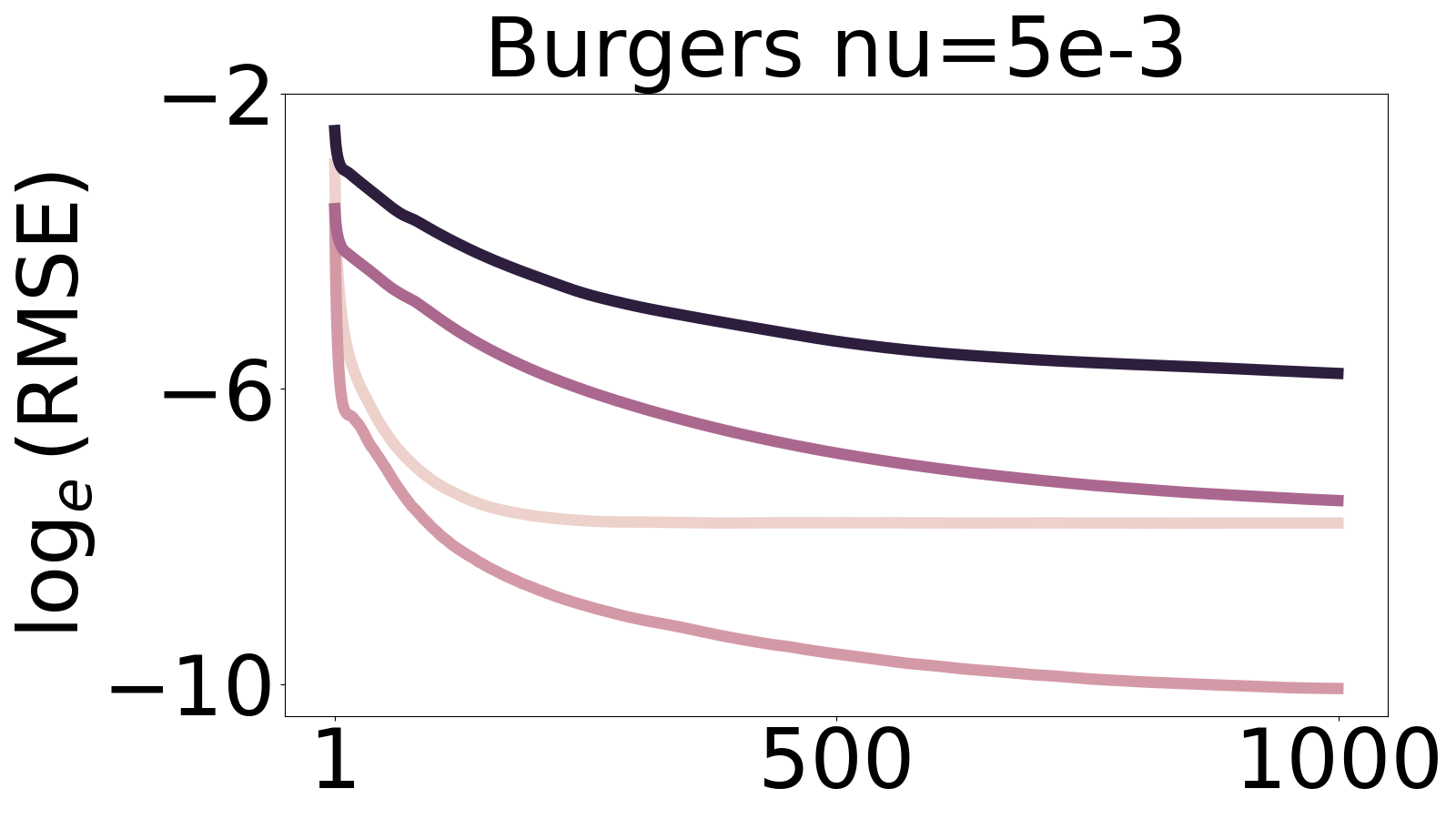}
            \end{subfigure}
            \hfill
            \begin{subfigure}
                \centering
                \includegraphics[width=0.19\linewidth]{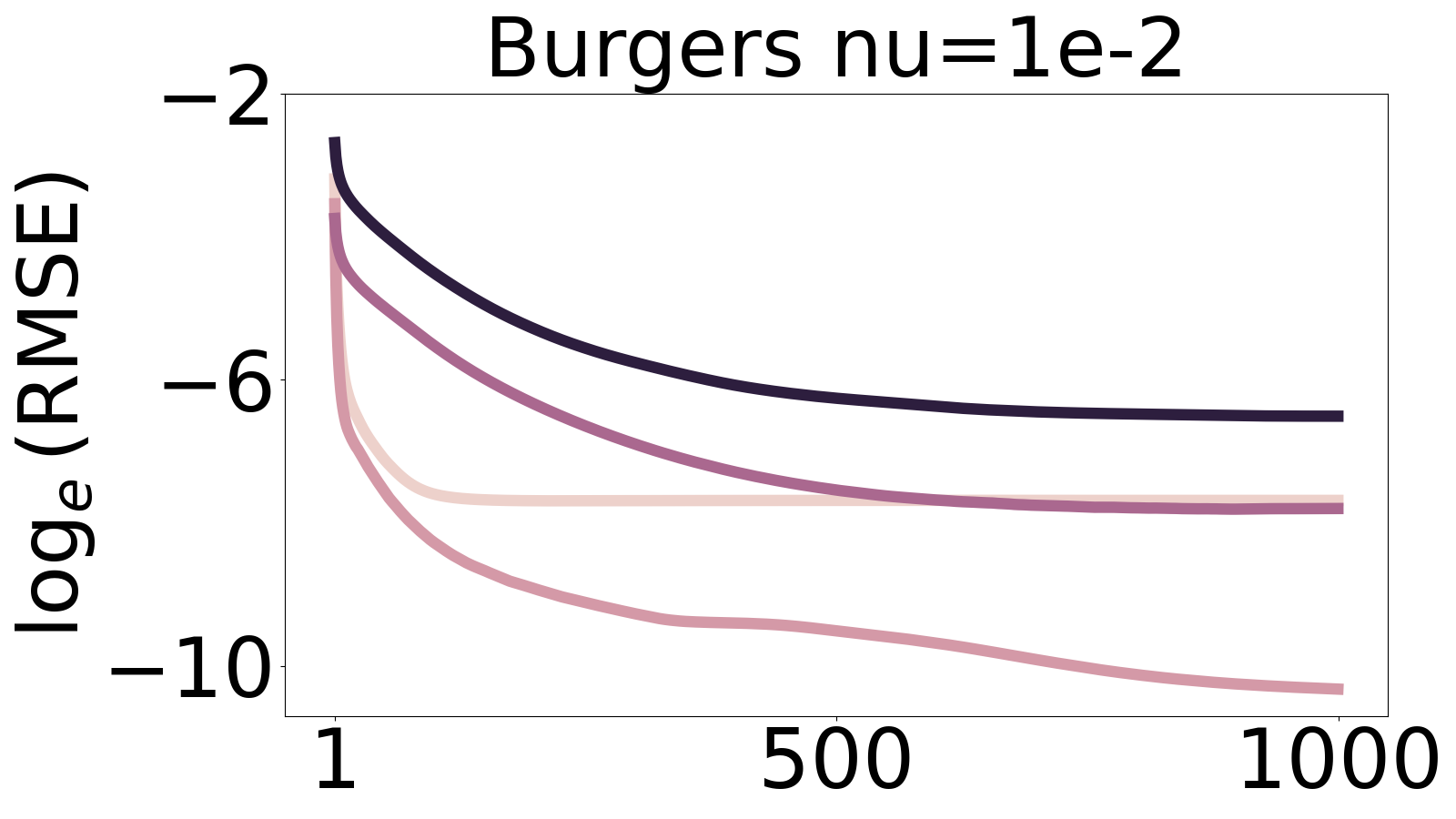}
            \end{subfigure}
            \hfill
            \begin{subfigure}
                \centering
                \includegraphics[width=0.19\linewidth]{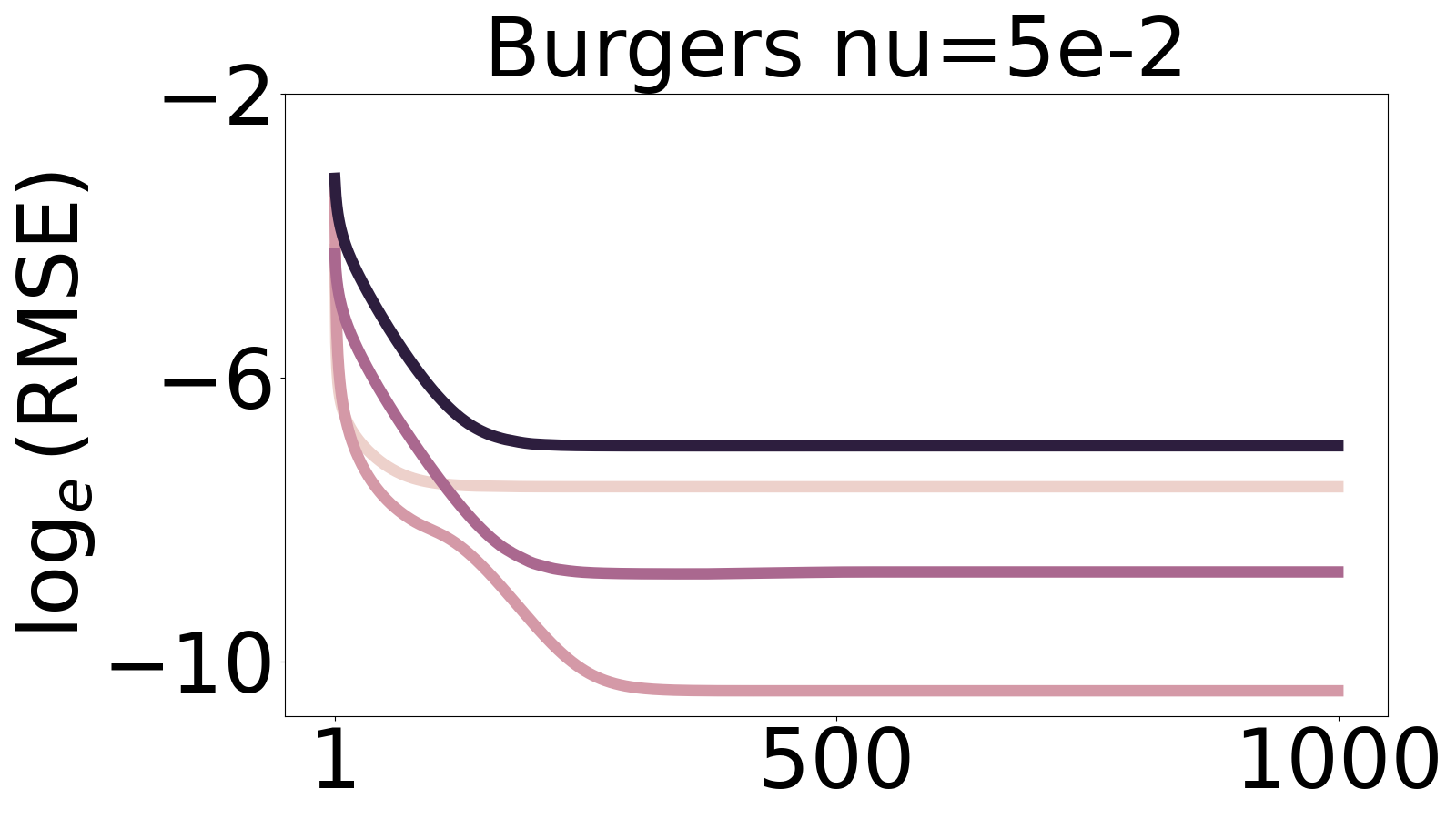}
            \end{subfigure}
        \end{minipage}
    }
    \caption{We compare FNO and Transformer against pseudo-spectral solver operating at lower resolutions on Burgers' equation across five kinematics viscosity. Solver $x$ implies pseudo-spectral solver operating at $x \times x$ resolution.}
    \label{fig:solver_burgers}
\end{figure*}

\begin{figure*}[htbp]
    \centering
    \scalebox{1.0}{
        \begin{minipage}{\textwidth}
            \centering
            \begin{subfigure}
                \centering
                \includegraphics[width=0.19\linewidth]{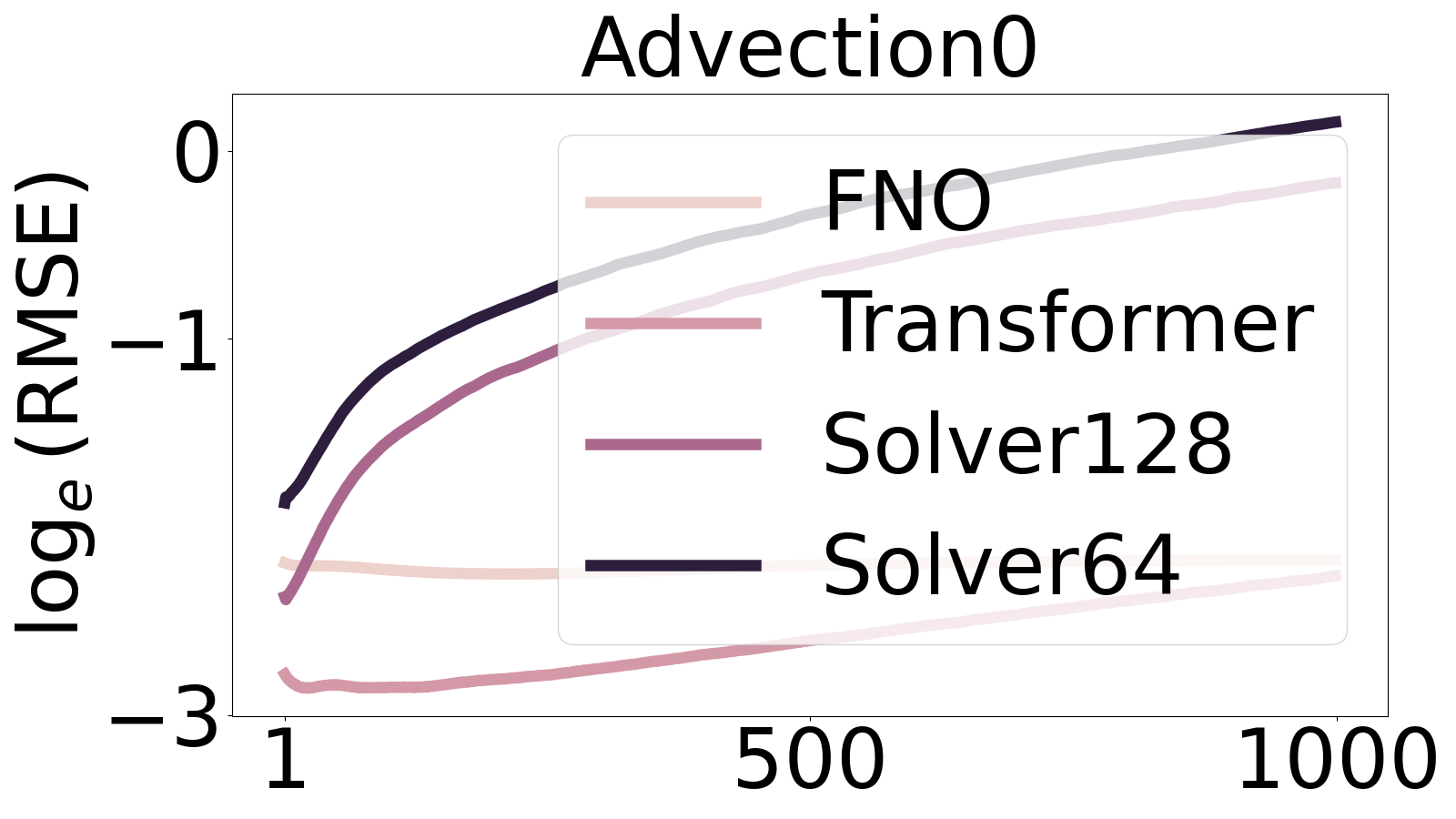}
            \end{subfigure}
            \hfill
            \begin{subfigure}
                \centering
                \includegraphics[width=0.19\linewidth]{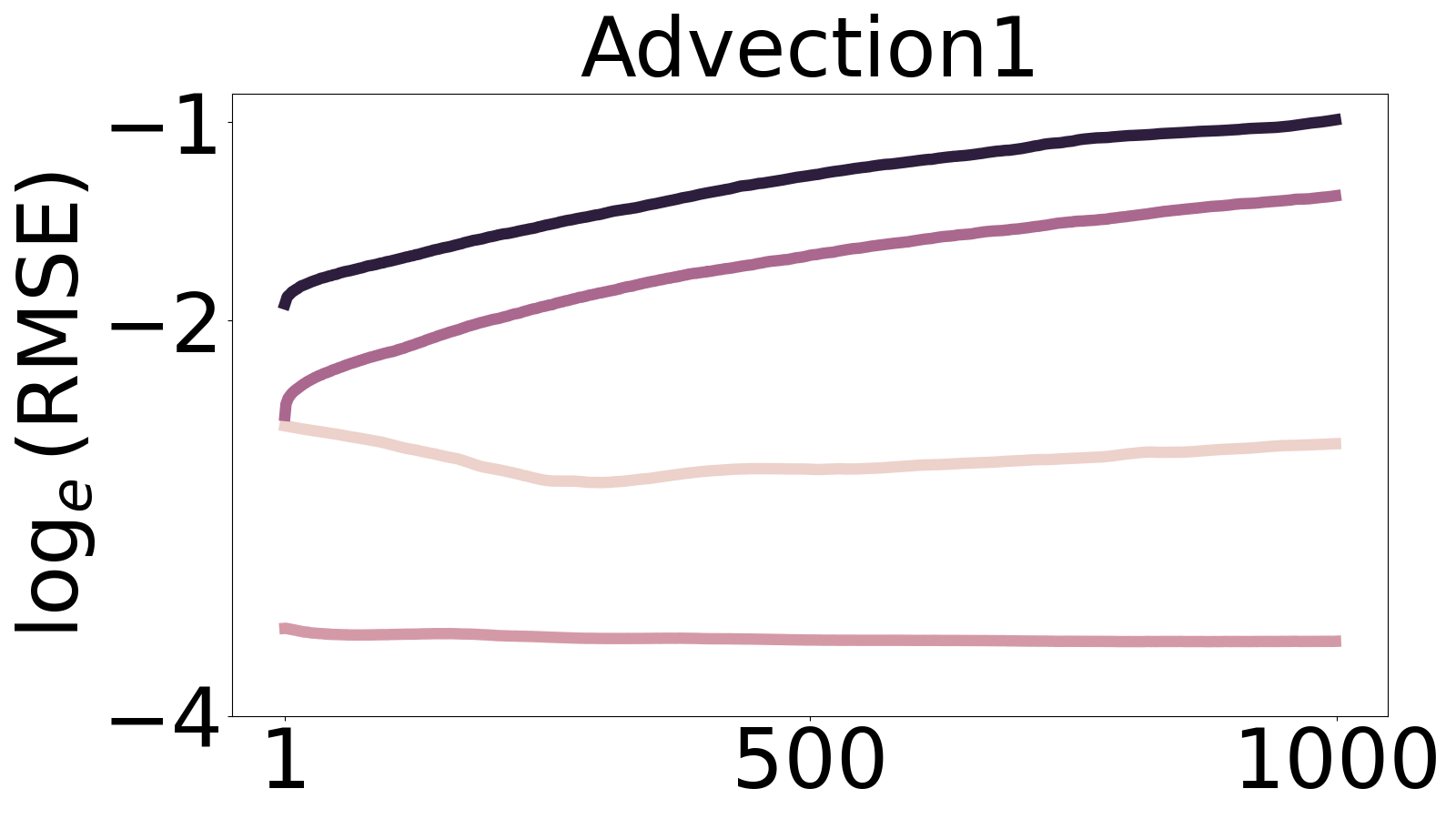}
            \end{subfigure}
            \hfill
            \begin{subfigure}
                \centering
                \includegraphics[width=0.19\linewidth]{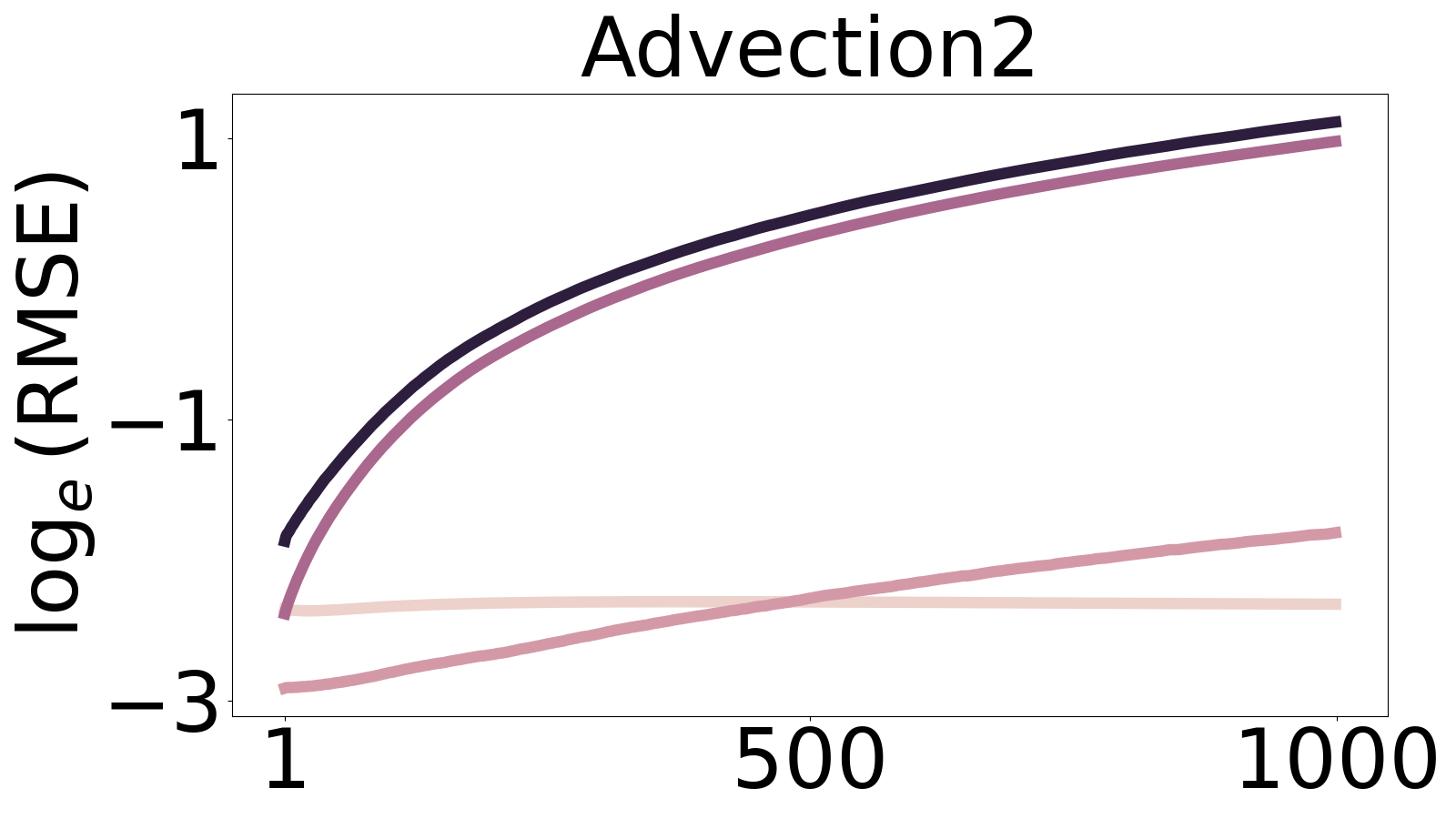}
            \end{subfigure}
            \hfill
            \begin{subfigure}
                \centering
                \includegraphics[width=0.19\linewidth]{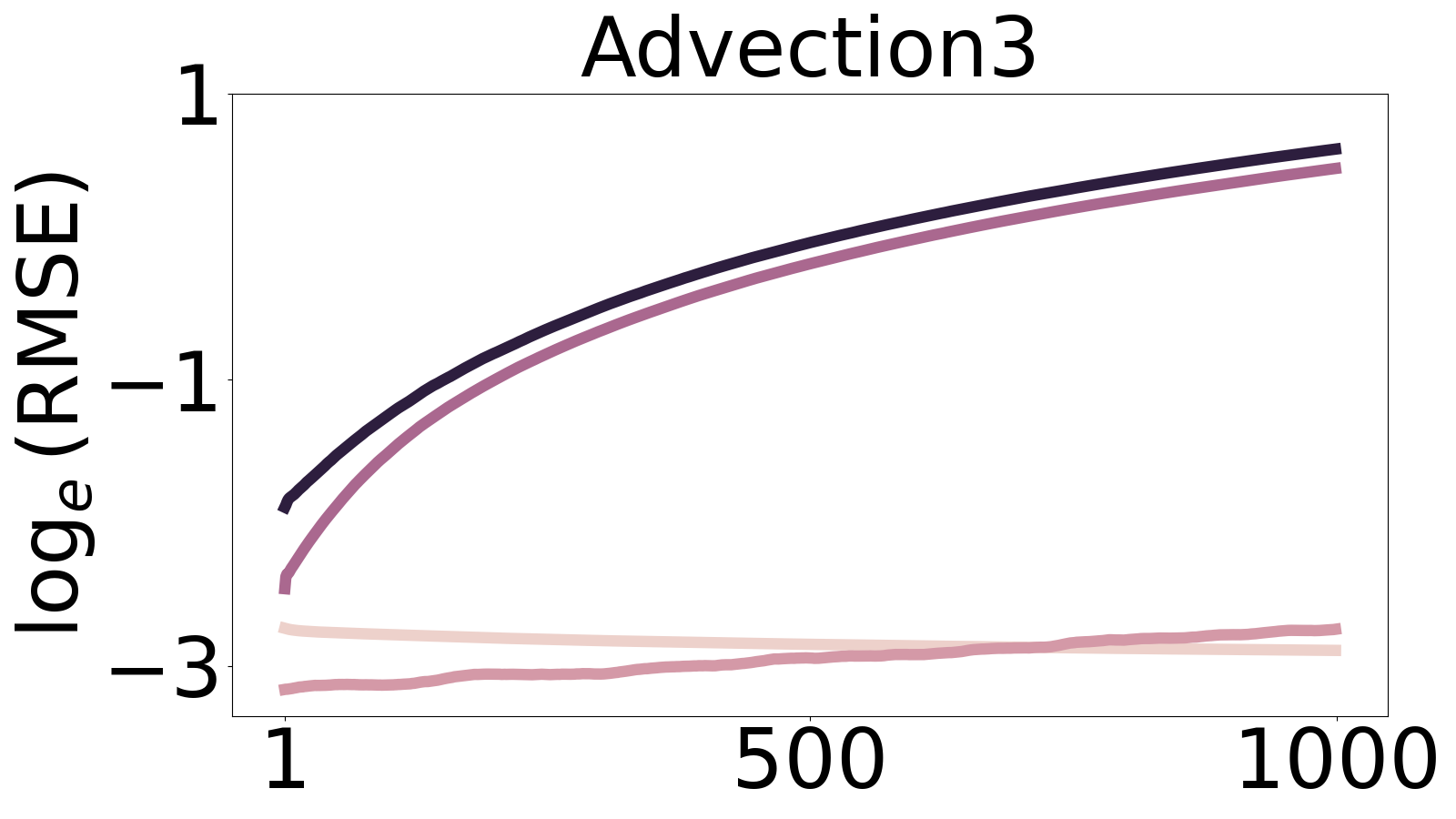}
            \end{subfigure}
            \hfill
            \begin{subfigure}
                \centering
                \includegraphics[width=0.19\linewidth]{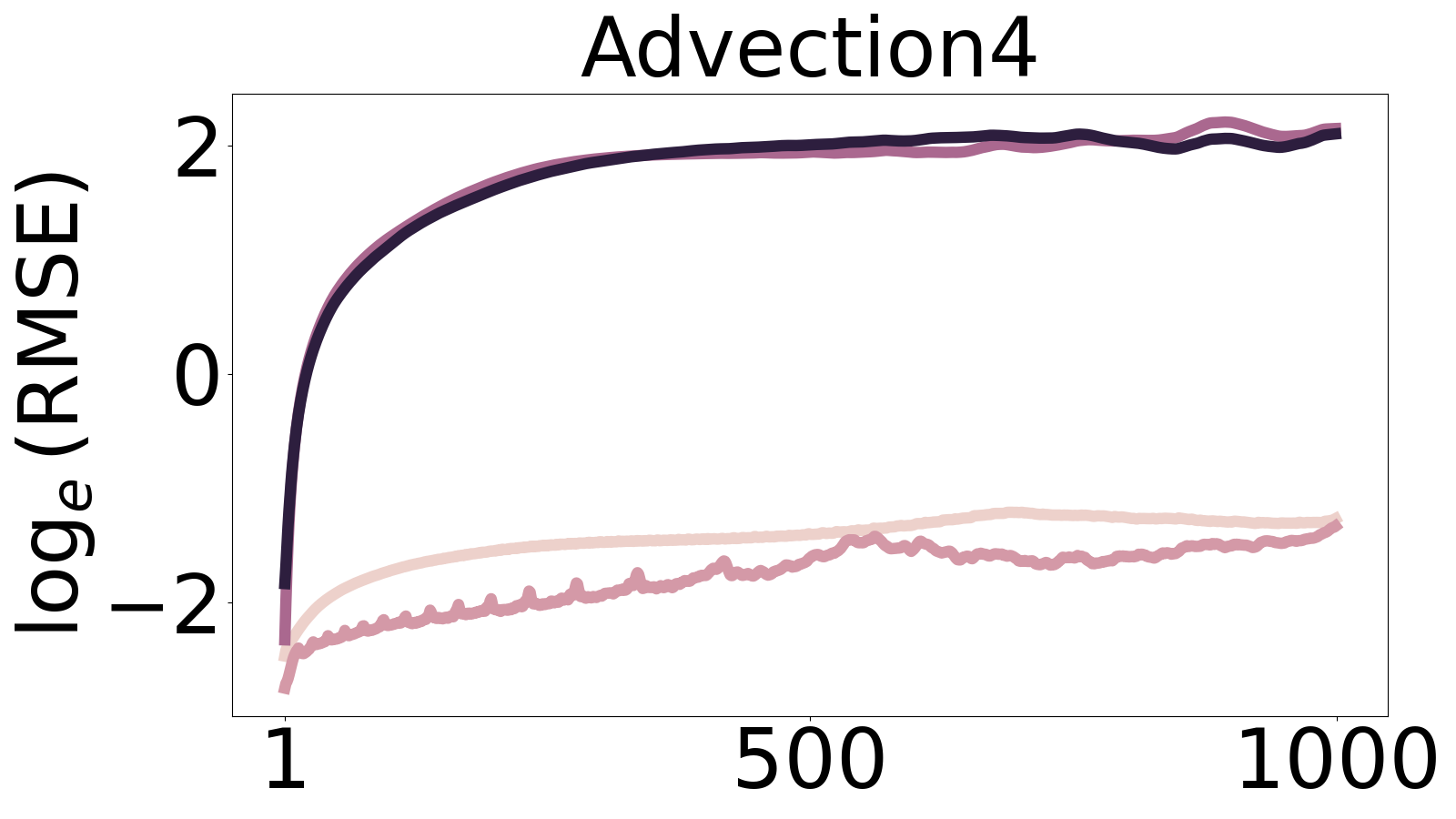}
            \end{subfigure}
        \end{minipage}
    }
    \caption{We compare FNO and Transformer against pseudo-spectral solver operating at lower resolutions on Advection equation across five advection speed types. Solver $x$ implies pseudo-spectral solver operating at $x \times x$ resolution.}
    \label{fig:solver_advection}
\end{figure*}

\section{Supplementary Results on Comparison with Traditional Solvers}
\label{appendix:comp_solvers}

\begin{figure}[h] % r=right, l=left; width controls text wrap
  \centering
  \vspace{-10pt} % adjust vertical spacing if needed
  \includegraphics[width=0.65\linewidth]{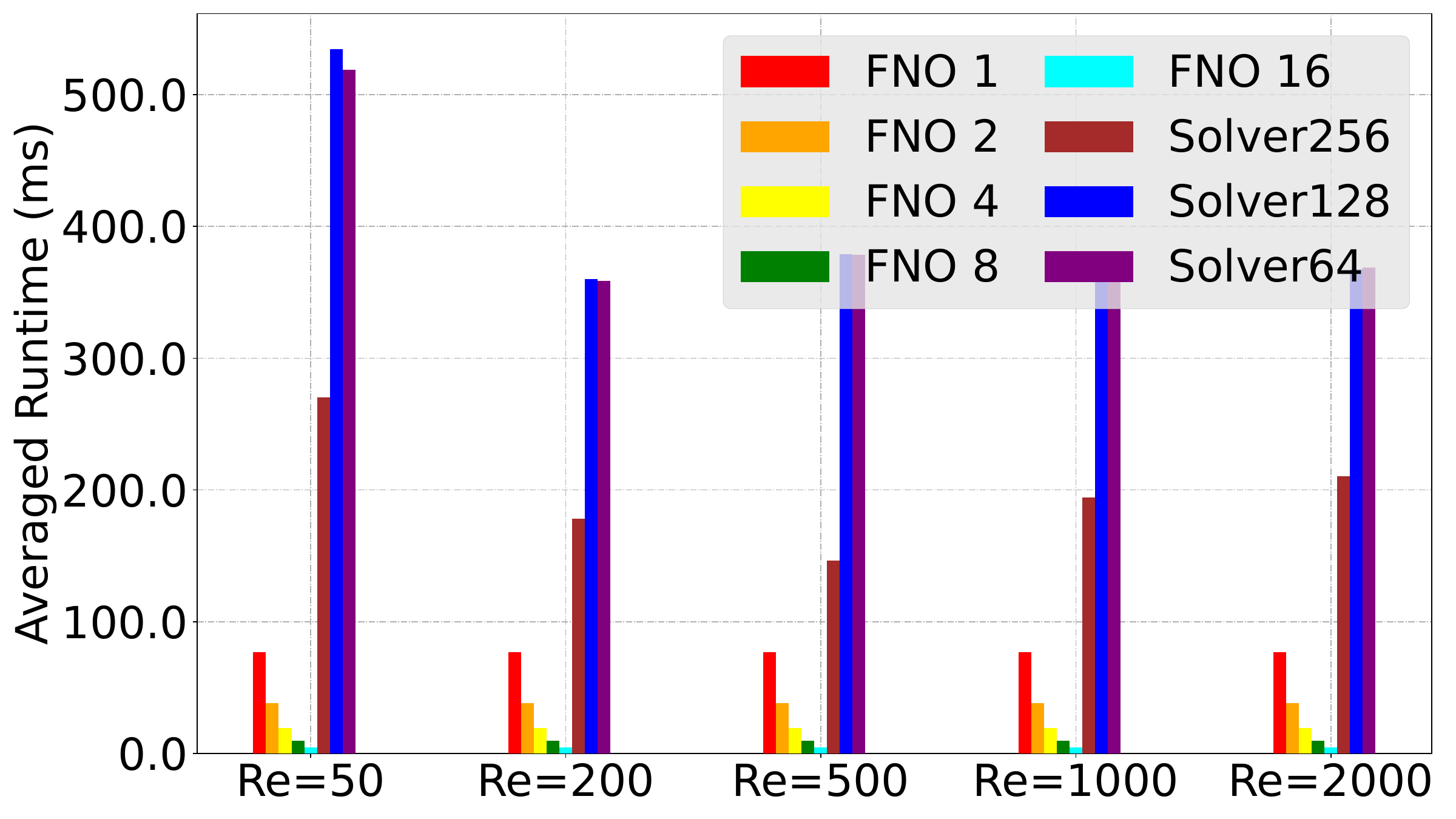}
  \caption{We evaluate the runtime performance of the FNO when trained to predict various stepsize against pseudo-spectral solver operating at lower resolutions for predicting up to $T=\frac{1}{2}$. Solver $x$ implies pseudo-spectral solver operating at $x \times x$ resolution. Data generated at $256 \times 256$ are considered ground truth. To ensure numerical stability, adaptive time-stepping governed by the CFL condition is adopted even for solvers at low resolution. \textit{FNO-$x$} denotes FNO trained to predict with stepsize $x \Delta t^*$, where $\Delta t^* = \tfrac{1}{32}$.}
  \label{fig:runtime_timesteps}
  \vspace{-10pt} % pull up text if there is too much white space
\end{figure}

To further demonstrate the performance of neural operators relative to traditional numerical solvers, we present additional rollout results and runtime comparisons. Figure~\ref{fig:solver_burgers} reports the evolution of the rollout error for Burgers' equation under five viscosity regimes, ranging from $\nu = 5 \times 10^{-4}$ to $\nu = 5 \times 10^{-2}$. We observe that neural operator models, particularly FNO, consistently achieve lower error over long rollouts compared to traditional finite-difference solvers at multiple resolutions (Solver64, Solver128). This advantage becomes more pronounced as the viscosity increases, where the numerical dissipation in coarse solvers leads to early error saturation. Similar trends are observed for the Advection equation (Figure~\ref{fig:solver_advection}), confirming the generality of this behavior across PDE types.

\begin{figure}[htbp]
  \centering
  %––––––––––––––––––––––––––––––––––––––––––––––––––––––––––––––––––––––––––––––
  \begin{minipage}[t]{0.48\textwidth}
    \centering
    \includegraphics[width=\linewidth]{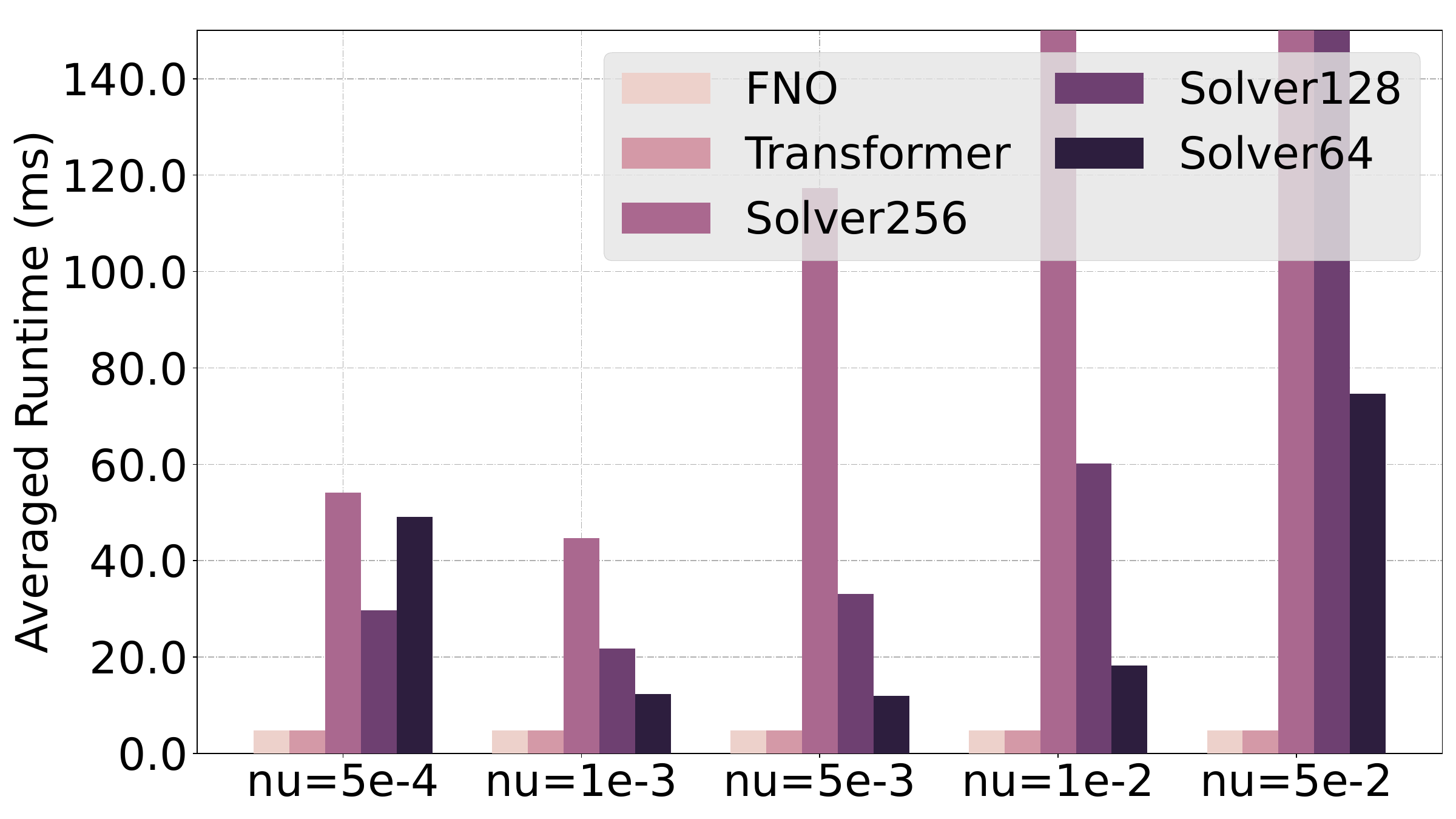}
    
    \vspace{0.3ex}
    {\small\itshape Burgers' Equation\label{fig:1}}
  \end{minipage}
  \hfill
  \begin{minipage}[t]{0.48\textwidth}
    \centering
    \includegraphics[width=\linewidth]{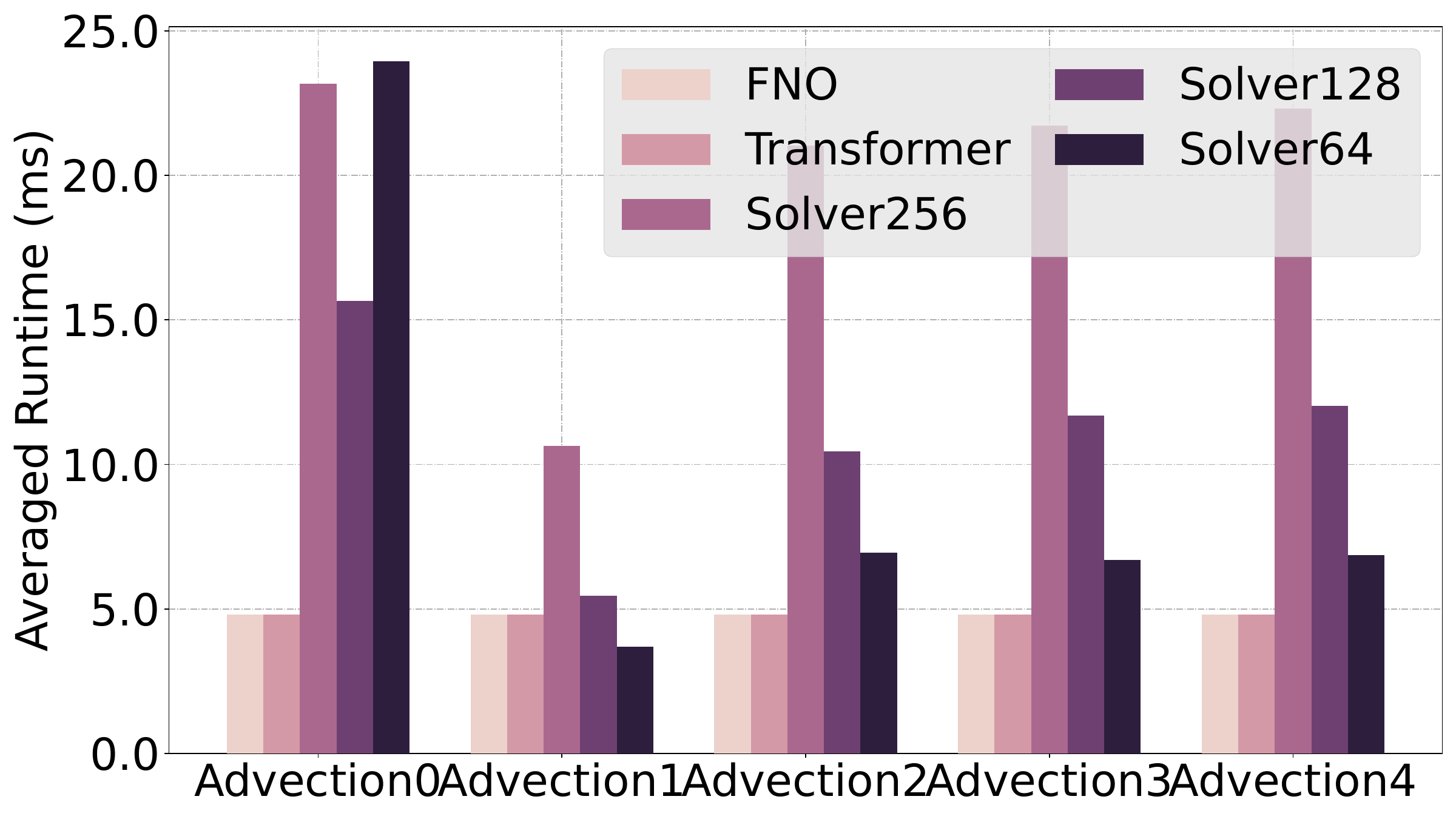}
    
    \vspace{0.3ex}
    {\small\itshape Advection Equation\label{fig:2}}
  \end{minipage}
  %––––––––––––––––––––––––––––––––––––––––––––––––––––––––––––––––––––––––––––––
  \caption{We compare the runtime of the FNO and Transformer against pseudo-spectral solver operating at lower resolutions for predicting up to $T=\frac{1}{32}$. Solver $x$ implies pseudo-spectral solver operating at $x \times x$ resolution. Data generated at $256 \times 256$ are considered ground truth. Note that to ensure numerical stability, adaptive time-stepping governed by CFL condition is adopted even for solvers at low-resolution.}
  \label{fig:runtime_solver}
\end{figure}

In terms of efficiency, Figure~\ref{fig:runtime_solver} compares the averaged per-step runtime across models and solver resolutions. FNO and Transformer-based models offer significant speedups over Solver 128 and Solver 256, especially when considering high-resolution solvers that incur large computational overhead. This speed advantage, combined with improved predictive accuracy, highlights the practical benefits of neural operators as surrogates for long-horizon PDE simulation. These results suggest that data-driven neural surrogates can provide a favorable trade-off between accuracy and computational cost, making them attractive for real-time or large-scale simulation scenarios. In conclusion, FNO not only achieves better predictive accuracy but also faster speed.

\subsection{Neural Solver with Various Step Sizes}
\label{appendix:comp_solvers_timesteps}
\begin{figure*}[htbp]
    \centering
    \scalebox{1.0}{
        \begin{minipage}{\textwidth}
            \centering
            \begin{subfigure}
                \centering
                \includegraphics[width=0.19\linewidth]{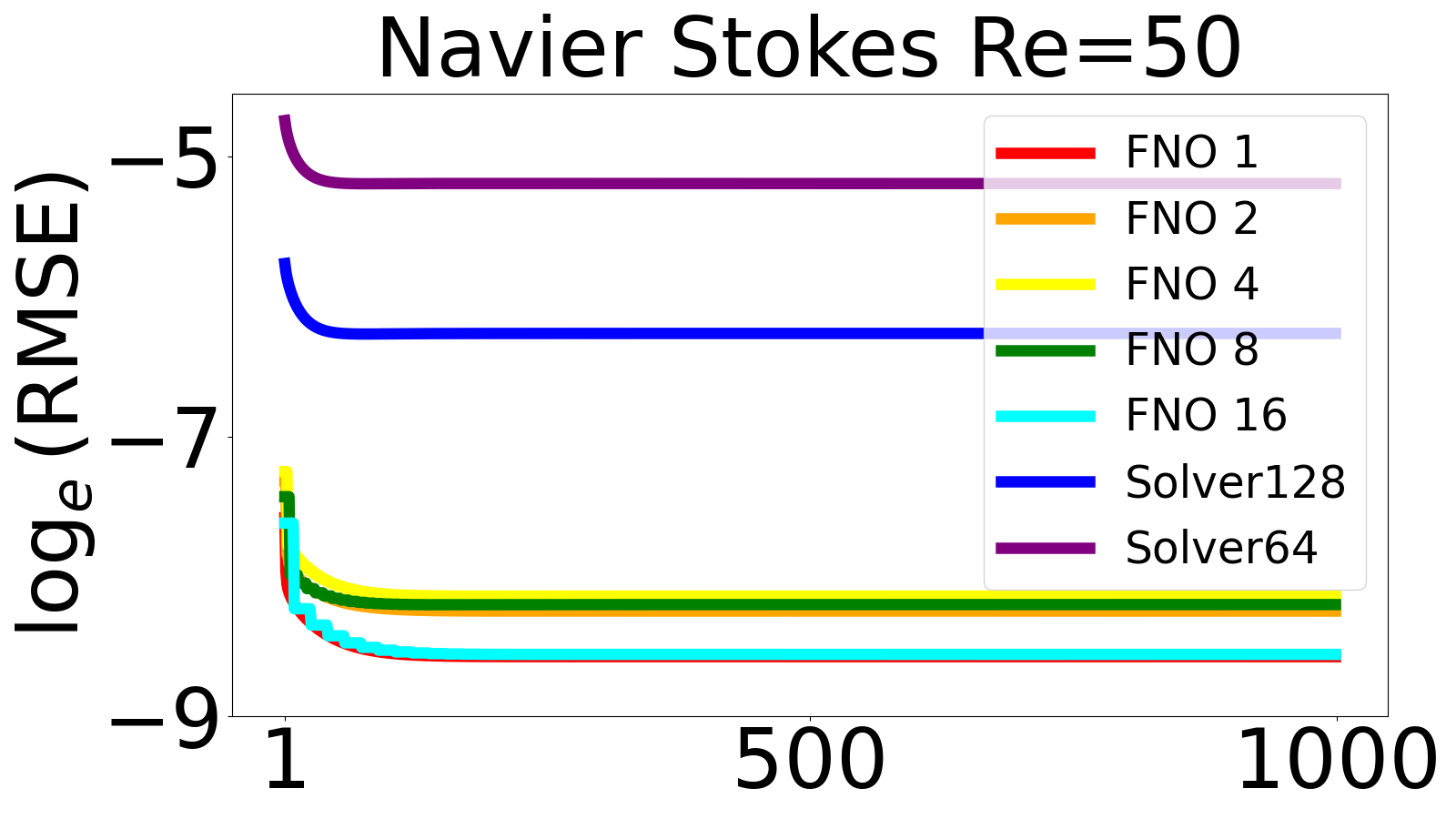}
            \end{subfigure}
            \hfill
            \begin{subfigure}
                \centering
                \includegraphics[width=0.19\linewidth]{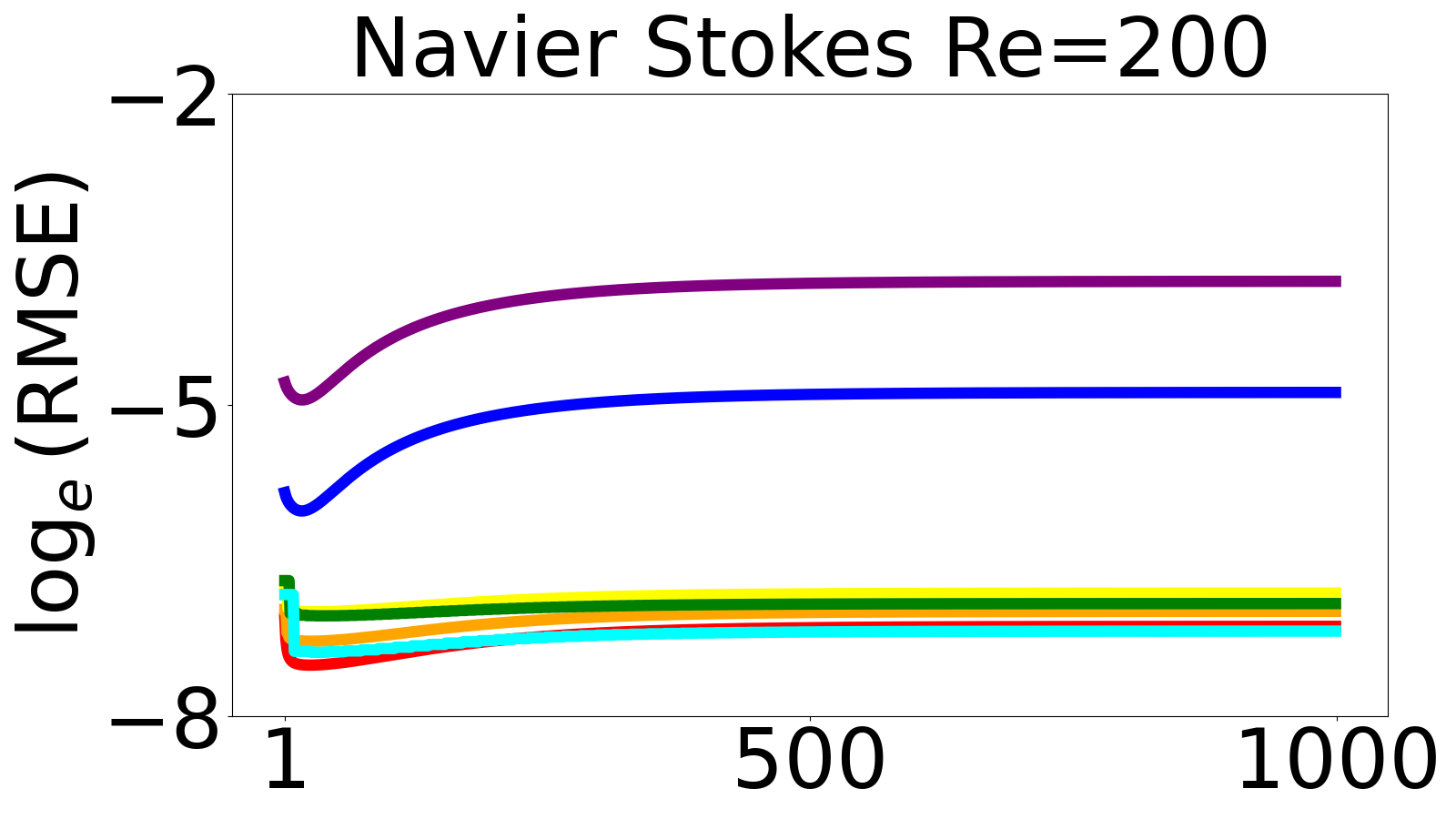}
            \end{subfigure}
            \hfill
            \begin{subfigure}
                \centering
                \includegraphics[width=0.19\linewidth]{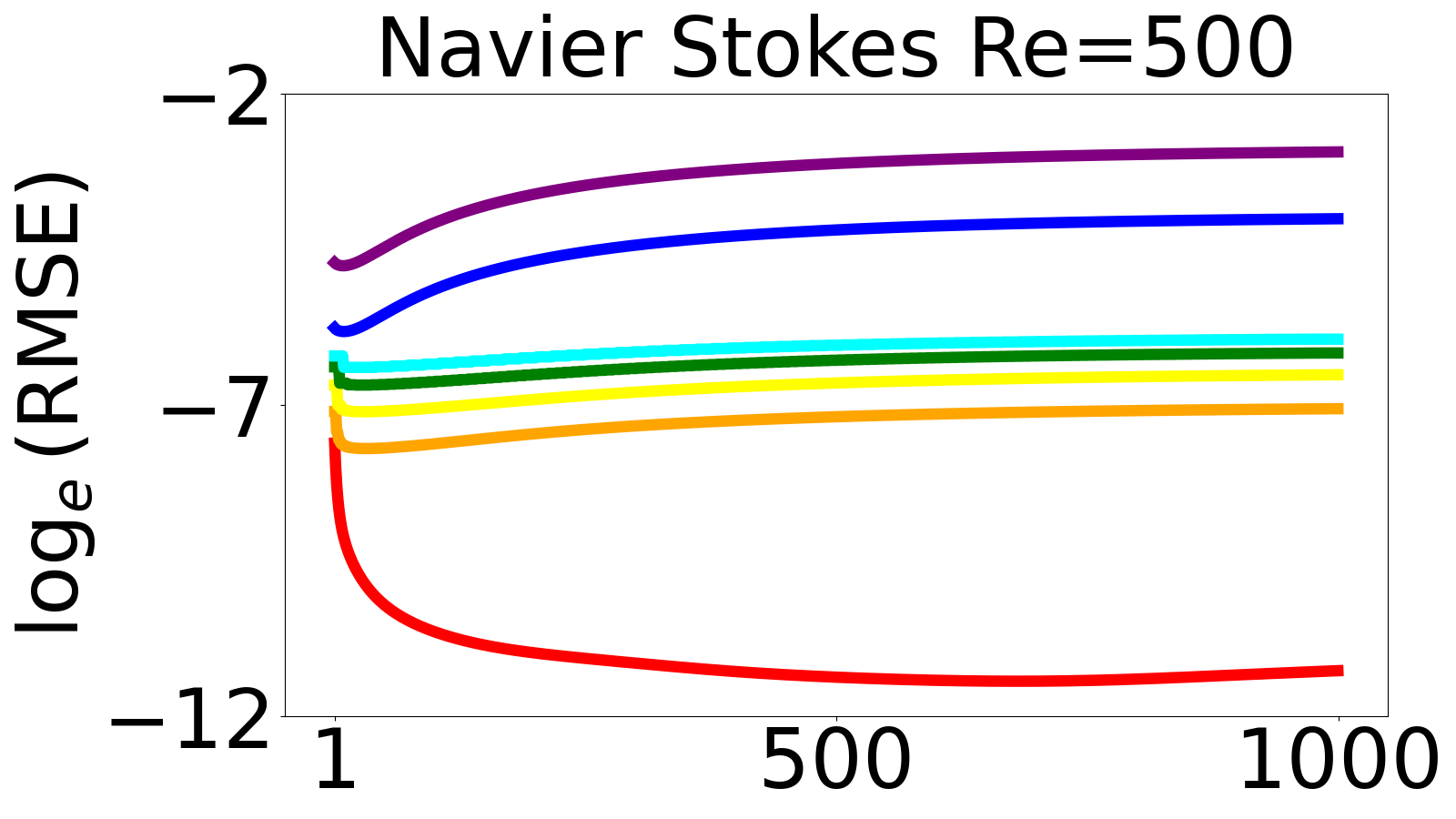}
            \end{subfigure}
            \hfill
            \begin{subfigure}
                \centering
                \includegraphics[width=0.19\linewidth]{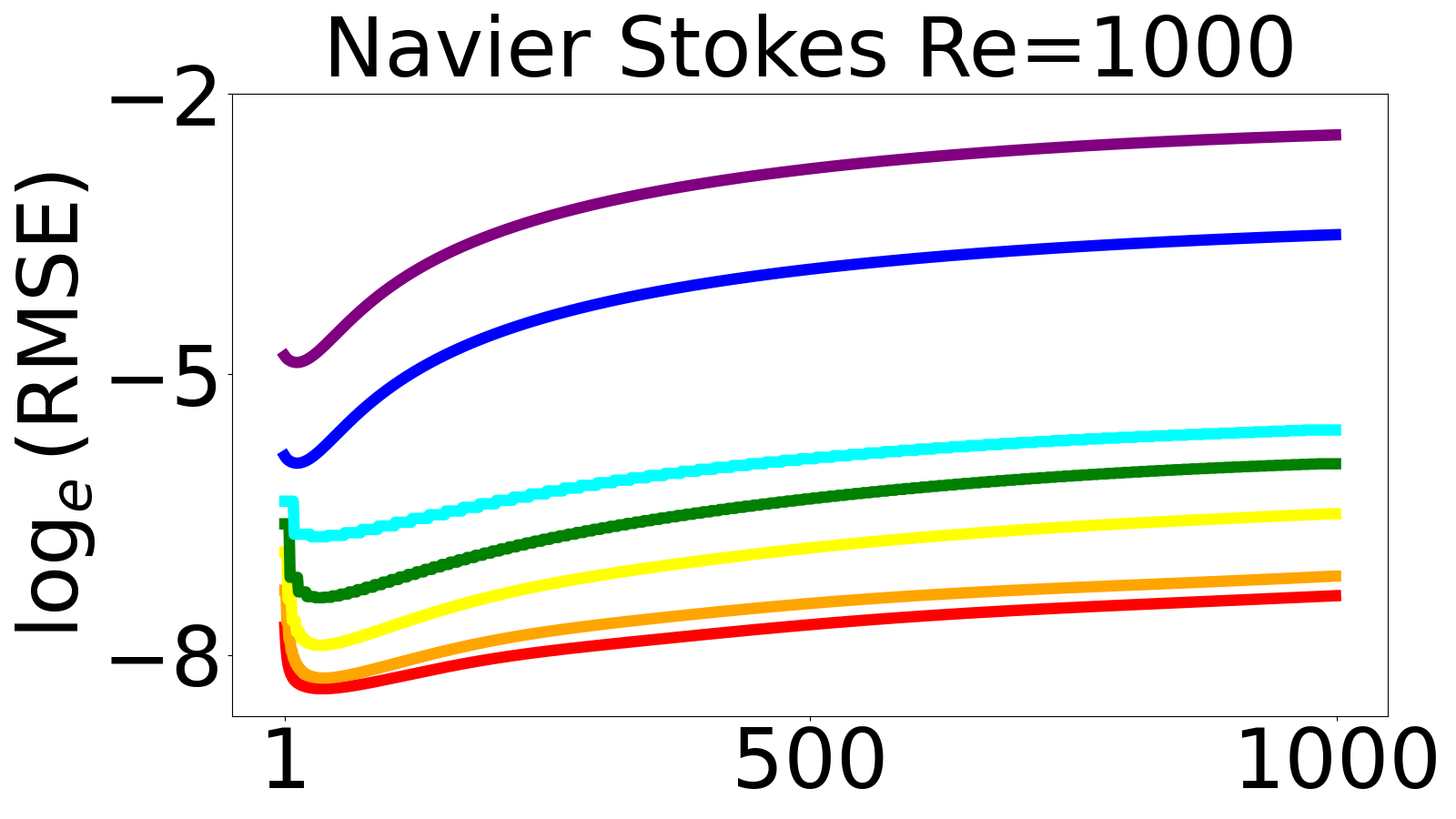}
            \end{subfigure}
            \hfill
            \begin{subfigure}
                \centering
                \includegraphics[width=0.19\linewidth]{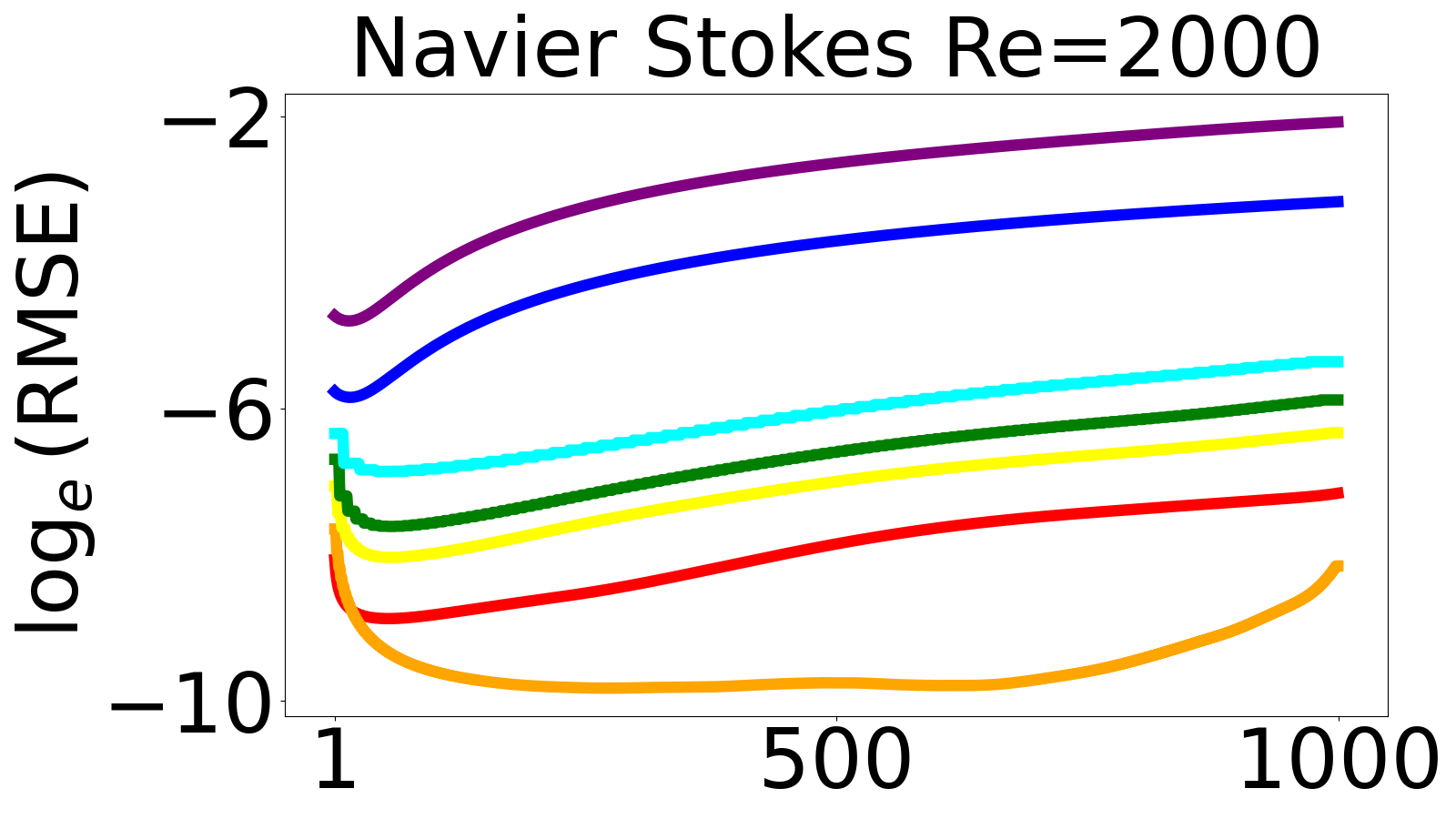}
            \end{subfigure}
        \end{minipage}
    }
    \caption{We compare FNO when trained to predict with various stepsize against pseudo-spectral solver operating at lower resolutions on incompressible N-S across five Reynold numbers. Solver $x$ implies pseudo-spectral solver operating at $x \times x$ resolution. \textit{FNO-$x$} denotes FNO trained to predict with stepsize $x \Delta t^*$, where $\Delta t^* = \tfrac{1}{32}$.}
    \label{fig:solver_timesteps}
    \vspace{-0.5cm}
\end{figure*}

We train FNO to make predictions with different step sizes. \textit{FNO-$x$} denotes FNO trained to predict with step size $x \Delta t^*$, where $\Delta t^* = \tfrac{1}{32}$. The results are shown in Figure~\ref{fig:solver_timesteps}, and they demonstrate that increasing the step sizes leads to only minor performance degradation. Figure~\ref{fig:runtime_timesteps} reports the runtime comparison for predicting up to $T = \tfrac{1}{2}$, where we observe speedups of several hundred times for \textit{FNO-16}. The experiment is conducted on incompressible N-S dataset.

\section{Supplementary Results on Different Modules}
\label{sec:module_more}

In this section, we provide additional experimental results on the evaluation of different modular designs on prediction performance and computational efficiency across Stochastic N-S. The quantitative results are summarized in Table~\ref{tab:main_spde}. 

From the first block of the table, which compares different spatial representations $\mathcal{X}$ under a fixed next-step predictor and physical variable decoding head, we observe that Fourier-based representations achieve the lowest RMSE (0.0398) despite having a relatively large parameter count (72M), demonstrating their strong inductive bias for capturing global structures in stochastic flows. Convolutional representations, while requiring fewer parameters (7.63M), exhibit competitive nRMSE and fRMSE, suggesting that local receptive fields are still effective when combined with sufficient depth. Interestingly, graph-based representations lag behind all other variants, highlighting their limited expressiveness in handling high-resolution stochastic turbulence.

In the second block, we fix the spatial representation to self-attention and vary the temporal modeling module $\mathcal{Y}$. Neural ODE-based temporal evolution provides the most stable long-horizon predictions, achieving the lowest errors across all three metrics, confirming its ability to model continuous-time dynamics more faithfully. Temporal bundling also improves over simple next-step prediction by reducing error accumulation over time, while pure self-attention temporal modeling performs the worst, potentially due to overfitting to short-term dependencies.

Overall, these results indicate that (1) Fourier-based spatial encoders are well-suited for stochastic PDE modeling when computational resources allow, (2) convolutional encoders strike a good balance between efficiency and accuracy, and (3) Neural ODE is the most robust choice for temporal evolution, especially under stochastic forcing. These findings justify the adoption of Fourier representation with Neural ODE temporal modeling in our final architecture.

\begin{table*}[h]
\centering
\setlength{\tabcolsep}{3mm}{
\begin{tabular}{lc|ccc}
\toprule[1.1pt]
 \multicolumn{1}{l}{}& \multicolumn{1}{l}{}& \multicolumn{3}{c}{Stochastic N-S} \\ \cmidrule(lr){3-5} 
\textsc{Model}
& \textsc{Param}
& \textsc{RMSE}$\downarrow$  & \textsc{nRMSE}$\downarrow$  &
\textsc{fRMSE}$\downarrow$  \\
\midrule
\rowcolor[RGB]{234, 238, 234} \multicolumn{5}{l}{\textit{\textbf{$\mathcal{X}$} + Next-step + Physical variable}} \\
$\mathcal{X}$ = Graph representaion & 3.13M &0.0917&0.2256&0.0067
   \\
$\mathcal{X}$ = Conv representation & 7.63M &0.0438&\textcolor{textpink}{0.0325}&\textcolor{textpink}{0.0042}
   \\
$\mathcal{X}$ = Fourier representation & 72.0M &\textcolor{textpink}{0.0398}&0.0463&0.0050
   \\
$\mathcal{X}$ = Latent representation & 89.0M &0.0560&0.0392&0.0072
   \\
$\mathcal{X}$ = Self-attention representation & 36.8M &0.0500&0.0430&0.0047
   \\
\midrule
\rowcolor[RGB]{234, 238, 234} \multicolumn{5}{l}{\textit{ Self-attention representation + $\mathcal{Y}$ + Physical variable}} \\
$\mathcal{Y}$ = Self-attention & 36.4M & 0.1094&0.0740&0.0104
  \\
$\mathcal{Y}$ = Temporal Bundling & 36.3M & 0.0998&0.0675&0.0089
  \\
$\mathcal{Y}$ = Neural ODE & 37.1M & \textcolor{textpink}{0.0913}&\textcolor{textpink}{0.0617}&\textcolor{textpink}{0.0083}
  \\
\bottomrule[1.1pt]
\end{tabular}
}
\caption{
Evaluation of different modular designs on prediction performance and computational efficiency across Stochastic N-S.
}
\label{tab:main_spde}
\end{table*}

\section{Supplementary Results on Generalization Study}
\label{sec:supp_generalization}

\subsection{Long-horizon Rollout for Diffusion and Deterministic Models}
\label{sec:supp_generalization_diff}

In our generalization experiments, we primarily select the optimal loss setting, i.e., the physical variable prediction setting. 
In addition, we compare deterministic models trained with physical variable prediction loss and those trained with a denoising loss. 
The generalization ability is evaluated on both out-of-distribution (OOD) data and at longer time horizons. 
The network architecture employs self-attention for both spatial and temporal representations, and all experiments are conducted on the Compressible Navier–Stokes equations.

\paragraph{Out-of-Distribution Generalization.} 
For the OOD generalization experiments, the model is trained only on the setting ``Mach = 0.1, $\zeta = 10^{-8}$'' and evaluated on unseen viscosity parameters. 
Table~\ref{tab:ood_generalization} summarizes the results, showing that models trained with denoising loss consistently outperform their deterministic counterparts.

\begin{table}[!t]
\centering
\setlength{\tabcolsep}{10pt}
\begin{tabular}{cccccc}
\toprule
\textbf{Mach} & \textbf{$\zeta$} & \textbf{Method} & \textbf{RMSE} $\downarrow$ & \textbf{nRMSE} $\downarrow$ \\
\midrule
0.1 & $1\text{e-8}$ & Deterministic & 0.2278 & 0.2033 \\
    &               & Denoising    & \textbf{0.1664} & \textbf{0.1530} \\ \hline
0.1 & $1\text{e-2}$ & Deterministic & 0.9129 & 0.3669 \\
    &               & Denoising    & \textbf{0.7540} & \textbf{0.3336} \\\hline
0.1 & $1\text{e-1}$ & Deterministic & 0.6660 & 0.2720 \\
    &               & Denoising    & \textbf{0.5427} & \textbf{0.2744} \\\hline
1.0 & $1\text{e-8}$ & Deterministic & 0.8693 & 0.7591 \\
    &               & Denoising    & \textbf{0.8156} & \textbf{0.7162} \\\hline
1.0 & $1\text{e-2}$ & Deterministic & 0.5098 & 0.6160 \\
    &               & Denoising    & \textbf{0.4924} & \textbf{0.6048} \\\hline
1.0 & $1\text{e-1}$ & Deterministic & 0.6801 & 0.2910 \\
    &               & Denoising    & \textbf{0.5679} & \textbf{0.2923} \\
\bottomrule
\end{tabular}
\caption{Generalization performance on out-of-distribution data. Models trained with denoising loss exhibit consistently lower RMSE and nRMSE across all conditions.}
\label{tab:ood_generalization}
\end{table}

\paragraph{Long-Horizon Generalization.}
For long-horizon tests, the model is trained with next-step prediction only, and rollout is performed autoregressively at test time. 
Table~\ref{tab:long_horizon} reports the RMSE and nRMSE for up to 16 rollout steps. 
The denoising-trained model consistently achieves lower errors and exhibits improved stability over longer horizons.

\begin{table}[h!]
\centering
\setlength{\tabcolsep}{5pt}
\begin{tabular}{c|cc|cc}
\toprule
\textbf{Time Step} & \textbf{Deterministic RMSE} & \textbf{Denoising RMSE} & \textbf{Deterministic nRMSE} & \textbf{Denoising nRMSE} \\
\midrule
1  & 0.1754 & \textbf{0.1601} & 0.1917 & \textbf{0.1611} \\
2  & 0.1753 & \textbf{0.1390} & 0.1940 & \textbf{0.1422} \\
3  & 0.1739 & \textbf{0.1313} & 0.1960 & \textbf{0.1355} \\
4  & 0.1755 & \textbf{0.1290} & 0.2012 & \textbf{0.1329} \\
5  & 0.2484 & \textbf{0.1742} & 0.2873 & \textbf{0.1869} \\
6  & 0.2470 & \textbf{0.1621} & 0.2915 & \textbf{0.1742} \\
7  & 0.2458 & \textbf{0.1593} & 0.2929 & \textbf{0.1734} \\
8  & 0.2530 & \textbf{0.1600} & 0.3043 & \textbf{0.1729} \\
9  & 0.3086 & \textbf{0.1936} & 0.3822 & \textbf{0.2164} \\
10 & 0.3135 & \textbf{0.1842} & 0.3956 & \textbf{0.2059} \\
11 & 0.3023 & \textbf{0.1812} & 0.3877 & \textbf{0.2065} \\
12 & 0.3280 & \textbf{0.1838} & 0.4160 & \textbf{0.2071} \\
13 & 0.3688 & \textbf{0.2116} & 0.4800 & \textbf{0.2414} \\
14 & 0.3754 & \textbf{0.2078} & 0.5083 & \textbf{0.2366} \\
15 & 0.3659 & \textbf{0.2071} & 0.4926 & \textbf{0.2407} \\
16 & 0.3979 & \textbf{0.2167} & 0.5273 & \textbf{0.2476} \\
\bottomrule
\end{tabular}
\caption{Long-horizon rollout performance. Denoising-trained models maintain lower error accumulation compared to deterministic models, demonstrating enhanced stability.}
\label{tab:long_horizon}
\vspace{-0.3cm}
\end{table}

Overall, these results confirm that training with a denoising loss improves both OOD generalization and long-horizon stability.

\subsection{Scaling Analysis}
\label{sec:supp_generalization_diff_scale}
\begin{figure}[h]
  \centering
  \includegraphics[width=.65\textwidth]{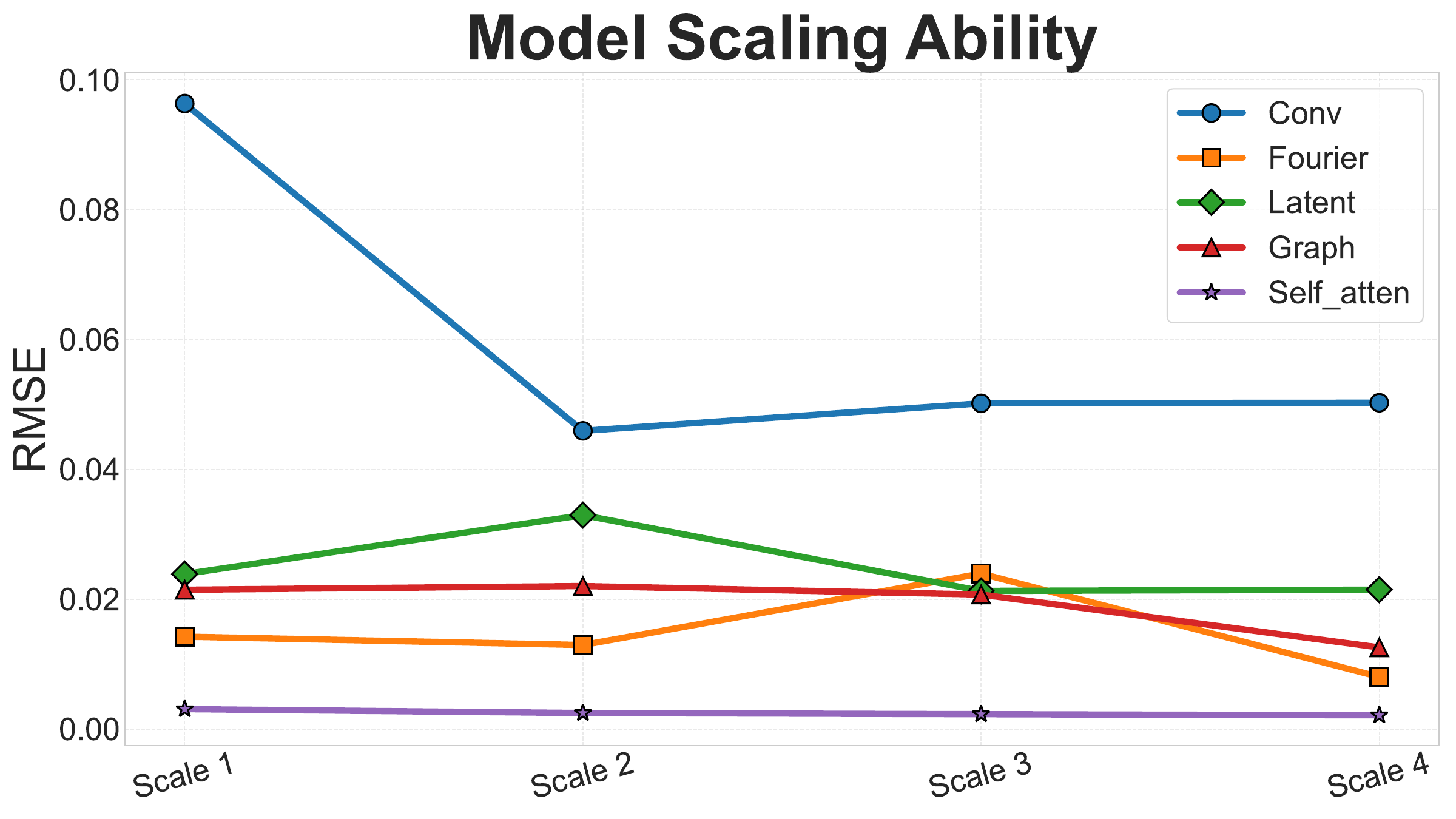}
  \caption{RMSE versus model capacity for five architectures at four scaling levels.}
  \label{fig:scaling_results}
\end{figure}

We systematically assessed the impact of model capacity on predictive accuracy across five representative architectures by defining four scaling levels for each:
\begin{itemize}[leftmargin=*]
  \item \textbf{Convolutional modules:} number of layers $\in \{2,3,4,5\}$
  \item \textbf{Fourier-based learning:} hidden dimension $ \in \{64,128,256,512\}$
  \item \textbf{Latent learning:} attention dimension $ \in \{64,128,256,512\}$
  \item \textbf{Self-attention learning:} hidden size $ \in \{96,192,384,768\}$
  \item \textbf{Graph learning:} hidden feature size $ \in \{64,128,256,512\}$
\end{itemize}
As shown in Figure~\ref{fig:scaling_results}, RMSE decreases monotonically from Scale 1 to Scale 4 for all five methods. The self-attention model achieves the largest absolute reduction (over $0.08$), demonstrating its superior ability to exploit increased representational capacity. The Fourier‐ and graph‐based models follow, with improvements of approximately $0.06$ and $0.05$, respectively, indicating diminishing returns at higher dimensions. The convolutional baseline exhibits the smallest gain (about $0.03$), suggesting that depth alone yields limited expressivity for this task, while the latent flow model attains an intermediate reduction (about $0.04$). These findings underscore that transformer‐style self‐attention mechanisms benefit most from scaling, making them particularly well‐suited for high‐fidelity modeling in our PDE‐driven benchmarks.

\section{Supplementary Results on Mixed Boundary Conditions}
\label{sec:mixedbc}

In this section, we aim to investigate whether \method{} can be extended beyond rectilinear grids to incorporate mixed boundary conditions (BCs) typical of wall-bounded flows, assessing whether the observed advantages of attention/Fourier methods persist in the presence of wall physics.We have introduced a new mixed-boundary task with complex geometry following~\cite{herde2024poseidon}. We additionally include a steady compressible Euler benchmark of flow past airfoils, a classical setting in aerodynamic shape optimization. This task inherently features mixed boundary conditions: (1) far-field Dirichlet conditions prescribing the free-stream state ($T^\infty=1,\; M^\infty=0.729,\; p^\infty=1,\; \alpha=2.31^\circ$); and (2) solid-wall boundary conditions imposed on the curved airfoil surface. The computational mesh used for generating the data is elliptic and body-fitted, which introduces strong spatial inhomogeneity and curvature effects before interpolation onto the Cartesian grid. This setting directly addresses the reviewer’s request by simultaneously incorporating non-rectilinear geometry, wall physics, and non-periodic far-field conditions. The results of mainstream neural operators ($\mathcal{Y}=\text{Next-step},\;\mathcal{Z}=\text{Variable}$) are reported in Table \ref{tab:mixedbc}.

\begin{table}[h]
\centering
\setlength{\tabcolsep}{8pt}
\begin{tabular}{lccc}
\toprule
\textbf{$\mathcal{X}$} & \textbf{RMSE} & \textbf{nRMSE} & \textbf{fRMSE} \\
\midrule
Graph & 0.2954 & 0.3071 & 0.0347 \\
ROM & 0.2179 & 0.2095 & 0.0197 \\
Conv & 0.2134 & 0.2126 & 0.0193 \\
Fourier & 0.2047 & 0.2096 & 0.0177 \\
Self-attention & 0.1843 & 0.1752 & 0.0119 \\
\bottomrule
\end{tabular}
\caption{Results on complex geometry and mixed boundary conditions.}
\label{tab:mixedbc}
\vspace{-0.3cm}
\end{table}

\subsection{Per-Boundary-Condition Error Analysis}
\label{sec:bc-analysis}

We further investigate whether the advantages of Fourier- and attention-based operators persist under \emph{realistic wall physics} and \emph{mixed boundary conditions}, rather than being artifacts of periodic or rectilinear domains.
To address this concern, we introduce a per-boundary-condition error analysis on the \textbf{compressible Euler airfoil} task. This problem naturally exhibits \emph{mixed boundary conditions}, including no-slip (solid wall) behavior on the airfoil surface, far-field boundary conditions, and complex wake dynamics downstream. The curved geometry and strong wall-normal gradients make this task particularly suitable for stress-testing spatial inductive biases beyond idealized settings.

\paragraph{Region-wise error decomposition.}
We partition the computational domain into three physically meaningful subsets:
\begin{itemize}
    \item $\Gamma_{\text{wall}}$: cells adjacent to the airfoil surface, capturing near-wall behavior and strong gradients;
    \item $\Gamma_{\text{far}}$: the outer band corresponding to far-field boundary regions;
    \item $\Omega_{\text{int}}$: the remaining interior flow region.
\end{itemize}

For each subset, we report the RMSE of all baseline models. In addition, we analyze the error as a function of distance to the airfoil wall, which reveals how prediction accuracy degrades in the most challenging near-wall regime.
Table~\ref{tab:bc_breakdown} summarizes the per-boundary-condition RMSE. As expected, all models incur larger errors near solid boundaries ($\Gamma_{\text{wall}}$) due to steep gradients and complex wall effects. Nevertheless, a clear and consistent trend emerges: \textbf{Fourier- and self-attention-based operators maintain their relative advantage} over convolutional and graph-based baselines across all regions, including $\Gamma_{\text{wall}}$, $\Gamma_{\text{far}}$, and $\Omega_{\text{int}}$.

\begin{table}[h]
    \centering
    \begin{tabular}{lccc}
        \toprule
        \textbf{Method} & $\boldsymbol{\Gamma_{\text{wall}}}$ & $\boldsymbol{\Gamma_{\text{far}}}$ & $\boldsymbol{\Omega_{\text{int}}}$ \\
        \midrule
        Graph          & 0.3554 & 0.2354 & 0.2954 \\
        ROM            & 0.2679 & 0.1679 & 0.2179 \\
        Conv           & 0.2634 & 0.1634 & 0.2134 \\
        Fourier        & 0.2447 & 0.1647 & 0.2047 \\
        Self-attention & \textbf{0.2243} & \textbf{0.1443} & \textbf{0.1843} \\
        \bottomrule
    \end{tabular}
    \caption{Per-boundary-condition RMSE on the compressible Euler airfoil task. Lower is better.}
    \label{tab:bc_breakdown}
\end{table}

These results demonstrate that the superior performance of Fourier- and attention-based operators is \emph{not} an artifact of periodic boxes or regular grids. Instead, their advantages extend robustly to curved geometries with mixed boundary conditions and wall-dominated aerodynamics. Even in the most challenging near-wall region, attention-based models achieve the lowest error, highlighting their ability to capture long-range interactions and complex boundary effects in realistic fluid dynamics scenarios.

\section{Supplementary Results on Non-Cartesian meshes}
\label{sec:supp_mesh}

In this section, we aim to investigate whether \method{} can be extended beyond rectilinear grids to stress spatial inductive biases on non-regular discretizations ($e.g.$, curved/non-rectilinear geometries and unstructured meshes), assessing whether the observed advantages of attention/Fourier methods persist in the presence of wall physics.

To directly address the geometry and meshing aspect, we extend our evaluation with a non-Cartesian mesh dataset, \textbf{Pipe}~\cite{li2022fourier}, which features incompressible flow in a pipe and thus introduces a \emph{complex geometry} in the computational domain. Unlike regular-grid settings, Pipe is discretized on a non-Cartesian mesh, providing a substantially different spatial structure and serving as a stronger test of spatial inductive bias beyond rectilinear grids.

We evaluate representative families of methods: (i) \textbf{graph-based} approaches (mesh connectivity as the primary inductive bias), (ii) \textbf{Fourier}-based models (spectral bias), and (iii) \textbf{self-attention} models (global token interactions). All methods are trained and evaluated under the same protocol for a fair comparison.
Table~\ref{tab:pipe_noncartesian} reports the results on Pipe. Self-attention achieves the best performance across both RMSE and normalized RMSE (nRMSE), with Fourier methods close behind, while the graph baseline lags significantly. Importantly, these results \emph{consistently corroborate} our central claim: \textbf{self-attention provides the strongest spatial representation}, and its advantage persists even when the discretization departs from regular grids and the domain geometry is non-trivial.

\begin{table}[h]
    \centering
    \begin{tabular}{lcc}
        \toprule
        \textbf{Method} & \textbf{nRMSE} $\downarrow$ & \textbf{RMSE} $\downarrow$ \\
        \midrule
        Graph          & 3.4678 & 0.0234 \\
        Fourier        & 0.7483 & 0.0050 \\
        Self-attention & \textbf{0.6499} & \textbf{0.0043} \\
        \bottomrule
    \end{tabular}
    \caption{Results on the non-Cartesian mesh dataset \textbf{Pipe}. Lower is better.}
    \label{tab:pipe_noncartesian}
    \vspace{-0.3cm}
\end{table}

These findings suggest that the benefit of attention is not an artifact of regular-grid convolutions or rectilinear geometries. Rather, attention's ability to form global interactions appears to provide a robust spatial representation that transfers to non-Cartesian mesh settings. Fourier methods also remain competitive, indicating that spectral biases can still be useful under irregular discretizations, whereas the graph baseline in this configuration is substantially less effective.

\section{Supplementary Results on 3D PDE}
\label{sec:supp_3d}

To directly test whether our conclusions regarding spatial representation quality transfer to 3D, we additionally evaluate the models on a 3D magnetohydrodynamics (MHD) compressible turbulence dataset from~\cite{ohana2024well}.
The dataset consists of isothermal MHD simulations without self-gravity (as in the diffuse ISM) at resolution $64^3$.
Table~\ref{tab:mhd3d} reports RMSE, normalized RMSE (nRMSE), and fRMSE. We follow the same evaluation protocol and metrics used throughout the paper for fair comparison across spatial encoders.

\begin{table}[h]
\centering
\begin{tabular}{lccc}
\toprule
\textbf{Method} & \textbf{RMSE} $\downarrow$ & \textbf{nRMSE} $\downarrow$ & \textbf{fRMSE} $\downarrow$ \\
\midrule
Latent              & 0.2976 & 0.6658 & 0.0246 \\
Conv                & 0.0596 & 0.1300 & 0.0061 \\
Fourier             & 0.0808 & 0.1815 & 0.0064 \\
Self-attention      & \textbf{0.0481} & \textbf{0.1094} & \textbf{0.0036} \\
\bottomrule
\end{tabular}
\caption{Results on 3D magnetohydrodynamics (MHD) compressible turbulence at $64^3$ resolution~\cite{ohana2024well}. Lower is better. Self-attention yields the best performance across all metrics, supporting our conclusion that attention provides the most effective spatial representation.}
\label{tab:mhd3d}
\end{table}

The 3D results further strengthen our main claim: \emph{self-attention provides the most effective spatial representation}, outperforming convolutional and Fourier alternatives on this 3D turbulence benchmark.
A global self-attention layer scales as $\mathcal{O}(N^2)$ in compute and memory, whereas Fourier operators typically scale closer to $\mathcal{O}(N\log N)$.
Therefore, purely from asymptotic complexity, Fourier methods are expected to retain a memory/runtime advantage as 3D resolution increases.

\section{Supplementary Results on Multi-physics PDE}
\label{sec:supp_mpde}

To explore a broader range of coupled physical phenomena beyond those covered in the original benchmark, and to assess whether the reported model rankings carry over to genuinely multiphysics PDE regimes, we conducted an additional evaluation on a challenging two-phase boiling dataset~\cite{hassan2023bubbleml}. This dataset explicitly features liquid–vapor phase change, moving interfaces, and tightly coupled heat–momentum dynamics.
These characteristics introduce strong nonlinearity and multi-scale behavior ($e.g.$, nucleation-triggered high-frequency transients), which are precisely the failure modes where conclusions from single-physics settings may not generalize.

The dataset comprises two representative and industrially relevant configurations: \emph{pool boiling} and \emph{flow boiling}. Pool boiling consists of a fluid volume above a bottom heater, with lateral walls and an outlet at the top, capturing buoyancy-driven plumes, interface deformation, and repeated nucleation/merging events. Flow boiling, in contrast, drives an FC-72 coolant through a heated channel with prescribed inlet velocity and thermal flux. This setting introduces strong convective transport, interfacial instabilities, and tight coupling between advection, heat transfer, and phase change. Together, these scenarios cover applications ranging from nuclear-waste cooling to data-center thermal management, and form a stringent testbed for coupled multiphysics operator learning.

We evaluate representative families of neural operators spanning graph-based, reduced-order or latent-operator baselines, convolutional operators, Fourier/spectral operators, and self-attention-based operators. We report standard error metrics, including RMSE and nRMSE, as well as fRMSE to assess performance on high-frequency interfacial dynamics.
Table~\ref{tab:multiphysics_boiling} summarizes the results. Across this two-phase setting, Fourier-type operators retain advantages in capturing interfacial curvature, vapor--liquid separation, and high-frequency dynamics associated with nucleation events, as reflected by their favorable fRMSE. Meanwhile, self-attention-based architectures continue to demonstrate strength in modeling long-range couplings, such as thermal plume formation and bubble--bubble interactions, achieving the best overall accuracy among the tested families. We also observe that attention-based models can be more sensitive to parameter-dependent dynamics ($e.g.$, inlet conditions and heat flux), consistent with the intuition that global interaction patterns may shift with operating regimes.

Crucially, these findings provide direct empirical evidence that the architectural insights identified by the benchmark do transfer to a genuinely coupled multiphysics regime involving moving interfaces and phase change. In particular, (i) spectral inductive biases remain effective for sharp, multi-scale interface phenomena and fast transients, and (ii) attention mechanisms remain competitive---and often superior---when long-range coupling dominates.

\begin{table}[h]
\centering
\begin{tabular}{lccc}
\toprule
Method & RMSE & nRMSE & fRMSE \\
\midrule
Graph-based  & 0.0784 & 0.1342 & 0.0743 \\
ROM / Latent  & 0.0846 & 0.1392 & 0.0756 \\
Convolutional  & 0.0572 & 0.0927 & 0.0512 \\
Fourier / Spectral  & 0.0562 & 0.0949 & 0.0481 \\
Self-attention & \textbf{0.0497} & \textbf{0.0784} & \textbf{0.0425} \\
\bottomrule
\end{tabular}
\caption{Results on the two-phase boiling multiphysics dataset (pool boiling $+$ flow boiling). Lower is better for all metrics.}
\label{tab:multiphysics_boiling}
\end{table}

\section{More Implementation Details}
\label{sec:imple_details}

\noindent\textbf{Modular Comparison.} All of our experiments are implemented on PyTorch on 8 $\times$ NVIDIA A6000 GPUs. For different experiments, we utilize different optimizers with different strategies, including weight decay, learning rates, and scheduler obtained from grid search for all experiments. 

\noindent\textbf{Traditional solvers.} FNO uses width$=$12 and width$=$32. We train the model for 10 epochs using a cosine annealing learning rate with a starting learning rate of $1e-4$ and an end learning rate of $1e-6$. We use the Adam optimizer.

\noindent\textbf{Discretization comparison.} FNO uses width=32 and width=128. MeshGraphnets uses latent size=128 and message passing steps=10. GNS uses latent size=128 and message passing steps=10. All models are trained for 100 epochs using a cosine annealing learning rate with a start learning rate of $1e-4$ and an end learning rate of $1e-5$. We use Adam optimizer for all model training.

\noindent\textbf{Hyperparameter Selection.} We first normalize the computational budgets by adjusting the GFLOPs of all methods to comparable levels. Subsequently, we conduct a systematic grid search over both model-level and training-level hyperparameters, exploring architecture choices, learning rates, regularization strengths, and optimization strategies. For each method, we select the configuration that achieves the best validation performance under the matched compute budget. While we strive to showcase the strongest performance of each baseline, we acknowledge that absolute fairness across all possible hyperparameter combinations is inherently challenging. Our procedure nonetheless provides a principled and reproducible way to tune each method to near-optimal performance.

\noindent\textbf{Protocol Template Availability.}
To further facilitate reproducibility and fair comparison, we also include the full protocol templates in the released codebase. These templates specify the optimizer type, search ranges for learning rate and weight decay, scheduler choices, and training length for each module family. Users can directly load these configurations to reproduce our reported results or extend them to new architectures with minimal modification. Please refer to the \texttt{config} folder in our anonymous GitHub codebase \url{https://anonymous.4open.science/r/FD-Bench-15BC} for more details.

\section{Easy Integration of Existing Leaderboard}
\label{appendix:leader}
Our benchmark is designed to be seamlessly extensible and easily integrated with public leaderboards and community-driven competitions. For example, users can directly build from widely used public libraries such as \texttt{neuraloperator} by adding them as a submodule, installing their dependencies, and defining a simple configuration file under \texttt{config$/$public}. The model can then be initialized in \texttt{src$/$train.py} and executed using our unified training scripts (\textit{e.g.}, \texttt{bash src$/$train$\_$public.sh}). This plug-and-play design significantly lowers the barrier to participation, enabling researchers to quickly evaluate their methods under standardized settings. This will encourage community participation, provide a living benchmark for tracking state-of-the-art progress, and foster reproducible and comparable research in data-driven fluid simulation.

\section{More Datasets Details}
\label{sec:supp_data}

\subsection{Datasets for Modular Comparison}
\label{sec:modular_data}
\noindent\textbf{Compressible N-S Equations.} The compressible N-S (CNS) equations govern the dynamics of fluid flows with variable density and internal energy. The compressible model accounts for variations in density $\rho$ and pressure $p$ due to thermodynamic effects, and captures phenomena such as shock waves, sound propagation, and high-speed turbulent mixing.

We utilize the 2D data from PDEBench~\citep{takamoto2022pdebench}. The general form of the CNS equations in conservative form is given by:
\begin{subequations}\label{eq:cns}
\begin{align}
    &\text{ Mass conservation (Continuity equation):} \quad 
    \partial_t \rho + \nabla \cdot (\rho \textbf{v}) = 0, \\
    &\text{ Momentum conservation:} \quad 
    \partial_t (\rho \textbf{v}) + \nabla \cdot (\rho \textbf{v} \otimes \textbf{v}) = -\nabla p + \nabla \cdot \sigma, \\
    &\text{ Total energy conservation:} \quad 
    \partial_t E + \nabla \cdot [(E + p)\textbf{v}] = \nabla \cdot (\sigma \cdot \textbf{v}) + \nabla \cdot (k \nabla T),
\end{align}
\end{subequations}

where $\rho$ is the mass density, $\textbf{v} \in \mathbb{R}^d$ is the velocity field, $p$ is the pressure, $E = \epsilon + \frac{1}{2}\rho \|\textbf{v}\|^2$ is the total energy per unit volume, $\epsilon$ is the internal energy, $T$ is the temperature, $k$ is the thermal conductivity, and $\sigma$ is the viscous stress tensor given by:
\begin{equation}
\sigma = \eta \left[ \nabla \textbf{v} + (\nabla \textbf{v})^T \right] + \left( \zeta - \frac{2}{3}\eta \right)(\nabla \cdot \textbf{v})\mathbf{I},
\end{equation}
with $\eta$ and $\zeta$ denoting shear and bulk viscosities, respectively.
To close the system, an equation of state (EOS) is required. A common choice for an ideal gas is:
\begin{equation}
    p = (\gamma - 1)\epsilon, \quad \text{or} \quad p = \rho R T,
\end{equation}
where $\gamma$ is the adiabatic index and $R$ is the specific gas constant.

\vspace{1mm}

The CNS equations describe a wide range of fluid phenomena where compressibility effects are non-negligible. In particular, they can capture shock and rarefaction waves, supersonic and hypersonic flows, acoustic propagation, atmospheric and weather systems, combustion and detonation and astrophysical flows.

\noindent\textbf{Diffusion-Reaction Flow.}
We consider a 2D diffusion-reaction system involving two nonlinearly coupled variables: the activator \( u = u(t,x,y) \) and the inhibitor \( v = v(t,x,y) \). We utilize the data from PDEBench~\citep{takamoto2022pdebench}.  The governing equations are:
\begin{align} \label{eq:2d_diff-react}
    \partial_t u  &= D_u (\partial_{xx} u + \partial_{yy} u) + R_u, \\
    \partial_t v  &= D_v (\partial_{xx} v + \partial_{yy} v) + R_v \, ,
\end{align}
where \( D_u \) and \( D_v \) denote the diffusion coefficients, and \( R_u(u,v) \) and \( R_v(u,v) \) represent the nonlinear reaction terms for the activator and inhibitor, respectively. The simulation is defined over the spatial domain \( x, y \in (-1,1) \) and temporal range \( t \in (0,5] \). This system is widely used to model biological pattern formation.
The reaction terms follow the FitzHugh–Nagumo model~\citep{Klaasen1984fhn}:
\begin{align}
R_u(u,v) &= u - u^3 - k - v, \\
R_v(u,v) &= u - v,
\end{align}
with $k = 5 \times 10^{-3}$, and diffusion coefficients $D_u = 1 \times 10^{-3}, D_v = 5 \times 10^{-3}$. The initial condition is sampled from a standard normal distribution: $u(0,x,y) \sim \mathcal{N}(0,1.0)$, for $x, y \in (-1,1)$. We provide simulation data discretized at high resolution $(N_x = N_y = 512, N_t = 501)$ and a downsampled version $(N_x = N_y = 128, N_t = 101)$ for training. Spatial discretization uses the finite volume method~\citep{Moukalled2016}, and time integration is performed via a fourth-order Runge-Kutta scheme from the \texttt{scipy} library.

\noindent\textbf{Kolmogorov Flow.} It is a novel physical scenario and a 2D variant of the classical Kolmogorov flow~\citep{majda2002vorticity}. We follow Poseidon~\citep{herde2024poseidon} for data settings. This flow is governed by the incompressible N-S equations with an external forcing term. The governing equations take the form:
\begin{equation}
    \label{eq:FNS}
    u_t +(u\cdot \nabla)u + \nabla p - \nu \Delta u = f, \quad {\rm div}~u =0,
\end{equation}
defined over the spatial domain $[0,1]^2$ with periodic boundary conditions imposed in both spatial directions. Here, $u = (u_x, u_y)$ denotes the velocity field, $p$ is the pressure, $\nu$ is the viscosity, and $f$ is the external forcing. The forcing term $f$ is smooth, spatially varying, and held constant over time. It is defined as:
\begin{equation}
\label{eq:fin}
    f(x, y) = 0.1\sin(2\pi(x+y)).
\end{equation}
This configuration introduces a diagonal sinusoidal excitation that drives the flow in a nontrivial manner, resulting in rich dynamics ideal for evaluating neural operator performance.
The initial conditions are based on the vorticity,
which is assumed to be constant along a uniform (square) partition of the underlying domain. Numerical simulations are conducted using the same discretization and time-stepping schemes adopted for other N-S-based flows.

This problem can be recast into the unified PDE framework by defining an augmented state variable $U = [u_x, u_y, f]$, and introducing the trivial evolution equation $\partial_t f = 0$ to reflect the time-invariance of the forcing. The initial data is augmented accordingly using Equation~\eqref{eq:fin}. The corresponding solution operator $\sol(t,\cdot)$ maps the initial state $U^0_{x,y}$ to the solution at time $t$ as:
\vspace{-0.1cm}
\begin{equation}
    \sol(t, U^0_{x,y}) = [u_x(t), u_y(t), f]
\end{equation}
\vspace{-0.1cm}

where $(u_x, u_y)$ evolve according to the forced N-S Equations~\eqref{eq:FNS}, and $f$ remains fixed over time.

\noindent\textbf{Stochastic N-S Equation}. We extend the existing 2D incompressible N-S equation by adding a stochastic forcing term, resulting in the 2D stochastic N-S equation, an emerging modeling class of fluid dynamics that aims to better handle uncertainty and study the mix-in behavior of fluid flows \citep{stochastic_navier_stokes}. In particular, we present semilinear form over the space-time interval $(t,x)\in [0,T]\times [0,1]^2$, similar to the dataset in \citep{salvi2022neuralstochasticpdesresolutioninvariant}:
\vspace{-0.3cm}
\begin{equation}
    \frac{\partial \mathbf{u}}{\partial t} = \nu \Delta \mathbf{u}  +\mathbf{f} + \sigma \xi, \quad \mathbf{u}(0,x) = \mathbf{u}_0(x),
\end{equation}
\vspace{-0.3cm}

where $\xi = \dot{W}$ for $W$ a $Q$-Wiener process, colored in space and with scale $\sigma = 0.05$ and viscosity $\nu=10^{-4}$. For our data generation, the initial condition is sampled according to $\mathbf{u}_0\sim \mathcal{N}(0, 3^{3/2}(-\Delta+49I)^{-3})$ with periodic boundary conditions.

\subsection{Datasets for comparison with numerical solvers}
\label{appendix:datasets_solver}
\textbf{2D Incompressible N-S Equation in Vorticity Form.}
This pseudo‐spectral N-S solver uses a semi‐implicit, two‐stage (Heun) time integrator and adaptive time stepping governed by CFL conditions. At each step, the vorticity field $\omega(x,y)$ is transformed to Fourier space, the stream‐function $\psi$ is obtained via multiplication by the precomputed inverse Laplacian, and the velocity $(u,v)=(\partial_y\psi,,-\partial_x\psi)$ is recovered by spectral differentiation. Nonlinear advection is computed in physical space and re‐transformed, while viscous diffusion is handled implicitly through Crank–Nicolson scaling. To suppress aliasing, high‐wavenumber modes are zeroed via a $2/3$‐rule mask.

We use a physical domain of $1 \times 1$ ($L_1 = L_2 = 1$) and a discretization grid size of $256 \times 256$ ($s_1 = s_2 = 256)$. The total time $T=20$, and we store vorticity data at each $\frac{1}{50}$ intervals. This means that the neural solvers make predictions with $\Delta t = \frac{1}{50}$. We generate 1200 sequences each with a sequence length of 1000. We use 1000 sequences for training, 100 sequences for validation, and 100 sequences for testing. The external forcing term is $f(x,y) = 0.1 (\sin (2\pi(x+y)) + \cos (2\pi(x+y)))$. The initial vorticity is generated from a 2D gaussian random field.

\begin{algorithm}[t]
\caption{Pseudo-Spectral Solver for 2D Incompressible N-S}
\textbf{Require: } initial vorticity $\omega^0(x,y)$, forcing $f(x,y)$ (optional), final time $T$, Reynolds number $\Re$, grid sizes $(s_1,s_2)$, domain $(L_1,L_2)$
\begin{algorithmic}[1]
\STATE Build 2/3‐rule dealiasing mask $M = \bigl(|K_X|<\tfrac23\,\max|k_x|\bigr)\land\bigl(|K_Y|<\tfrac23\,\max|k_y|\bigr)$
\STATE $t\gets 0$
\WHILE{$t < T$}
  \STATE Compute velocity $\mathbf u = (u,v)$ from $\hat\omega$ via Poisson‐solve + spectral derivatives
  \STATE Determine time step $\Delta t$ from CFL:  
  \[
    \Delta t = \min\bigl(\mathrm{cfl }\,h/\|\mathbf u\|_\infty,\;\mathrm{cfl }\,h^2/\nu\bigr),
    \quad \nu = 1/\Re
  \]
  \STATE $\mathcal N^{(1)} \gets -\,\mathcal{F}\{\mathbf u\cdot\nabla\omega\} + \hat f$
  \STATE Predictor: 
  \[
    \hat\omega^* \gets \frac{\hat\omega + \Delta t\bigl(\mathcal N^{(1)} - \tfrac{\nu}{2}\,G\,\hat\omega\bigr)}
                                {1 + \tfrac{\nu}{2}\,G\,\Delta t}
  \]
  \STATE Compute $\mathcal N^{(2)}$ at $\hat\omega^*$ similarly
  \STATE Corrector: 
  \[
    \hat\omega \gets \frac{\hat\omega + \Delta t\bigl(\tfrac{\mathcal N^{(1)}+\mathcal N^{(2)}}{2} 
                             - \tfrac{\nu}{2}\,G\,\hat\omega\bigr)}
                               {1 + \tfrac{\nu}{2}\,G\,\Delta t}
  \]
  \STATE Dealias: $\hat\omega \gets \hat\omega \;\cdot\; M$
  \STATE $t \gets t + \Delta t$
\ENDWHILE
\STATE $\omega(x,y) \gets \mathcal{F}^{-1}\{\hat\omega\}$
\STATE Recover $\psi$ and then $\mathbf u = (\partial_y\psi,\,-\partial_x\psi)$
\end{algorithmic}
\end{algorithm}

\noindent\textbf{2D Burgers' Equation.}
This solver advances the 2D Burgers equations as follows:
\begin{equation}
\begin{aligned}
u_t + u\,u_x + v\,u_y &= \nu\!\left(u_{xx} + u_{yy}\right),\\
v_t + u\,v_x + v\,v_y &= \nu\!\left(v_{xx} + v_{yy}\right).
\end{aligned}
\end{equation}
using a pseudo‐spectral RK4 integrator. Spatial derivatives are computed in Fourier space, while nonlinear advection is formed in physical space and then dealiased with a 2/3‐rule mask. Viscous diffusion appears as a spectral multiplier $-\nu,K^2$ and is treated in the same explicit stages. Crucially, the time step $\Delta t$ is chosen adaptively via a CFL constraint.

We use a physical domain of $1 \times 1$ ($L_1 = L_2 = 1$) and a discretization grid size of $256 \times 256$ ($s_1 = s_2 = 256)$. The total time $T=20$, and we store velocity data at each $\frac{1}{50}$ intervals. We generate 1200 sequences each with a sequence length of 1000. We use 1000 sequences for training, 100 sequences for validation, and 100 sequences for testing. The initial velocity is generated from a 2D gaussian random field.

\begin{algorithm}[t]
\caption{Pseudo‐Spectral Solver for 2D Burgers Equation}
\textbf{Require: } initial velocity fields $(u^0, v^0)$, final time $T$, viscosity $\nu$, grid sizes $(s_1, s_2)$, domain $(L_1, L_2)$
\begin{algorithmic}[1]
\STATE Compute $K^2 = K_X^2 + K_Y^2$, replace zero‐mode by $\varepsilon$ to avoid division by zero
\STATE Build 2/3‐rule dealiasing mask $M = \bigl(|K_X|<\tfrac23\,\max|k_x|\bigr)\land\bigl(|K_Y|<\tfrac23\,\max|k_y|\bigr)$
\STATE $t\gets 0$
\WHILE{$t<T$}
  \STATE Determine time step $\Delta t$ from CFL:  
  \[
    \Delta t = \min\bigl(\mathrm{cfl}\,h/U_{\max},\;\mathrm{cfl}\,h^2/\nu\bigr)
  \]
  \STATE // RK4 stages
  \STATE Compute $(k_1^u,k_1^v)\gets\textsc{ComputeRHS}(\hat u,\hat v)$
  \STATE Compute $(k_2^u,k_2^v)\gets\textsc{ComputeRHS}(\hat u + \tfrac{\Delta t}{2}k_1^u,\;\hat v + \tfrac{\Delta t}{2}k_1^v)$
  \STATE Compute $(k_3^u,k_3^v)\gets\textsc{ComputeRHS}(\hat u + \tfrac{\Delta t}{2}k_2^u,\;\hat v + \tfrac{\Delta t}{2}k_2^v)$
  \STATE Compute $(k_4^u,k_4^v)\gets\textsc{ComputeRHS}(\hat u + \Delta t\,k_3^u,\;\hat v + \Delta t\,k_3^v)$
  \STATE $\hat u\gets \hat u + \tfrac{\Delta t}{6}(k_1^u + 2k_2^u + 2k_3^u + k_4^u)$
  \STATE $\hat v\gets \hat v + \tfrac{\Delta t}{6}(k_1^v + 2k_2^v + 2k_3^v + k_4^v)$
  \STATE Apply dealiasing: $\hat u\gets \hat u\cdot M,\;\hat v\gets \hat v\cdot M$
  \STATE $t\gets t + \Delta t$
\ENDWHILE
\STATE $(u,v)\gets \mathcal F^{-1}\{\hat u,\hat v\}$
\end{algorithmic}
\end{algorithm}

\noindent\textbf{ComputeRHS}($\hat u,\hat v$):  
Compute nonlinear advection and diffusion in spectral space, i.e.  
\[
R_u = -\,\mathcal F\{u\,u_x + v\,u_y\}\,M \;-\;\nu\,K^2\,\hat u,\quad
R_v = -\,\mathcal F\{u\,v_x + v\,v_y\}\,M \;-\;\nu\,K^2\,\hat v.
\] 

\noindent\textbf{2D Advection Equation.}
This solver advances the linear advection equation
\begin{equation}
    \begin{aligned}
    u_t + v_x(x,y)\,u_x + v_y(x,y)\,u_y = 0
\end{aligned}
\end{equation}
via a pseudo‐spectral Runge–Kutta 4 scheme.  At each step, the field is FFT‐transformed, the spatial‐dependent velocity is sampled, and an adaptive CFL‐based time step is chosen to ensure stability. Nonlinear advection fluxes are computed in physical space, re‐transformed, differentiated spectrally, and then combined in the RK4 stages; dealiasing enforces the 2/3‐rule before the next step.

We use a physical domain of $1 \times 1$ ($L_1 = L_2 = 1$) and a discretization grid size of $256 \times 256$ ($s_1 = s_2 = 256)$. The total time $T=20$, and we store field data at each $\frac{1}{50}$ intervals. We generate 1200 sequences each with a sequence length of 1000. We use 1000 sequences for training, 100 sequences for validation, and 100 sequences for testing. The field velocity is generated from a 2D gaussian random field.

In the following, we specify the spatial dependent velocity.

\textbf{Advection0: }
\begin{equation}
    \begin{aligned}
        v_x(x,y) &= \sin\!\bigl(2\pi y + t\bigr), \\
v_y(x,y) &= \cos\!\bigl(2\pi x + t\bigr).
    \end{aligned}
\end{equation}

\textbf{Advection1: }
\begin{equation}
    \begin{aligned}
        v_x(x,y) &= U_0 \frac{y}{L_y},\\
v_y(x,y) &= 0.
    \end{aligned}
\end{equation}

\textbf{Advection2: }
\begin{equation}
    \begin{aligned}
        v_x(x,y) &= \sin\!\Bigl(\frac{\pi x}{L_1}\Bigr)\,\cos\!\Bigl(\frac{\pi y}{L_2}\Bigr),\\
v_y(x,y) &= -\cos\!\Bigl(\frac{\pi x}{L_1}\Bigr)\,\sin\!\Bigl(\frac{\pi y}{L_2}\Bigr).
    \end{aligned}
\end{equation}

\textbf{Advection3: }
\begin{equation}
    \begin{aligned}
        v_x(x,y) &= -\bigl(y - \tfrac{L_2}{2}\bigr),\\
v_y(x,y) &= x - \tfrac{L_1}{2}.
    \end{aligned}
\end{equation}

\textbf{Advection4: }
\begin{equation}
    \begin{aligned}
        v_x(x,y,t) &= a\,x,\\
v_y(x,y,t) &= -a\,y.
    \end{aligned}
\end{equation}

\begin{algorithm}[t]
\caption{Pseudo‐Spectral RK4 Solver for 2D Linear Advection}
\textbf{Require: } initial scalar field $u^0$, , final time $T$, advection speed function $\text{VELOCITY}(x,y)$, grid sizes $(s_1, s_2)$, domain $(L_1, L_2)$
\begin{algorithmic}[1]
\STATE Precompute wavenumbers $K_X,K_Y$ and 2/3‐rule dealias mask $M$
\STATE $\hat u \gets \mathcal F\{u\}$
\STATE $t\gets 0$
\WHILE{$t < T$}
  \STATE Obtain velocity field $(v_x,v_y)\gets\textsc{Velocity}(x,y, t)$
  \STATE Determine time step $\Delta t$ from CFL: $\Delta t\gets 0.5\min\bigl(L_x/(N_x\|v_x\|_\infty),\,L_y/(N_y\|v_y\|_\infty)\bigr)$
  \STATE // RK4 stages
  \STATE $k_1 \gets \textsc{RHS}(\hat u, v_x, v_y)$
  \STATE $k_2 \gets \textsc{RHS}(\hat u + \tfrac{\Delta t}{2}k_1,\;v_x(t+\tfrac{\Delta t}{2}),\,v_y(t+\tfrac{\Delta t}{2}))$
  \STATE $k_3 \gets \textsc{RHS}(\hat u + \tfrac{\Delta t}{2}k_2,\;v_x(t+\tfrac{\Delta t}{2}),\,v_y(t+\tfrac{\Delta t}{2}))$
  \STATE $k_4 \gets \textsc{RHS}(\hat u + \Delta t\,k_3,\;v_x(t+\Delta t),\,v_y(t+\Delta t))$
  \STATE Update spectral solution:
    \[
      \hat u \gets \hat u + \tfrac{\Delta t}{6}\,(k_1 + 2k_2 + 2k_3 + k_4)
    \]
  \STATE Apply dealiasing: $\hat u \gets \hat u \cdot M$
  \STATE $t\gets t + \Delta t$
\ENDWHILE
\STATE $u\gets \mathcal F^{-1}\{\hat u\}$
\end{algorithmic}
\end{algorithm}

\noindent\textbf{ComputeRHS}($\hat u,v_x,v_y$):  
Compute the spectral right‐hand side of $u_t + \nabla\!\cdot\!(\mathbf v\,u)=0$ by  
1. transforming $\hat u$ to $u$,  
2. forming fluxes $u\,v_x$ and $u\,v_y$ and re‐transforming,  
3. differentiating in $x,y$ via $iK_X,iK_Y$,  
4. summing $-(\partial_x+\partial_y)$ in spectral space.

\subsection{Datasets for comparison with discretization strategy}
\label{appendix:datasets_discretization}
We employ Smoothed Particle Hydrodynamics (SPH) \citep{gingold1977smoothed} to generate three Lagrangian particle subsets: the Taylor–Green vortex (TGV), the lid-driven cavity flow (LDC), and the reverse Poiseuille flow (RPF). Each subset is governed by the compressible N-S equations.
For Taylor–Green vortex, we set the Reynolds number $\mathrm{Re}=100$ and simulate on a $1\times1$ periodic domain with spatial resolution $\Delta x = 0.01$ and time step $\Delta t = 4\times10^{-4}$. We generate 100 sequences for training and 50 sequences for testing, each comprising 126 time steps, using $N = 10,000$ particles. For Lid-driven cavity, we also use the same domain and discretization parameters ($\Delta x = 0.01$, $\Delta t = 4\times10^{-4}$) at $\mathrm{Re}=100$, producing 10k frames for training and 5k frames for testing with $N = 11,236$ particles. For Reverse Poiseuille flow, $\mathrm{Re}$ is set as 10 on a $1.12\times1.12$ periodic domain with $\Delta x = 1.25\times10^{-2}$ and $\Delta t = 5\times10^{-4}$. We generate 20k training frames and 5k testing frames with $N = 12,800$ particles. For all three datasets, particle positions are recorded every 100 time steps, so that the neural models are trained to predict the flow evolution over intervals of $100 \Delta t$.

% \noindent\textbf{Main results.}

\section{Visualization Study}
\label{sec:vis}

\begin{figure*}[h]
\centering
\includegraphics[width=\linewidth]{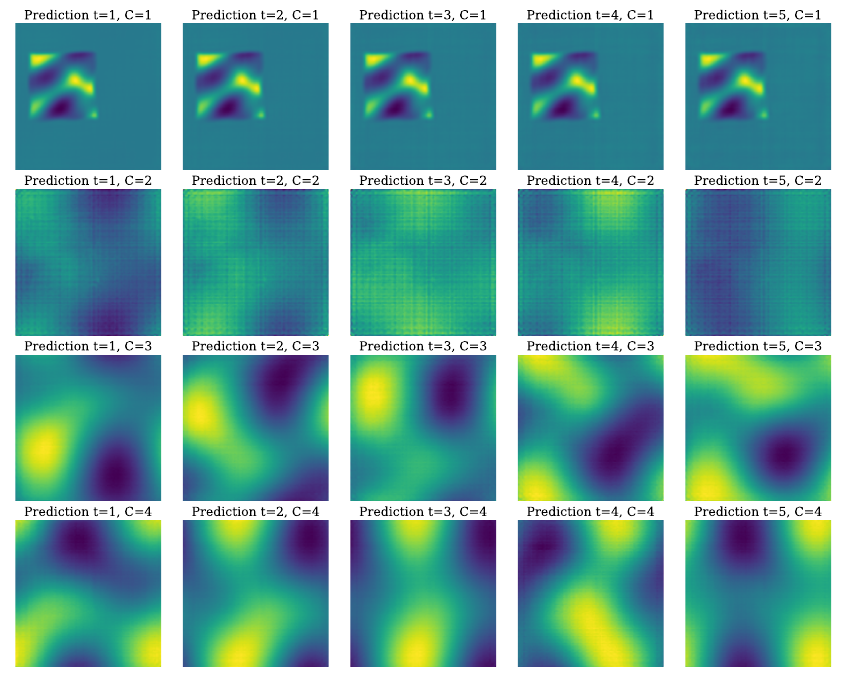}
\caption{Visualization of Fourier$+$Next-step$+$Variable on Compressible N-S.}
\label{fig:vis_cns_1}
\end{figure*}

\begin{figure*}[h]
\centering
\includegraphics[width=\linewidth]{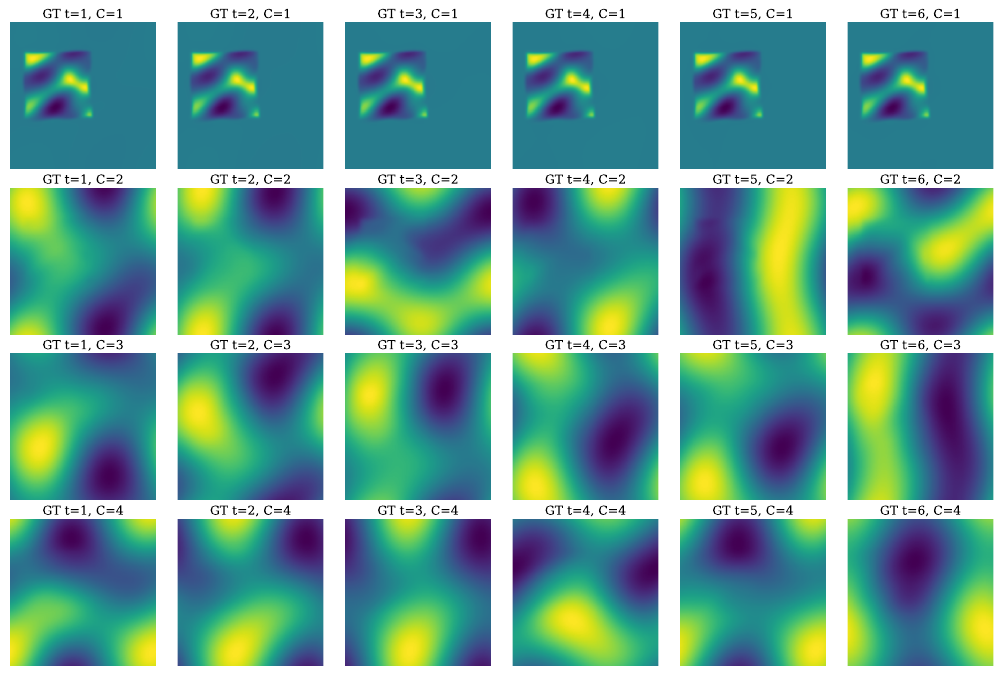}
\caption{Visualization of Fourier$+$Next-step$+$Variable on Compressible N-S.}
\label{fig:vis_cns_2}
\end{figure*}

\begin{figure*}[h]
\centering
\includegraphics[width=\linewidth]{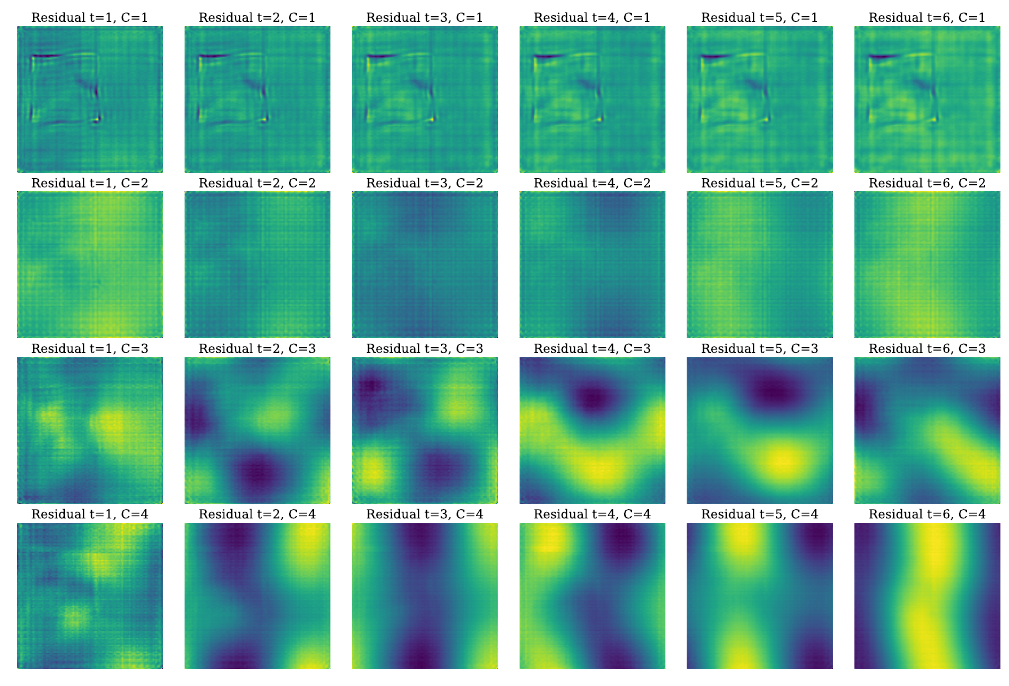}
\caption{Visualization of Fourier$+$Next-step$+$Variable on Compressible N-S.}
\label{fig:vis_cns_3}
\end{figure*}

\begin{figure*}[h]
\centering
\includegraphics[width=\linewidth]{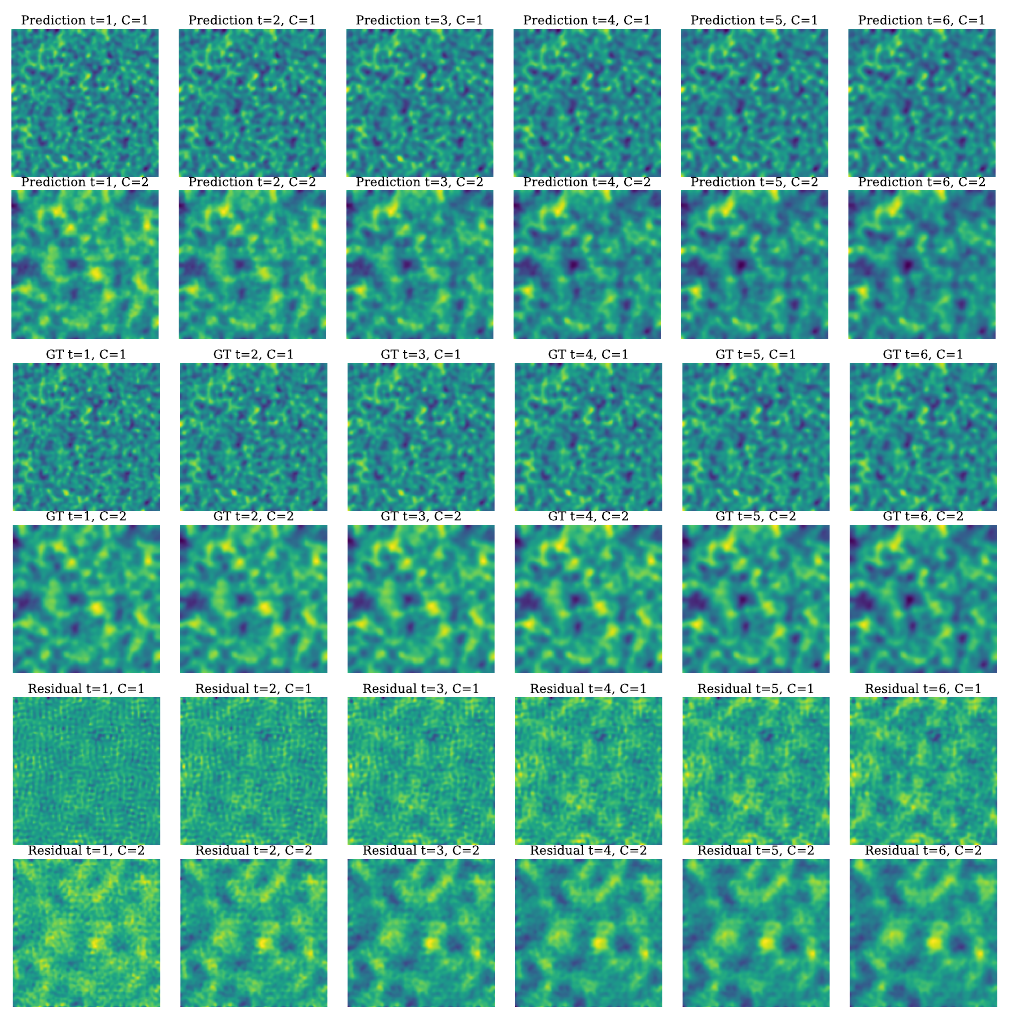}
\caption{Visualization of Fourier$+$Next-step$+$Variable on Diffusion-Reaction.}
\label{fig:vis_dr}
\end{figure*}

\begin{figure*}[h]
\centering
\includegraphics[width=\linewidth]{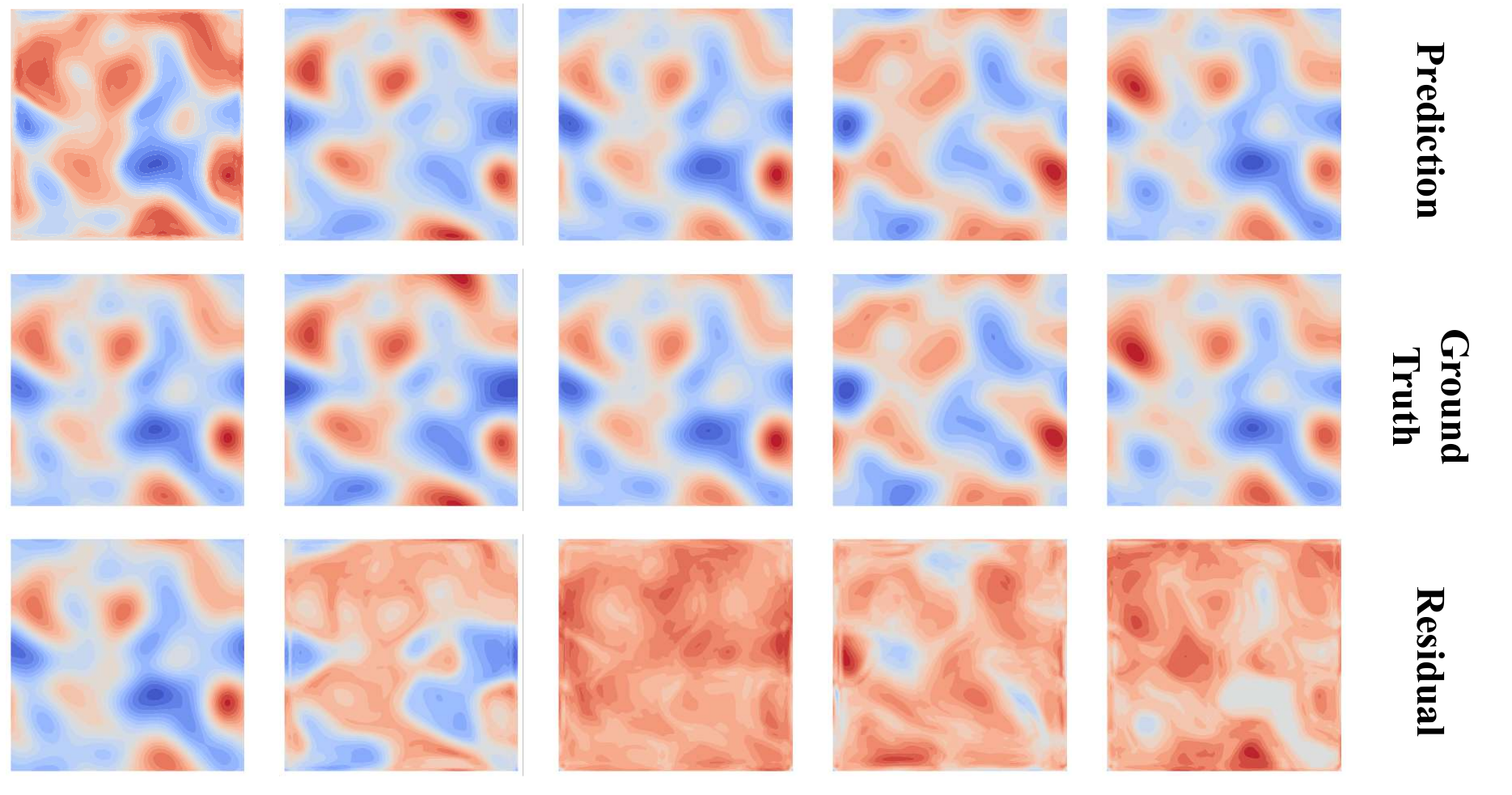}
\caption{Visualization of Conv$+$Next-step$+$Variable on Stochastic N-S .}
\label{fig:vis1}
\vspace{-0.3cm}
\end{figure*}

\begin{figure*}[h]
\centering
\includegraphics[width=\linewidth]{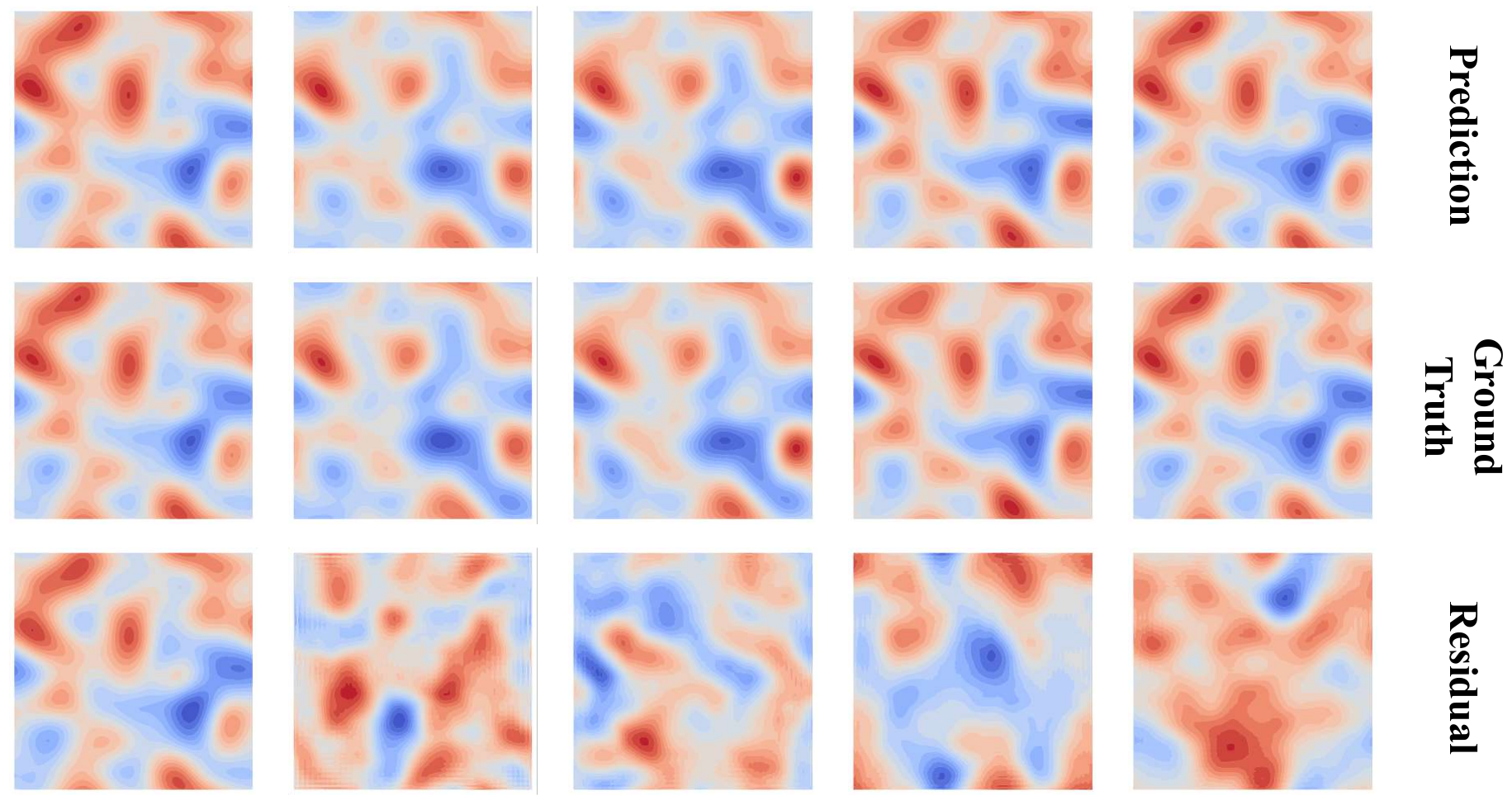}
\caption{Visualization of Fourier$+$Next-step$+$Variable on Stochastic N-S .}
\label{fig:vis2}
\vspace{-0.3cm}
\end{figure*}

\begin{figure*}[h]
\centering
\includegraphics[width=\linewidth]{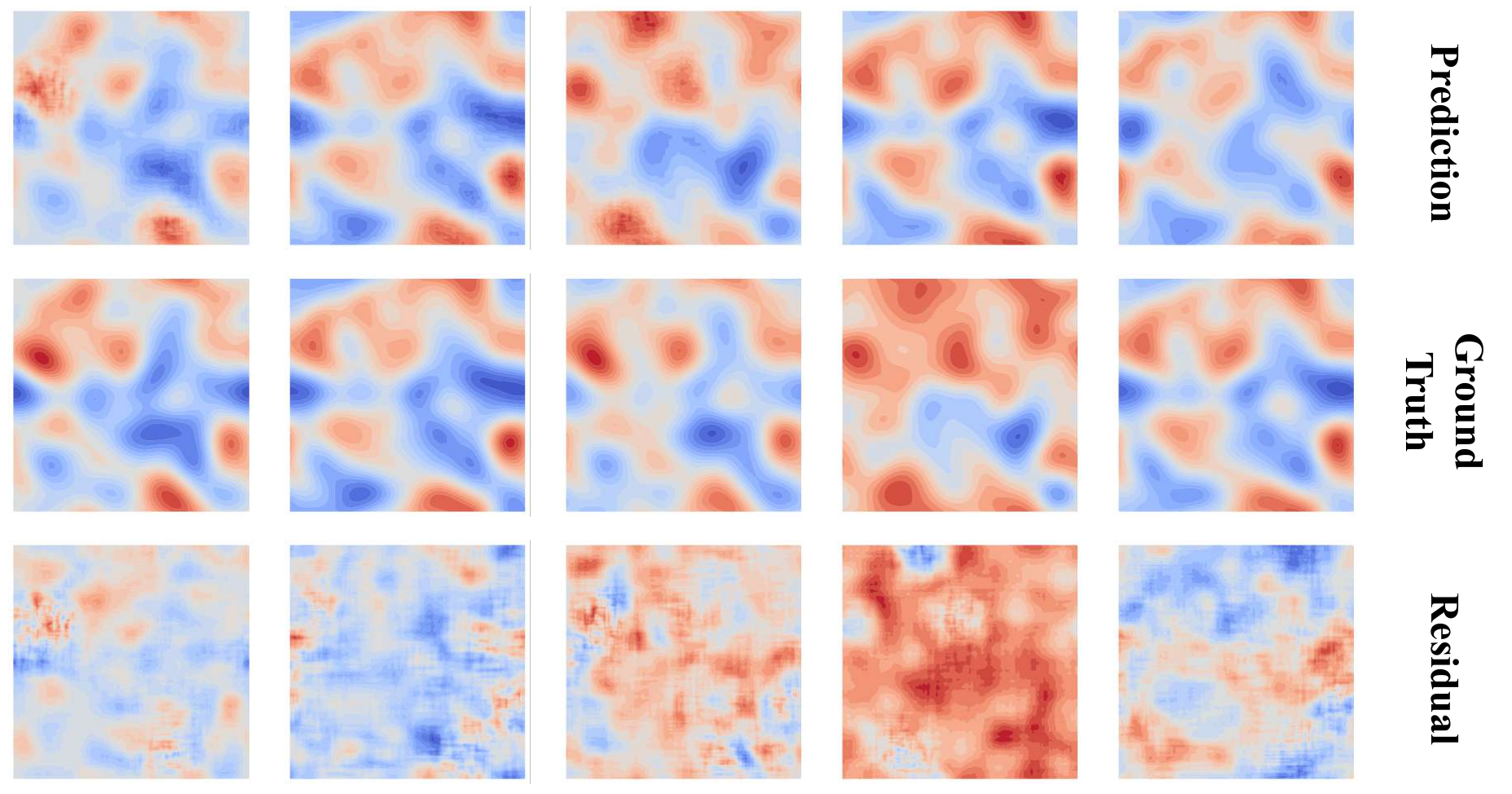}
\caption{Visualization of Latent$+$Next-step$+$Variable on Stochastic N-S .}
\label{fig:vis3}
\end{figure*}

In this section, we provide more visualizations for the simulated fluid and showcases as a supplement. We choose some specific methods and fluid types for demonstration. For example, we show the visualization results of Fourier-based methods on Compressible N-S in Figure \ref{fig:vis_cns_1}, \ref{fig:vis_cns_2} and \ref{fig:vis_cns_3} for prediction, ground truth and residual gap respectively. Besides, we also provide results of the Fourier-based method on Diffusion-Reaction in Figure \ref{fig:vis_dr}. And visualizations of three different methods on Stochastic N-S are shown in Figure \ref{fig:vis1}, \ref{fig:vis2}, and \ref{fig:vis3}.

\section{Limitations and Broad Impact}
\label{appendix:limit}
Despite our efforts to cover a diverse set of architectures, our benchmark is limited by time constraints and thus does not yet include all possible representation paradigms (\textit{e.g.}, higher-order spectral methods, graph-wavelet embeddings, or learned Lagrangian bases). Expanding to these and other emerging modalities remains an important direction for future work.

Another limitation is that our experiments are mainly based on 2D PDEs. Prior works~\citep{lifourier,li2023fourier} have shown that the same operator-based architectures scale to 3D turbulence, but at significantly higher costs. Our focus here is to first establish a fair, reproducible 2D benchmark, which we view as a necessary prerequisite for systematic 3D evaluation. Although our current experiments focus on 2D systems, the modular decomposition and evaluation pipeline are dimension-agnostic and can be directly extended to 3D settings, subject to available computational resources.

Nonetheless, we have released our full evaluation pipeline as a modular, well-documented codebase that makes it straightforward to incorporate new model classes, data modalities, or loss functions. This design ensures that researchers can readily extend the benchmark, plugging in novel representations and comparators with minimal effort. In this way, our framework lays the groundwork for a fair, reproducible, and easy-to-use codebase and standard for comparing data-driven fluid simulation methods going forward.

\clearpage

\end{document}